\documentclass{aa}

\usepackage{graphicx}
\usepackage{txfonts}
\usepackage{lscape}
\usepackage{color,xcolor,xspace,pdflscape,dblfloatfix}

\newcommand{\equ}[1]{Eq.~\ref{eq:#1}}

\newcommand{\fig}[1]{Fig.~\ref{fig:#1}}

\newcommand{\tab}[1]{Table~\ref{tab:#1}}

\newcommand{\sect}[1]{Sect.~\ref{sec:#1}}
\newcommand{\app}[1]{Appendix~\ref{app:#1}}

\newcommand{\lcdm}[0]{$\Lambda$CDM\xspace}
\newcommand{\rbr}[0]{\ensuremath{r_\mathrm{br}}\xspace}
\DeclareMathOperator\erf{erf}
\newcommand{\reqM}[0]{\ensuremath{r_{\!a_0\mathrm{, M}}}\xspace}
\newcommand{\reqN}[0]{\ensuremath{r_{\!a_0\mathrm{, N}}}\xspace}
\newcommand{\req}[0]{\ensuremath{r_{\!a_0}}\xspace}

\makeatletter
\AddToHook{env/tabular/begin}{\let\input\@@input}
\makeatother

\begin{document} 

  \title{The galactic acceleration scale is imprinted on  globular cluster systems of early-type galaxies of most masses and on red and blue globular cluster subpopulations}
    
\titlerunning{GCS breaks in Fornax Cluster}

   \author{Michal B\'ilek\inst{1,2,3}
   \and Michael Hilker\inst{3}
   \and Florent Renaud\inst{4,5}
   \and Tom Richtler\inst{6}
   \and Avinash Chaturvedi\inst{3}
   \and Srdjan Samurovi\'c\inst{7} }

   \institute{LERMA, Observatoire de Paris, CNRS, PSL Univ., Sorbonne Univ., 75014 Paris, France\\
    \email{michal.bilek@obspm.fr}
        \and
        Coll\`ege de France, 11 place Marcelin Berthelot, 75005 Paris, France
        \and
        European Southern Observatory, Karl-Schwarzschild-Strasse 2, 85748 Garching bei M\"unchen, Germany
        \and
        Department of Astronomy and Theoretical Physics, Lund Observatory, Box 43, SE-221 00 Lund, Sweden
        \and
        University of Strasbourg Institute for Advanced Study, 5 all\'ee du G\'en\'eral Rouvillois, F-67083 Strasbourg, France
        \and
        Departamento de Astronomia, Universidad de Concepci\'on, Concepci\'on, Chile
        \and
        Astronomical Observatory of Belgrade, Volgina 7, 11060 Belgrade, Serbia
             }

   \date{Received ...; accepted ...}

 
   \abstract
   {Globular clusters (GCs) carry information about the formation histories and gravitational fields of their host galaxies. B\'ilek et al. (2019, BSR19 hereafter) reported that the radial profiles of the volume number density of GCs in GC systems (GCSs) follow broken power laws, while the breaks occur approximately at the $a_0$ radii. These are the radii at which the gravitational fields of the galaxies equal the galactic acceleration scale $a_0 = 1.2\times10^{-10}\,$m\,s$^{-2}$ known from the radial acceleration relation or the MOND theory of modified dynamics. }
   {Our main goals here are to explore whether the results of BSR19 hold true for galaxies of a wider mass range and for the red and blue GC subpopulations.}
   {We exploited catalogs of photometric GC candidates in the Fornax galaxy cluster based on ground and space observations  and a new catalog of  spectroscopic GCs of NGC\,1399, the central galaxy of the cluster. For every galaxy, we obtained  the parameters of the broken power-law density by fitting the on-sky distribution of the GC candidates, while allowing for a constant density of contaminants. The logarithmic stellar masses of our galaxy sample span $8.0-11.4\,M_\sun$.}
   {All investigated GCSs with a sufficient number of members show broken power-law density profiles. This holds true for the total GC population and the blue and red subpopulations. The inner and outer slopes and the break radii agree well for the different GC populations. The break radii agree with the $a_0$ radii typically within a factor of two for all GC color subpopulations. The outer slopes correlate better with the $a_0$ radii than with the galactic stellar masses. The break radii of NGC\,1399 vary in azimuth, such that they are greater toward and against the direction to NGC\,1404, which tidally interacts with NGC\,1399. }
   {}

               
   \maketitle
%
\section{Introduction}
Globular clusters (GCs) are compact (a few parsecs), massive ($10^4-10^6\,M_\sun$) star systems found in nearly all galaxies. A galaxy similar to the Milky Way has a few hundreds of them, while giant ellipticals can have more than ten thousand GCs. {The colors of many galaxies form a bimodal distribution, with rather universal positions for the two peaks. Therefore, GCs are divided into two types -- the metal-poor ``blue GCs'' and the metal-rich ``red GCs'' \citep{Brodie2006, cantiello20}. 
Red GCs generally follow the kinematics of the stars in a galaxy, with a similar rotational velocity and velocity dispersion. In contrast, blue GCs often show complex kinematics \citep{schuberth10, Coccato2013, Chaturvedi22}. The spatial distribution of GCs around galaxies is more centrally concentrated for the red GCs than for the blue GCs. The distinct properties of the red and blue GCs point toward their different formation pathways \citep{ashman95, peng06, Brodie2006}. It seems that the blue GCs are added to massive galaxies via accretion of low-mass galaxies, while most red GCs form in situ, together with the stars of the host galaxy {\citep{cote98,harrissaasfee,tonini13,renaud17}}. Globular cluster systems thus carry information about the assembly history of their host galaxies \citep{Peng2008, Brodie2014, Harris2016}. The number of GCs that a galaxy hosts is proportional to the expected mass of its dark matter halo \citep{spitler09,harris15}. The kinematics and distribution of GC systems (GCSs) reflect the profiles of the gravitational fields of their host galaxies  \citep{samur14,samur16,alabi17,bil19}. } 

In the paper by \citet{bil19b} (BSR19 hereafter), an interesting new property of GCSs of early-type galaxies was noted. They parametrized\footnote{{This parametrization originates from \citet{bil19}. In that paper, it was used for practical reasons. It could be implemented easily in the numerical solver of the Jeans equation and it described the observed projected profiles of density of GCSs well.}} the volume number density of GCs in a GCS, $\rho$ by a broken power law as 
\begin{equation}
\begin{aligned}
\rho(r) &  = \rho_0\ r^a &\quad\textrm{ for }\quad r<r_\mathrm{br},\\
\rho(r) &  = \rho_0\ r_\mathrm{br}^{a-b}\ r^b &\quad\textrm{ for }\quad r\geq r_\mathrm{br},
\end{aligned}
\label{eq:bpl}
\end{equation} 
where $r$ is the galactocentric radius. The parameter \rbr was called the break radius.
The authors found that the break radius coincides well with the $a_0$ radius, that is the radius at which the {expected} gravitational acceleration generated by the baryons of the galaxy equals the galactic acceleration scale $a_0 = 1.2\times10^{-10}\,$m\,s$^{-2}$. {The values of the break radii did not agree with the values of the other characteristic lengths of the galaxies, such as stellar effective radii or dark halo scale radii. These other lengths were either several times bigger or smaller than the break radii, at least for a substantial fraction of the galaxy sample (see their Table 1).}

{The galactic acceleration scale is known best from the behavior of the observed gravitational fields of galaxies \citep[e.g.,][]{lelli17,li17}. In the regions of galaxies where the gravitational acceleration expected by Newtonian gravity from the distribution of baryons  $g_\mathrm{N}$ is greater than $a_0$, the observed gravitation acceleration equals  $g_\mathrm{N}$, meaning that the Newtonian dynamics does not require dark matter. On the other hand, in the regions where $g_\mathrm{N}$ is lower than $a_0$, the observed gravitational acceleration is very close to $\sqrt{g_\mathrm{N}a_0}$. The same rules apply even for many, or perhaps all, GCs \citep{scarpa10,scarpa11,ibata11,sanders12,hernandez12,hernandez20}. }

{This behavior was initially predicted by the modified Newtonian dynamics (MOND), which is a class of modified gravity and inertia theories \citep{milg83a}. Here we assume MOND to be a modified gravity theory. It predicts that the gravitational acceleration in spherical isolated objects is} \citep{milg83a,qumond,famaey12}
\begin{equation}
    g_\mathrm{M} = g_\mathrm{N}\,\nu(g_\mathrm{N}/a_0).
    \label{eq:mond}
\end{equation}
The function $\nu$ is not known exactly, but it must have the limit behavior $\nu(x)\sim x^{-1/2}$ for $x\ll 1$, and $\nu(x)\sim 1$ for $x\gg 1$. This gives rise to two regimes of a gravitational field around a galaxy: the strong field, the so-called Newtonian regime, and the weak field, the so-called deep-MOND regime. {The observed counterpart of \equ{mond} is known as the radial acceleration relation \citep{mcgaugh16}.}

Apart from the radial acceleration relation, {MOND predicted or explained many other observational laws \citep{milg83a, milg83b, milg83c}, all of which contain the constant $a_0$.} This is the case for the baryonic Tully-Fisher relation \citep{mcgaugh20,lelli19}, the Faber-Jackson relation \citep{faber76,famaey12}, and the radial acceleration relation, which connect the mass or mass distribution of galaxies to the velocities of stars and gas in them. The Fish law \citep{fish64,allen79} and Freeman limit \citep{freeman70,mcgaugh95,fathi10,famaey12} give upper limits on the surface brightness for elliptical and spiral galaxies, respectively, above which galaxies are rare. Recently, there appeared a MOND explanation  \citep{milg21} of the Fall relation \citep{fall83,posti18}, which connects the mass and specific angular momentum of galaxies. {The law of the universal surface density of the cores of the putative dark matter halos \citep{kormendy04,donato09,salucci12} can be explained by MOND as well \citep{milg09c}.}

Finally, there are interesting numerical coincidences of $a_0$ with the constants of cosmology   \citep{milg83a,milg20}. If we denote $H_0$ the Hubble constant, $c$ the speed of light, $G$ the gravitational constant, $R_H$ the size of the  cosmic horizon, and $M_H$ the total mass inside the cosmic horizon, then we find the order-of-magnitude equalities $a_0 \approx cH_0 \approx c^2\Lambda^{1/2} \approx c^2/R_H \approx c^4/GM_H$. There is no clear explanation of these coincidences yet \citep{navarro17,milg20}. 

{The finding of BSR19, of the equality of the break and $a_0$ radii, is thus another case of the many occurrences of the constant $a_0$ in extragalactic astronomy.} { More precisely, in this work we consider two types of $a_0$ radii: i) the one where the acceleration calculated from the distribution of baryons and  Newtonian gravity equals $a_0$ and ii) where the acceleration calculated for MOND gravity via \equ{mond} equals $a_0$. For most galaxies, the two $a_0$ radii are numerically similar. Therefore, in this paper, if we do not specify whether we are referring to  Newtonian or MOND $a_0$, then we mean that the statement is valid for both options.}

\begin{table*}
\caption{Parameters of the investigated galaxies.}          
\label{tab:gals} 
\centering            
\begin{tabular}{lllllll}
\hline\hline
Name & FDS ID & $\log_{10}\frac{M_*}{M_\sun}$  & $R_\mathrm{e}$  & $n$ & \reqM & \reqN \\
 & & & [arcmin] &  & [kpc] & [kpc] \\
\hline
ESO358-006  &  FDS19\_0001  &  $9.40 \pm 0.09$  &  0.25  &  1.1  &  $1.44^{+0.5}_{\mbox{~~--}}$  &  --  \\
 ESO358-050  &  FDS4\_0001  &  $9.5 \pm 0.1$  &  0.32  &  1.8  &  $1.50^{+0.4}_{-0.4}$  &  $0.43^{+0.3}_{\mbox{~~--}}$  \\
 NGC1316  &  FDS26\_0001  &  $11.40 \pm 0.06$  &  1.0  &  4.3  &  $21.5^{+2}_{-2}$  &  $14.3^{+1}_{-1}$  \\
 NGC1336  &  FDS20\_0000  &  $9.7 \pm 0.1$  &  0.70  &  4.6  &  $1.63^{+0.3}_{-0.3}$  &  $0.89^{+0.2}_{-0.2}$  \\
 NGC1351  &  FDS19\_0000  &  $10.30 \pm 0.07$  &  0.75  &  5.9  &  $4.65^{+0.5}_{-0.5}$  &  $2.91^{+0.3}_{-0.3}$  \\
 NGC1373  &  FDS16\_0002  &  $9.40 \pm 0.09$  &  0.16  &  3.9  &  $1.95^{+0.3}_{-0.3}$  &  $1.24^{+0.2}_{-0.2}$  \\
 NGC1379  &  FDS11\_0002  &  $10.40 \pm 0.06$  &  0.47  &  2.7  &  $6.57^{+0.7}_{-0.6}$  &  $4.11^{+0.5}_{-0.4}$  \\
 NGC1380  &  FDS11\_0006  &  $10.90 \pm 0.05$  &  0.77  &  3.0  &  $11.79^{+0.9}_{-0.9}$  &  $7.51^{+0.7}_{-0.6}$  \\
 NGC1380B  &  FDS11\_0005  &  $9.7 \pm 0.1$  &  0.28  &  1.9  &  $2.65^{+0.5}_{-0.5}$  &  $1.40^{+0.4}_{-0.3}$  \\
 NGC1381  &  FDS11\_0004  &  $10.20 \pm 0.07$  &  0.31  &  2.6  &  $5.48^{+0.6}_{-0.5}$  &  $3.53^{+0.4}_{-0.4}$  \\
 NGC1387  &  FDS11\_0001  &  $10.70 \pm 0.05$  &  0.61  &  5.5  &  $8.90^{+0.7}_{-0.6}$  &  $5.84^{+0.5}_{-0.4}$  \\
 NGC1399  &  FDS11\_0003  &  $11.40 \pm 0.06$  &  2.6  &  8.1  &  $16.9^{+2}_{-1}$  &  $10.9^{+1}_{-1}$  \\
 NGC1404  &  FDS11\_0166  &  $11.00 \pm 0.06$  &  0.54  &  4.0  &  $14.1^{+1}_{-1}$  &  $9.49^{+0.8}_{-0.7}$  \\
 NGC1419  &  FDS13\_0000  &  $9.90 \pm 0.09$  &  0.16  &  4.0  &  $3.96^{+0.5}_{-0.5}$  &  $2.65^{+0.4}_{-0.3}$  \\
 NGC1427  &  FDS6\_0001  &  $10.50 \pm 0.06$  &  0.83  &  5.0  &  $6.11^{+0.6}_{-0.5}$  &  $3.80^{+0.4}_{-0.4}$  \\
 NGC1428  &  FDS6\_0002  &  $9.6 \pm 0.1$  &  0.19  &  1.7  &  $2.72^{+0.4}_{-0.4}$  &  $1.64^{+0.3}_{-0.3}$  \\
 Stack\_8.0  &  --  &  $8.0 \pm 0.1$  &  0.24  &  1.1  &  --  &  --  \\
 Stack\_8.5  &  --  &  $8.40 \pm 0.08$  &  0.23  &  1.3  &  --  &  --  \\
 Stack\_9.0  &  --  &  $9.00 \pm 0.08$  &  0.26  &  2.0  &  $0.36^{+0.2}_{-0.2}$  &  --  \\
 
\hline      
\end{tabular}
\tablefoot{Column 1: Common name of the galaxy. Column 2: Designation of the object in the FDS. Column 3: Decadic logarithm of the stellar mass of the galaxy in solar units. Column 4: Effective radius of the galaxy. Column 5: S\'ersic index of the galaxy. Column 6: MOND $a_0$ radius. Column 7: Newtonian $a_0$ radius. }
\end{table*}

The theoretical explanation for why the $a_0$ radii coincide with the break radii of GCSs has not been clarified yet, even if some initial proposal were given in BSR19. Importantly, according to one of the proposed explanations that involves the Newtonian gravity and dark matter, the match of the $a_0$ and break radii is of practical importance. It had been found before that the number of GCs that a galaxy has is proportional to the mass of its dark matter halo \citep{spitler09,harris15}. The new finding allows one to estimate the scale radius of the halo: the break radius should be located at the radius where the gravitational attraction of the stars of the galaxy equals that of the dark matter halo. One can thus solve the equation of the equity of the accelerations to obtain the scale radius of the halo.

The paper BRS19 left several important questions open, which we aim to answer here. We investigate the distribution of GCs primarily using photometric data for early-type galaxies in the Fornax galaxy cluster, but we also analyze new spectroscopic data for two galaxies.  The galaxy sample of BSR19 spanned only about one order of magnitude in stellar mass. The data investigated here allow us to verify that the match between the $a_0$ and break radii holds true for early-type galaxies spanning three orders of magnitude -- from  dwarfs of the mass of the Magellanic clouds to the brightest cluster galaxies. The distribution of GCs was investigated in BSR19 on the basis of the catalog of spectroscopically confirmed GCs. This approach can lead to distorted results, because spectroscopic surveys usually are spatially incomplete. This could have affected the derived parameters of the broken power-law profile. The verification of the match of the $a_0$ and break radii in the new data thus removes the shade of doubt that was left about the results of the previous work. The data in BSR19 do not allow the density profiles of the red and blue GC subpopulations to be investigated. We do this here and find that there are no statistically significant trends of the profile parameters with the color of the GCs. We exploit the new data for a further exploration of the profiles of GCSs. In particular, we investigate whether the parameters of the broken power-law profiles correlate with each other and with the parameters of the host galaxy. We also investigate in detail the GCS of NGC\,1399, that is the central galaxy of the Fornax cluster. We find that  its break radius depends on the relative velocity of the GCs with respect to the center of the galaxy, and that the break radius varies as a function of the position angle.  In this paper, we also aim to explain the reason of why the GCSs of our galaxies have the broken power-law density profiles and why the break radii coincide with the $a_0$ radii. Several explanations were proposed in BSR19, and here we add a few more. Then we make the first steps toward finding which of them is correct. None seem perfect at this point.

This paper is organized as follows. In \sect{data} we describe the {observational} data that we analyze here. The methods to extract and fit the radial profiles of volume number densities of GCs in the GCSs of the investigated galaxies are detailed in \sect{fitting}. The derivation of the $a_0$ radii for our sample galaxies is explained in \sect{ao}. We present our results in \sect{results}. In particular, in \sect{colors} we compare the structural parameters of GCSs for the total GC population and the red and blue subpopulations. In \sect{corr} we investigate the correlations of the structural parameters between each other and with the characteristics of the host galaxies and, finally, we explore the relation between the break and $a_0$ radii in \sect{equality}. We explore the details of the distribution of GCs of NGC\,1399 in \sect{n1399}. We explore the credibility of some potential explanations of the approximate coincidence of the $a_0$ and break radii in \sect{interpretation}. Finally we synthesize and summarize our findings in \sect{sum}. In this work, we denote the natural logarithm by $\log$ and the logarithm of the base $m$ by $\log_{m}$.  We assume the distance of the Fornax cluster and all the investigated galaxies to be 20.0\,Mpc \citep{blakeslee09}. This corresponds to the angular scale of 5.8\,kpc per arcminute.

\section{Data analyzed}
\label{sec:data}

The galaxies investigated in this paper are member galaxies of the Fornax galaxy cluster. They are listed in \tab{gals}. Low-mass Fornax cluster galaxies ($M_*<10^{9.5} M_\sun$) have too few GCs for constructing the density profiles of their GCSs individually. Therefore, we stacked the GC candidatess of many low-mass galaxies in three mass bins in order to get their average GCS density profiles. These are the three ``Stack'' entries in \tab{gals}. The details of the stacking procedure are explained in \sect{fds}.

All structural parameters of the galaxies in \tab{gals} were taken from \citet{su21}. They are based on GALFIT profile fitting \citep{penggalfit}, using S\'ersic functions,  to photometric $g'$-band data of the Fornax Deep Survey (FDS) \citep{iodice16,venhola18}. Stellar masses also come from that work. They were derived from empirical relations between colors and stellar mass-to-light ratio. We use archival photometric and spectroscopic catalogs of GC candidates for investigating their spatial distribution. Here follows a brief description of the datasets investigated in our work.

\subsection{ACS Fornax cluster survey data}

The ACS Fornax cluster survey (ACSFCS), taken using the Advanced Camera for Surveys (ACS) of the Hubble Space Telescope (HST),  imaged 43 early-type galaxies of the Fornax cluster. Full details of the ACSFCS, scientific motivations and data reduction techniques, are given in \citet{jordan07}. Each galaxy in ACSFCS was imaged in the F475W ($g$) and F850LP ($z$) bands. For studying the GCs, each image was sufficiently deep such that 90\% of the GCs within the ACS FOV can be detected.
The selection and identification of bonafide GCs of the ACSFCS galaxies are performed in the size-magnitude plane \citep{jordan15}. The resulting catalog of GCs in the ACSFCS provides the probability of an object being a GC, denoted by Pgcs, where Pgcs is a function of half-light radius, apparent magnitude and local background. For our analysis, we selected GCs with Pgcs larger than 0.5. {This leaves the faintest GC candidates of $m_g=26.3$\,mag.} To separate the GCs into red and blue subpopulations, we adopted a dividing $g-z$ color of 1.1 mag \citep{fahrion20}.

\subsection{Spectroscopically confirmed GCs}

We {studied} the spectroscopically confirmed sample of GCs of the Fornax cluster from the recent catalog produced by \cite{Chaturvedi22}. 
Reanalyzing data of the Fornax cluster taken using the Visible Multi-Object Spectrograph (VIMOS) at the Very Large Telescope (VLT) \citep{pota18} and adding literature work, they have produced the most extensive GC radial velocity catalog of the Fornax cluster \citep[see][for details]{Chaturvedi22}, comprising more than 2300 confirmed GCs. {The faintest GC has $m_g=24.2$\,mag.} They used a Gaussian modeling mixture technique to divide the GC population into red and blue GCs, with $g-i \sim 1.0$\,mag as separating color, which we adopt in our analysis.

\subsection{Fornax Deep Survey data}
\label{sec:fds}

The Fornax Deep Survey (FDS) is a joint project based on a guaranteed time observation of the FOCUS (P.I. R. Peletier) and VEGAS (P.I. E. Iodice, \citealp{Capaccioli15}) surveys. It consists of deep multiband ($u$, $g$, $r$ and $i$ ) imaging data from OmegaCAM at the VST \citep{kuijken11, Schipani12} and covers an area of  30 square degrees out to the virial radius of the Fornax cluster. 

We applied morphological and photometric selection criteria to the photometric compact sources catalog of the FDS  \citep{cantiello20} to decrease the fraction of contaminant objects that are not GCs. The criteria on the colors of the selected objects $g-i>0.5$
and $g-i<1.6$ were chosen according to the colors of spectroscopically confirmed GCs around the central cluster galaxy NGC\,1399. 
Further criteria were inspired by cross-matching the FDS and 
ACS catalogs of GCs of NGC\,1399. They were chosen such
that we do not exclude too many real GCs and, at the same time, exclude as many contaminants as possible. We found a good balance when using the following criteria: CLASS\_STAR>0.031, $m_g>20$ mag, and Elongation<3.
The meaning of the parameters is explained in \citet{cantiello20}. {After applying the selection criteria, the faintest GC candidate in this calatog has $m_g=27.0$\,mag.}   

\begin{figure}
        \resizebox{\hsize}{!}{\includegraphics{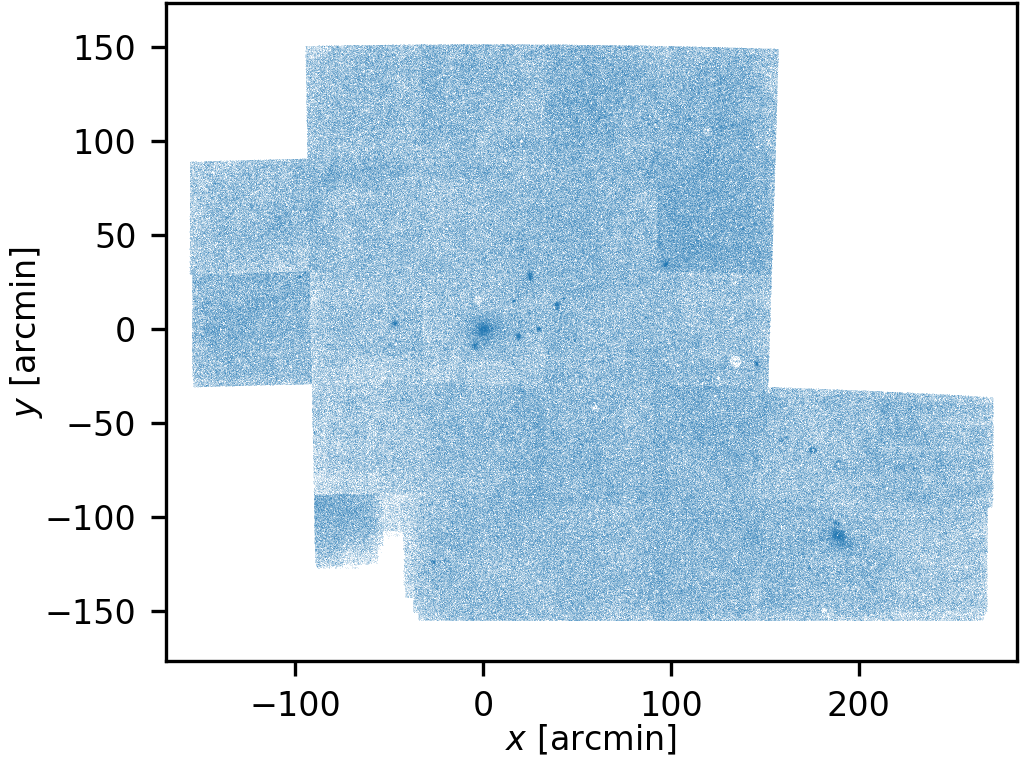}}
        \caption{Positions of GC candidates in the FDS catalog, after applying selection criteria. The coordinate system is centered on NGC\,1399. North is up, and west is to the right.}
        \label{fig:tiles}
\end{figure}

It turned out that the FDS catalog, after applying the GC candidate selection criteria, shows systematic variations of the surface density of sources that have a tile-like pattern, see \fig{tiles}. The tiles corresponds to the OmegaCAM imaging tiles of the FDS survey. They probably result from varying observing conditions, like seeing, sky transparency, etc. When fitting the surface density profiles of GCSs of individual galaxies, we had to be careful that the tile borders do not introduce any kinks in the profiles. We did this first by visual inspection of plots of positions of the GC candidates  in the wide neighborhood of the investigated galaxies and second by visual inspection of the plots of the radial profiles of density of the GC candidates around the target galaxies.

\subsubsection{Stacking of faint galaxies}
As mentioned above, we decided to stack the faint galaxies of similar stellar mass in order to have enough sources to extract the density profiles of their GCSs. We used only the FDS data for this, since most faint galaxies were not covered by the ACSFCS. If the hypothesis of this paper, that the break radii coincide with the acceleration radii, is correct, then the break radii of all GCSs stacked in this way should be roughly equal in each mass bin (an acceleration radius nevertheless depends also on the particular distribution of  mass in the given galaxy). We then treated the stacks as single galaxies. We created three such artificial objects. Their logarithmic masses were centered on the values of 8.0, 8.5 and 9.0\,$M_\sun$. The widths of all logarithmic mass bins were 0.5\,$M_\sun$. Stacking of galaxies of even lower masses did not provide sufficiently clear profiles of the projected density of GCs.

We stacked all galaxies from the \citet{su21} catalog that met the following criteria. First we excluded spiral galaxies, because our data do not allow to distinguish between GCs and star forming clumps or young star clusters in their disks.
Early-type galaxies were identified by requiring their asymmetry parameter, stated in \citet{su21}, to be lower than 0.06. We further excluded galaxies that are located close to the borders of  the tiles of the FDS mosaic, and galaxies whose GCSs surface density profiles might be affected by interloping GCSs of other nearby galaxies. These were identified by visual inspection of the positions of the GC candidates in the neighborhood of the galaxies being stacked. The list of stacked objects in every mass bin is stated in \tab{fits} in \app{fits}. 

We assigned to every stacked galaxy a stellar mass and S\'ersic parameters. These were obtained as the median values for all galaxies included in the corresponding stack.

\section{Fitting the surface density profiles of the GCSs by analytic functions}
\label{sec:fitting}

Here we describe how we determined for the investigated galaxies the parameters of the GCSs density profiles. 

\subsection{Extracting the observed profiles of surface density}
\label{sec:extracting}

For a given galaxy, we divided the GC candidates in radial bins according to their galactocentric distance, and for each bin, we calculated the surface density of sources in it. The bins had the shape of circular annuli, that is we ignored the possible ellipticity of the GCSs. 
We demonstrate in \app{ellipticity} that this simplification has no appreciable impact on the derived profile of the GCS density.
We chose the widths of the radial bins such that 
\begin{equation}
    N = n-A\overline{\gamma}
    \label{eq:bincond}
\end{equation} 
was constant in each bin. Here $n$ stands for the total number of sources falling in the annulus, $A$ the area of the annulus  and $\overline{\gamma}$ an estimate of the surface density of contaminants. The number $N$ is thus the estimate of the number of GCs in the aperture. The parameter $N$ was chosen to be big enough so  that the profile did not appear too noisy (as judged by eye) and small enough so that the two straight parts of the broken power law profile were resolved by at least three data points.  The purpose of subtracting the expected number of contaminants was to increase the signal to noise in the outer bins. This condition was used only for choosing the bin widths; the final profile parameters of the GCSs (including the contaminants level) were derived from fitting the surface density of all sources, that is $n$.

For the datasets that are not expected to contain many contaminants, that is the spectroscopic and ACS catalogs, we used simply $\overline{\gamma}=0$. Otherwise the value of $\overline{\gamma}$ was found iteratively in the following way. We first chose a low value of $\overline{\gamma}$  and constructed the observed surface density profile.  That was then fitted by some of the analytic model profiles described below. One of the parameters  of the fitted profile was the density of the background sources, $\gamma$.  In the next iteration we used a value of $\overline{\gamma}$ that lied between the value of $\overline{\gamma}$ used in the  previous iteration and the fitted value of $\gamma$. We had to stop increasing $\overline{\gamma}$ at some point, because when it is too large, the extracted surface density profile would be truncated or distorted, because the expression on the right-hand-side of the condition \equ{bincond} would be lower than the assigned value of $N$ at large radii. In some cases, the final $\overline{\gamma}$ was chosen such that the extracted profile did not show any small bumps arising from small clusters of contaminants (e.g., distant galaxy clusters in the background). 
We note that if the definitive $\overline{\gamma}$ was chosen to be somewhat below the true value, the resulting extracted profile would not be affected substantially; only the bins would not have the optimal widths. 

{The contamination by the light of the host galaxy makes it difficult to detect GCs near its center. In order to detect them, we have to subtract a model of the light of the galaxy from the image. The model is typically not perfect, and the fit residuals in the difference image still complicate the detection of the GCs. Furthermore, the light of the galaxy introduces photon noise, that decreases the signal-to-noise ratio of the GCs. Fainter GCs are affected more. This can deform the observed GCS density profile.  The ground-based FDS data are more sensitive to these problems than the HST data: the GCs in HST images appear sharper because of no atmospeheric seeing effects, and therefore reach a higher signal-to-noise ratio.  We indeed found the signature of this in our data: the inner slopes of the profiles of surface density of GC candidates were shallower in the FDS data than in the HST data. Examples can be seen in \fig{g14} (NGC\,1427) or \fig{g8} (NGC\,1380B). }

We were using two strategies to mitigate the {flattening
of the inner GC density profile because of the problems with the contamination by the light of the host galaxy}. If there were HST data available in the central region, we adopted the inner slope $a$ derived from these data. {For galaxies without HST data, we constructed the GCS surface density profiles only from bright-enough sources in FDS. We determined the magnitude cut using a plot magnitude - projected galactocentric distance of the sources. The limiting magnitude was set as that of the faintest sources that were still detectable in the very center of the galaxy.
This approach has the downside that it reduces substantially the number of GC candidates, such that it becomes harder to trace the GC density profile, particularly at large galactocentric distances. In \app{magcut} we demonstrate on the example of NGC\,1399 that the break radius does not depend significantly on the magnitude cut.}

The surface density profiles of the GCSs were extracted and analyzed only within a restricted range of the projected galactocentric radius defined by the limits $r_\mathrm{min}$ and $r_\mathrm{max}$.  The upper limit, $r_\mathrm{max}$, was usually enforced either by the proximity to the GCSs of other galaxies, or because of the fluctuations of the density of contaminants, or because there was a border of tiles of the FDS mosaic. {The lower limit, $r_\mathrm{min}$, was usually taken to be zero, but in some galaxies, it was used to reduce the problems with the contamination by the light of the host galaxy.} The radii $r_\mathrm{min}$ and $r_\mathrm{max}$ were determined by a visual inspection of both the map of the GC candidates near the inspected galaxy  and the surface density profile of the GC candidates.

When constructing the GCS surface density profiles for a galaxy, we had to exclude the areas occupied by the GCSs of neighboring galaxies. We excluded sources close to the following major galaxies (unless studying the GCSs of these galaxies themselves): ESO\,358-33, NGC\,1396, NGC\,1317, NGC\,1373, NGC\,1374, NGC\,1375, NGC\,1379, NGC\,1380, NGC\,1381, NGC\,1382, NGC\,1386, NGC\,1387, NGC\,1389, NGC\,1404, NGC\,1427 and NGC\,1427A. We marked all sources closer than 3\arcmin\xspace to these galaxies as excluded. This distance was chosen by visual inspection of the map of positions of all sources. We note that we did not exclude the area occupied by the GCS of NGC\,1399 because it is very extended. If we did so, it would not be possible to construct the observed surface density profiles of several smaller galaxies in the vicinity of NGC\,1399. Instead, whenever reasonable, we just assumed that the GCs of NGC\,1399 are distributed with a constant surface density in the vicinity of the investigated smaller galaxies, so that we could treat them as additional contaminants. This approach was not applied to NGC\,1404, which has a rather extended GCS and is located very close in projection to NGC\,1399. Due to the changing number density of NGC\,1399 GCs across the area of NGC\,1404, we could not consider them as uniformly distributed contaminants and we had to use another method, see \sect{n1404}.

For a few galaxies, we had to mark as excluded sources in the regions of blemishes of the FDS survey. Most often these were regions around bright stars that appear as holes in the maps of the positions of sources from the FDS catalog.

In general, when calculating surface density in a given annulus, we did not consider sectors in which there was at least one excluded source. In total, the observed surface density of sources in a given annulus was calculated as
\begin{equation}
    \Sigma_\mathrm{obs} = \frac{2\pi}{(2\pi-\alpha)}\frac{n}{A}, 
\end{equation}
where $\alpha$ stands for  the sum of angular extents of all excluded sectors, $n$ for the number of sources in the not-excluded sectors, and $A$ for the total area of the inspected annulus. Given that the number of sources in a given area follows the Poisson distribution, we estimated the uncertainty of the surface density of sources as
\begin{equation}
    \Delta\Sigma_\mathrm{obs} = \frac{2\pi}{(2\pi-\alpha)}\frac{\sqrt{n}}{A}.
\end{equation}

\subsubsection{The special case of NGC\,1404}
\label{sec:n1404}
The galaxy NGC\,1404 lies in projection close to the central cluster galaxy NGC\,1399. Their GCSs overlap. The GCS of NGC\,1399 has a strong gradient of surface density at the position of NGC\,1404. The method described above would not allow us to produce a reliable GC density profile of this galaxy. 
{Instead, we use the catalog of spectroscopic GCs by \cite{Chaturvedi22} for NGC\,1404 and assume that it covers the  galaxy  homogeneously. }
We could then make use of not only the information about the spatial positions of the sources, but also the information about their radial velocities, since the radial velocity of NGC\,1404 is $\sim$520 km\,s$^{-1}$ larger than that of NGC\,1399.   
We first applied spatial criteria on the sources to be used for constructing the observed profile of the GCS. They are depicted in the upper-right panel of \fig{g13}. We avoided regions that are closer than 8$\arcmin$ to NGC\,1399. We also avoided the region that is closer than $10\arcmin$ to the  point with the J2000 coordinates (54.86017, -35.75988), because this region seemed to suffer from geometrical incompleteness  of the spectroscopic survey because the density of sources in this region was lower than in its surroundings. Just as for all other galaxies, we applied a limit for the maximum distance of the used sources from the galaxy. We also applied a radial-velocity limit: all sources that had radial velocities lower than the center of NGC\,1404, 1947\,km\,s$^{-1}$, were excluded, as shown in the bottom-right panel of \fig{g13}. This helped us to reduce substantially the contamination by the GCs of NGC\,1399. While the observed number density profile was constructed from only 23 sources and there were only 2-3 sources per bin, the break in the profile is clearly visible and the break radius follows the same correlations as the break radii of the other galaxies.

\subsection{Models of density and surface density profiles of GCSs}
The extracted profiles of surface density of GCs candidates were fitted by one of the analytic functions described in this section.  We made use of the fact that for a spherically symmetric GCS, the 3-dimensional density profile $\rho$  corresponds to the projected surface density profile $\Sigma$ given by the Abel transform:
\begin{equation}
    \Sigma(R)=2\int_R^\infty \frac{\rho(r)r}{\sqrt {r^2-R^2}}\,\mathrm{d}r.
    \label{eq:abel}
\end{equation}
Here  $r$ stands for actual galactocentric distance and $R$ for the projected galactocentric distance.

We considered several types of volume density profiles. The first was a pure broken power law given by \equ{bpl}. Its corresponding surface density profile is given by the equations

\begin{equation}
\Sigma_\mathrm{BPL}(R)  = 
\begin{cases}
~2\rho_0 \left\{ f_0\left(u_\mathrm{br}, a\right) - f_0\left(0, a\right) +  \vphantom{\rbr^{a-b}} \right. & \\
~~~~~~~~~~~~~ \left. \rbr^{a-b}\left[ f_0\left(\infty, b\right) - f_0\left(u_\mathrm{br}, b\right) \right]  \right\}  & \textrm{for} ~~ R<\rbr  \\
\\
~2\rho_0 \rbr^{a-b}\left[ f_0(\infty, b) - f_0(0, b) \right]   & \textrm{for} ~~  R\geq\rbr,
\end{cases}
\label{eq:sigbpl}
\end{equation}
where
\begin{equation}
f_0(u,\alpha)  = R^{\alpha+1}  u\,  _2F_1\left(\frac{1}{2}, \frac{-\alpha}{2}; \frac{3}{2}; -u^2 \right),
\end{equation}
\begin{equation}
u = \sqrt{\left(\frac{r}{R}\right)^2-1}, 
\end{equation}
the symbol $_2F_1$ denotes the Gaussian hypergeometric function, and $u_\mathrm{br} = u(\rbr)$. This form of surface density profile was used to fit only the datasets that contain a negligible fraction of contaminants, that is usually for the GC candidates coming from the ACS catalogs.

For the datasets that contain contaminants, we had to add one more free parameter in the model profile, $\gamma$, that expresses the surface density of contaminants:
\begin{equation}
\Sigma_\mathrm{BPLB}(R)  = \Sigma_\mathrm{BPL}(R) +\gamma.
\label{eq:bplb}
\end{equation}
The contaminants were thus assumed to be distributed homogeneously, that is with no density gradient. This profile was mostly used for fitting of the profiles extracted from the FDS catalog.

For some galaxies, it was not possible to fit the surface density profile of GCS by a broken power law. This happened either when the GCS was so extended that the ACS field captured only the inner part of the broken power law profile, or the broken power law would be only poorly constrained because the galaxy had too few GCs.  In these situations, we only made a fit by a simple power-law density profile. The density profile of a single power law was parametrized as
\begin{equation}
\rho(r) = \rho_0r^a,
\label{eq:spl}
\end{equation}
which corresponds to the surface density profile
\begin{equation}
\Sigma_\mathrm{SPL}(R) =  \sqrt{\pi}\rho_0 R^{a+1}\frac{\Gamma\left(\frac{-a-1}{2}\right)}{\Gamma\left(\frac{-a}{2}\right)}.
\end{equation}
When contaminants were expected to contribute to the surface density profile substantially, we added the background term to the single power law:
\begin{equation}
\Sigma_\mathrm{SPLB}(R)  = \Sigma_\mathrm{SPL}(R) +\gamma.         
\end{equation}

\subsection{The fitting method}
We found the best-fit parameters of the models by the maximum likelihood method. Here we assumed the uncertainty of the number of sources in each bin follows the Gaussian distribution. Then the likelihood reads:
\begin{equation}
\ln \mathcal{L}(\vec{p}) = \sum_i -0.5\ln\left(2\pi\right)-\ln\left(\Delta\Sigma_{\mathrm{obs},i}\right)-\frac{\left[\Sigma_\mathrm{m}(r_i, \vec{p})-\Sigma_{\mathrm{obs},i}\right]^2}{2\Delta\Sigma_{\mathrm{obs},i}^2}.
\end{equation}
Here $\vec{p} = (p_1, p_2, p_3, \ldots, p_\nu)$ denotes the vector of the free parameters of the model, and $r_i$ the central radius of the $i$-th bin, that is the arithmetic average of the inner and outer radius of the given annulus, $\Sigma_\mathrm{m}$ denotes the surface density predicted by the fitted model and $\Sigma_{\mathrm{obs},i}$ the observed surface density of GC candidates in the given bin. 

We used the following method to estimate the uncertainty of the fitted parameters. It is sometimes called the method of support. It is based on the statistical likelihood-ratio test. Let $\vec{p}_\mathrm{max}$ denote the vector of the best-fit free parameters and   $\mathcal{L}_\mathrm{max} = \mathcal{L}(\vec{p}_\mathrm{max})$ the maximum value of the likelihood function. Then the upper (lower) limit on the value of the $j$-th free parameter, $p_j$, was estimated by maximizing (minimizing) $p_j$ over the region of the parameter space satisfying the condition
\begin{equation}
    \ln \mathcal{L}_\mathrm{max} - \ln \mathcal{L}(\vec{p})<0.5.
\end{equation}
The examples in \app{ellipticity} and \sect{attempt} demonstrate on artificial data that our methods are able to recover correctly the intrinsic parameters of the density of the GCSs.

\subsection{Resulting parameters of the fits}
\label{sec:resultingpars}
For every galaxy, we eventually obtained density profiles of their GCSs and their fits from up to three types of data (ACS, FDS, spectroscopic, see \sect{data}). All the fitted parameters are provided in \tab{fits}. In \app{fits} in Figs.~\ref{fig:g1}-\ref{fig:g19} we show plots of the observed density profiles together with the fitted models. For
\onecolumn
\begin{landscape}
\begin{table*}
	\caption{Final set of the fitted parameters of the volume density profiles (\equ{bpl} or \equ{spl}) of the GCSs of the investigated galaxies.}
	\label{tab:finalfits}
	\centering
	\begin{tabular}{p{2cm}|lll|lll|lll}
		\hline\hline
  & \multicolumn{3}{c|}{All}  &  \multicolumn{3}{c|}{Red} & \multicolumn{3}{c}{Blue}\\
   Galaxy & $a$ & $b$ & $r_\mathrm{br}$ &  $a$ & $b$ & $r_\mathrm{br}$ & $a$ & $b$ & $r_\mathrm{br}$  \\
  & & & [arcmin]  & & & [arcmin] & & & [arcmin] \\
\hline
ESO\,358-006  &  $-1.86^{+0.2, \mathrm{\textit{ h*}} }_{-0.2}$  &  $-3.70^{+1, \mathrm{\textit{ a}} }_{-0.7}$  &  $0.228^{+0.06, \mathrm{\textit{ h*}} }_{-0.1}$  &  $-3.24^{+0.3, \mathrm{\textit{ a}} }_{-0.3}$  &    &    &  $-2.60^{+0.4, \mathrm{\textit{ a*}} }_{-0.4}$  &    &  \\
ESO\,358-050  &  $-1.67^{+0.2, \mathrm{\textit{ h}} }_{-0.1}$  &  $-3.57^{+0.6, \mathrm{\textit{ a}} }_{-0.4}$  &  $0.469^{+0.09, \mathrm{\textit{ h}} }_{-0.1}$  &  $0.12^{+0.6, \mathrm{\textit{ h*}} }_{-0.5}$  &  $-3.71^{+0.8, \mathrm{\textit{ h}} }_{-1}$  &  $0.331^{+0.09, \mathrm{\textit{ h}} }_{-0.1}$  &  $-1.14^{+0.2, \mathrm{\textit{ h}} }_{-0.2}$  &  $-4.23^{+0.7, \mathrm{\textit{ h}} }_{-1}$  &  $0.463^{+0.08, \mathrm{\textit{ h}} }_{-0.08}$\\
NGC\,1316  &  $-2.387^{+0.03, \mathrm{\textit{ f}} }_{-0.03}$  &  $-4.82^{+0.7, \mathrm{\textit{ f*}} }_{-2}$  &  $9.6^{+1, \mathrm{\textit{ f}} }_{-1}$  &  $-1.52^{+0.2, \mathrm{\textit{ f}} }_{-0.1}$  &  $-3.32^{+0.5, \mathrm{\textit{ f*}} }_{-20}$  &  $5.23^{+3, \mathrm{\textit{ f}} }_{-0.8}$  &  $-2.350^{+0.02, \mathrm{\textit{ f}} }_{-0.04}$  &  $-4.37^{+0.8, \mathrm{\textit{ f*}} }_{-2}$  &  $12.2^{+2, \mathrm{\textit{ f}} }_{-2}$\\
NGC\,1336  &  $-1.39^{+0.1, \mathrm{\textit{ h}} }_{-0.1}$  &  $-2.59^{+0.2, \mathrm{\textit{ a}} }_{-0.2}$  &  $0.389^{+0.09, \mathrm{\textit{ h}} }_{-0.09}$  &  $-1.56^{+0.2, \mathrm{\textit{ h}} }_{-0.1}$  &  $-3.65^{+0.5, \mathrm{\textit{ a}} }_{-0.4}$  &  $0.483^{+0.08, \mathrm{\textit{ h}} }_{-0.1}$  &  $-1.92^{+0.1, \mathrm{\textit{ h}} }_{-0.1}$  &  $-4.54^{+0.9, \mathrm{\textit{ f}} }_{-0.6}$  &  $1.30^{+0.3, \mathrm{\textit{ m}} }_{-0.3}$\\
NGC\,1351  &  $-1.03^{+0.1, \mathrm{\textit{ h}} }_{-0.1}$  &  $-2.446^{+0.09, \mathrm{\textit{ a}} }_{-0.09}$  &  $0.252^{+0.05, \mathrm{\textit{ h}} }_{-0.05}$  &  $-1.65^{+0.1, \mathrm{\textit{ h}} }_{-0.1}$  &  $-2.80^{+0.2, \mathrm{\textit{ a}} }_{-0.2}$  &  $0.36^{+0.1, \mathrm{\textit{ h}} }_{-0.1}$  &  $-1.04^{+0.2, \mathrm{\textit{ h}} }_{-0.2}$  &  $-2.43^{+0.1, \mathrm{\textit{ a}} }_{-0.1}$  &  $0.213^{+0.07, \mathrm{\textit{ h}} }_{-0.06}$\\
NGC\,1373  &  $-0.18^{+0.3, \mathrm{\textit{ h}} }_{-0.2}$  &  $-3.76^{+0.6, \mathrm{\textit{ h}} }_{-0.7}$  &  $0.333^{+0.05, \mathrm{\textit{ h}} }_{-0.06}$  &  $-2.79^{+0.8, \mathrm{\textit{ h}} }_{-0.7}$  &    &    &  $2.59^{+0.5, \mathrm{\textit{ h*}} }_{-0.4}$  &  $-4.07^{+0.5, \mathrm{\textit{ h}} }_{-0.6}$  &  $0.321^{+0.04, \mathrm{\textit{ h}} }_{-0.04}$\\
NGC\,1379  &  $-1.37^{+0.2, \mathrm{\textit{ f}} }_{-0.2}$  &  $-3.74^{+0.5, \mathrm{\textit{ f}} }_{-0.7}$  &  $0.85^{+0.1, \mathrm{\textit{ f}} }_{-0.1}$  &  $-0.94^{+0.3, \mathrm{\textit{ f}} }_{-0.2}$  &  $-3.38^{+0.4, \mathrm{\textit{ f}} }_{-0.5}$  &  $0.474^{+0.08, \mathrm{\textit{ f}} }_{-0.1}$  &  $-0.98^{+1, \mathrm{\textit{ f}} }_{-0.8}$  &  $-3.41^{+0.6, \mathrm{\textit{ f}} }_{-1}$  &  $0.83^{+0.2, \mathrm{\textit{ f}} }_{-0.2}$\\
NGC\,1380  &  $-1.76^{+0.2, \mathrm{\textit{ h}} }_{-0.2}$  &  $-5.15^{+0.8, \mathrm{\textit{ f}} }_{-2}$  &  $1.40^{+0.4, \mathrm{\textit{ m}} }_{-0.2}$  &  $-1.54^{+0.2, \mathrm{\textit{ h}} }_{-0.1}$  &  $-4.05^{+0.3, \mathrm{\textit{ f}} }_{-0.2}$  &  $1.14^{+0.4, \mathrm{\textit{ h}} }_{-0.3}$  &  $-2.10^{+0.3, \mathrm{\textit{ h}} }_{-0.3}$  &  $-3.23^{+0.2, \mathrm{\textit{ f}} }_{-0.4}$  &  $1.50^{+0.3, \mathrm{\textit{ m}} }_{-0.6}$\\
NGC\,1380B  &  $-1.51^{+0.1, \mathrm{\textit{ h}} }_{-0.1}$  &  $-3.13^{+0.2, \mathrm{\textit{ h}} }_{-0.2}$  &  $0.274^{+0.05, \mathrm{\textit{ h}} }_{-0.06}$  &  $2.34^{+0.6, \mathrm{\textit{ h*}} }_{-0.6}$  &  $-3.42^{+0.3, \mathrm{\textit{ a}} }_{-0.3}$  &  $0.139^{+0.03, \mathrm{\textit{ h*}} }_{-0.04}$  &  $-1.706^{+0.1, \mathrm{\textit{ h}} }_{-0.09}$  &  $-3.15^{+0.3, \mathrm{\textit{ h}} }_{-0.3}$  &  $0.294^{+0.06, \mathrm{\textit{ h}} }_{-0.07}$\\
NGC\,1381  &  $-2.80^{+0.3, \mathrm{\textit{ h}} }_{-0.3}$  &    &    &  $-4.05^{+0.9, \mathrm{\textit{ h}} }_{-1}$  &    &    &  $-2.65^{+0.3, \mathrm{\textit{ h}} }_{-0.3}$  &    &  \\
NGC\,1387  &  $-3.08^{+0.2, \mathrm{\textit{ h}} }_{-0.2}$  &  $-4.6^{+2, \mathrm{\textit{ f}} }_{-20}$  &  $2.4^{+700, \mathrm{\textit{ f}} }_{-2}$  &  $-2.96^{+0.1, \mathrm{\textit{ a}} }_{-0.1}$  &    &    &  $-2.73^{+0.5, \mathrm{\textit{ h}} }_{-0.5}$  &  $-3.62^{+0.9, \mathrm{\textit{ f}} }_{-6}$  &  $0.86^{+400, \mathrm{\textit{ f*}} }_{-0.9}$\\
NGC\,1399  &  $-1.855^{+0.01, \mathrm{\textit{ a}} }_{-0.01}$  &  $-4.49^{+0.6, \mathrm{\textit{ f*}} }_{-0.9}$  &  $11.03^{+0.6, \mathrm{\textit{ f}} }_{-0.6}$  &  $-2.002^{+0.03, \mathrm{\textit{ a}} }_{-0.02}$  &  $-7.5^{+3, \mathrm{\textit{ f*}} }_{-7}$  &  $11.51^{+0.4, \mathrm{\textit{ f}} }_{-0.5}$  &  $-1.642^{+0.08, \mathrm{\textit{ a}} }_{-0.05}$  &  $-3.58^{+0.6, \mathrm{\textit{ f*}} }_{-0.8}$  &  $10.79^{+0.9, \mathrm{\textit{ f}} }_{-1}$\\
NGC\,1404  &  $-0.5^{+6, \mathrm{\textit{ s*}} }_{-1}$  &  $-4.0^{+2, \mathrm{\textit{ s*}} }_{-20}$  &  $2.67^{+1, \mathrm{\textit{ s*}} }_{-0.9}$  &    &    &    &    &    &  \\
NGC\,1419  &  $-1.50^{+0.1, \mathrm{\textit{ h}} }_{-0.1}$  &  $-2.92^{+0.2, \mathrm{\textit{ a}} }_{-0.2}$  &  $0.394^{+0.09, \mathrm{\textit{ h}} }_{-0.1}$  &  $-2.30^{+0.5, \mathrm{\textit{ h}} }_{-0.5}$  &  $-3.83^{+0.9, \mathrm{\textit{ a}} }_{-0.5}$  &  $0.51^{+0.8, \mathrm{\textit{ h}} }_{-0.3}$  &  $-0.93^{+0.2, \mathrm{\textit{ h}} }_{-0.2}$  &  $-2.78^{+0.3, \mathrm{\textit{ a}} }_{-0.3}$  &  $0.342^{+0.08, \mathrm{\textit{ h}} }_{-0.09}$\\
NGC\,1427  &  $-1.97^{+0.2, \mathrm{\textit{ h}} }_{-0.2}$  &  $-3.23^{+0.4, \mathrm{\textit{ f}} }_{-0.5}$  &  $1.32^{+0.2, \mathrm{\textit{ a}} }_{-0.2}$  &  $-1.87^{+0.1, \mathrm{\textit{ h}} }_{-0.2}$  &  $-3.45^{+0.5, \mathrm{\textit{ f}} }_{-1}$  &  $0.97^{+0.2, \mathrm{\textit{ a}} }_{-0.1}$  &  $-2.15^{+0.2, \mathrm{\textit{ h}} }_{-0.1}$  &  $-4.01^{+0.8, \mathrm{\textit{ f}} }_{-7}$  &  $1.64^{+0.8, \mathrm{\textit{ f}} }_{-0.4}$\\
NGC\,1428  &  $-1.81^{+0.4, \mathrm{\textit{ h}} }_{-0.4}$  &  $-3.77^{+0.8, \mathrm{\textit{ h}} }_{-1}$  &  $0.47^{+0.1, \mathrm{\textit{ h}} }_{-0.2}$  &  $-2.85^{+1, \mathrm{\textit{ h}} }_{-0.6}$  &  $-8.4^{+7, \mathrm{\textit{ h*}} }_{-8}$  &  $0.71^{+400, \mathrm{\textit{ h*}} }_{-0.2}$  &  $2.82^{+1, \mathrm{\textit{ h*}} }_{-0.8}$  &  $-5.1^{+1, \mathrm{\textit{ h}} }_{-2}$  &  $0.507^{+0.06, \mathrm{\textit{ h}} }_{-0.09}$\\
Stack\_8.0  &  $-1.54^{+0.2, \mathrm{\textit{ f}} }_{-0.3}$  &  $-3.90^{+0.9, \mathrm{\textit{ f}} }_{-10}$  &  $0.292^{+0.1, \mathrm{\textit{ f}} }_{-0.05}$  &  $-0.73^{+0.2, \mathrm{\textit{ f*}} }_{-0.1}$  &  $-2.99^{+0.4, \mathrm{\textit{ f*}} }_{-0.6}$  &  $0.088^{+0.02, \mathrm{\textit{ f}} }_{-0.02}$  &  $-0.36^{+0.2, \mathrm{\textit{ f*}} }_{-0.2}$  &  $-3.06^{+0.4, \mathrm{\textit{ f*}} }_{-0.6}$  &  $0.140^{+0.03, \mathrm{\textit{ f}} }_{-0.03}$\\
Stack\_8.5  &  $-1.286^{+0.07, \mathrm{\textit{ f}} }_{-0.08}$  &  $-3.50^{+0.4, \mathrm{\textit{ f}} }_{-0.6}$  &  $0.260^{+0.04, \mathrm{\textit{ f}} }_{-0.02}$  &  $-1.42^{+0.2, \mathrm{\textit{ f}} }_{-0.1}$  &  $-3.24^{+0.5, \mathrm{\textit{ f}} }_{-0.8}$  &  $0.276^{+0.07, \mathrm{\textit{ f}} }_{-0.07}$  &  $-1.214^{+0.1, \mathrm{\textit{ f}} }_{-0.09}$  &  $-3.60^{+0.5, \mathrm{\textit{ f}} }_{-0.7}$  &  $0.261^{+0.03, \mathrm{\textit{ f}} }_{-0.03}$\\
Stack\_9.0  &  $-1.65^{+0.2, \mathrm{\textit{ f}} }_{-0.2}$  &  $-3.22^{+0.6, \mathrm{\textit{ f}} }_{-1}$  &  $0.58^{+0.3, \mathrm{\textit{ f}} }_{-0.3}$  &  $-0.59^{+0.2, \mathrm{\textit{ f}} }_{-0.2}$  &  $-4.24^{+0.9, \mathrm{\textit{ f}} }_{-2}$  &  $0.358^{+0.07, \mathrm{\textit{ f}} }_{-0.07}$  &  $-1.75^{+0.6, \mathrm{\textit{ f*}} }_{-0.5}$  &  $-2.37^{+0.3, \mathrm{\textit{ f*}} }_{-10}$  &  $0.31^{+500, \mathrm{\textit{ f*}} }_{-0.3}$\\

		\\ \hline
	\end{tabular}
\tablefoot{\textit{(h)}: Parameter was derived from the HST ACS data. \textit{(f)}: Parameter was derived from the FDS data. \textit{(a)}: Parameter was derived as the weighted average from the HST ACS and FDS data. \textit{(s)}: Parameter was derived  from the spectroscopic data. \textit{(m)}: Parameter was estimated manually by visual inspection of FDS and ACS data, when fitting was not possible. Asterisks mark the suspicious measurements that could be catastrophic failures. }
\end{table*}
\end{landscape}
\twocolumn

\noindent
all galaxies, we show the background-subtracted profiles -- this means that the fitted value of the surface density of contaminant sources was subtracted from both the measured and fitted profiles. We also show profiles without background subtraction  for a few galaxies, namely for NGC\,1352 (\fig{g5}), NGC\,1387 (\fig{g11}) and NGC\,1427 (\fig{g15}).
In all plots showing the background-subtracted FDS profiles we also indicated by the dashed horizontal lines the fitted values of the background.

It was necessary to accept for every galaxy one final set of the parameters of the density profile of the GCS. We had to choose for every galaxy the values of the inner slope $a$, outer slope $b$ and the break radius \rbr. {We did not make any final choice of the surface density of contaminant sources and the normalization of the density profile of the GCS -- they are strongly influenced by the method of observation and were not necessary for the subsequent analysis.}

\begin{table*}[h!]
\caption{Number of galaxies for which the given parameter is larger for the red GCs, for the blue GCs, or for which the values are consistent. The last column gives the total number of galaxies for which this comparison was possible.}
\label{tab:numbers}    
\centering                              
\begin{tabular}{lllll}
\hline\hline                
Parameter & red>blue & red<blue & consistent &
 total \\
\hline
$a$  &  3 (27\%)  &  4 (36\%)  &  4 (36\%)  &  11\\
$b$  &  0 (0\%)  &  2 (22\%)  &  7 (78\%)  &  9\\
$a-b$  &  1 (14\%)  &  0 (0\%)  &  6 (86\%)  &  7\\
$r_\mathrm{br}$  &  0 (0\%)  &  5 (45\%)  &  6 (55\%)  &  11\\

\hline                                  
\end{tabular}
\end{table*}

\begin{table*}
\caption{Statistics of parameters of GCSs density profiles.}
\label{tab:stat}    
\centering                              
\begin{tabular}{l|ll|ll|ll}
\hline\hline                
 GC population & $a$ & $\sigma_{\mathrm{int}, a}$ &
 $b$ &   $\sigma_{\mathrm{int}, b}$ &
 $a-b$ &   $\sigma_{\mathrm{int}, a-b}$ \\
\hline
All  &  $-1.7^{+0.2}_{-0.2}$  &  $0.6^{+0.1}_{-0.1}$  &  $-3.4^{+0.2}_{-0.2}$  &  $0.6^{+0.2}_{-0.2}$  &  $1.5^{+0.1}_{-0.1}$  &  $0.02^{+0.2}_{-0.02}$  \\
Red  &  $-1.9^{+0.2}_{-0.2}$  &  $0.7^{+0.2}_{-0.1}$  &  $-3.5^{+0.2}_{-0.2}$  &  $0.3^{+0.2}_{-0.1}$  &  $1.9^{+0.2}_{-0.2}$  &  $0.4^{+0.2}_{-0.2}$  \\
Blue  &  $-1.7^{+0.2}_{-0.2}$  &  $0.5^{+0.1}_{-0.1}$  &  $-3.5^{+0.2}_{-0.3}$  &  $0.5^{+0.2}_{-0.2}$  &  $1.6^{+0.2}_{-0.1}$  &  $0.01^{+0.3}_{-0.01}$  \\

\hline                                  
\end{tabular}
\end{table*}

The way we selected the final set of parameters of the GCS profile density was different for different galaxies. We chose the parameters derived from the spectroscopic data only if no other data were available for the given galaxy, because spectroscopic data can easily be degraded by geometric incompleteness of the survey. Whenever possible, we based the final parameters on the FDS or ACS data. If data from both surveys were available  for a given galaxy, the strategy of accepting the final parameters was decided on the basis of a  visual inspection of the surface density profiles. If the inner slope, $a$, was different in the FDS and ACS data, we accounted the difference to {the problems with contamination by the light of the host galaxy in the FDS data (see \sect{extracting}) and preferred the ACS value}. For many galaxies, the ACS data did not fully cover the outer parts of the broken power law profiles. Therefore, if the outer slopes of the profile came out differently, we adopted $b$ from the FDS data. Regarding the break radii, if the FDS profiles appeared affected by the {contamination by the light of the host galaxy}  (i.e., the inner slope was shallower for the FDS data than for the ACS data), we preferred the break radius estimated from the ACS data. If the break was close to the outer limit of the ACS data, we preferred \rbr from the FDS data. 

In the cases that a given parameter $p_j$ appeared consistent between the two datasets for a given galaxy, we combined the measurements using the following form of the weighed average. It can account for the fact that our estimates of the uncertainty limits on $p_j$ were asymmetric.  Let $\Delta^+ p_j$ and $\Delta^- p_j$ denote the upper and lower errorbars of the parameter $p_j$, respectively. Next, let us introduce the ``joint Gaussian distribution''
\begin{equation}
    g(x, \overline{x}, \sigma_+, \sigma_-) =
    \begin{cases}
    \frac{\sqrt{2}}{\sqrt{\pi}(\sigma_+ +\sigma_-)}  \exp{\left[-\frac{(x-\overline{x})^2}{2\sigma_+^2}\right]}~~~~\textrm{for}~~~~x\geq\overline{x}\\  \\
    \frac{\sqrt{2}}{\sqrt{\pi}(\sigma_+ +\sigma_-)}  \exp{\left[-\frac{(x-\overline{x})^2}{2\sigma_-^2}\right]}~~~~\textrm{for}~~~~x<\overline{x},\\
    \end{cases}
    \label{eq:joint}
\end{equation}
which is subsequently used on several occasions in this paper. We approximated the probability distribution of $p_j$ by $g(p_j, \overline{p_j},  \Delta^+p_j, \Delta^+p_j)$.
Here $\overline{p_j}$ stands for the best-fit value of $p_j$. If $g_\mathrm{FDS}$ denotes the probability distribution for the FDS data and $g_\mathrm{ACS}$ that for the ACS data, then the function
\begin{equation}
    g_\mathrm{tot}(p_j) = g_\mathrm{FDS}(p_j)g_\mathrm{ACS}(p_j)
\end{equation}
represents the total likelihood function of $p_j$. We got the final estimate of $p_j$ as the argument $p_{j,\mathrm{jointmax}}$ maximizing the function $g_\mathrm{tot}(p_j)$. The uncertainty limits were obtained through the method of support, that is as the border values of the interval of $p_j$ meeting
\begin{equation}
\ln g_\mathrm{tot}(p_{j,\mathrm{jointmax}}) - \ln g_\mathrm{tot}(p_j)<0.5.    
\end{equation}

For some galaxies, the GCS break radius was close to the border of the ACS data and the inner reliable limit of the FDS data. For such galaxies, we had only the fits of the inner slope from ACS data and the outer slope from the FDS data. In such cases, we opted for a ``manual'' method when we estimated the break radii and their uncertainties by visual inspection of both profiles in the background-subtracted surface density plots. That was, for example, the case for the profile of the blue subsample of the sources around NGC\,1336, as shown in \fig{g4}. In these cases, when we accepted this subjective method, we stated very conservative estimates of the uncertainty limits.

The final parameters are listed in \tab{finalfits}.  The notes indicate the data or joining method that was used to obtain the value: $f$ indicates the FDS data, $h$ the HST ACS data, $s$ the spectroscopic data, $a$ the weighed average, and $m$ the manual joining method. There are also some values marked by an asterisk. They indicate suspicious measurements. Those were identified by visual inspections of the observed and fitted profiles of surface density in \app{fits}. The measurements were usually deemed suspicious if one of the parts of the broken power law profile was resolved by only one or two data points. This corresponds to the galaxies that have few GCs. Next, we marked all parameters of the density profile of NGC\,1404 suspicious, because this galaxy is known to be undergoing a tidal interaction with NGC\,1399, their GCSs overlap and we used spectroscopic data to fit the profile, that can suffer from incomplete spatial coverage of the spectroscopic surveys. Finally, we deemed suspicious the fitted values of $b$ for the two galaxies with the most extended GCS, that is NGC\,1399 and NGC\,1316, because they might have been affected by the large-scale sensitivity variations of the FDS survey (\sect{fds}). This is discussed in detail in \app{sensitivity}. {It turned out that all of the suspicious measurements actually do not deviate noticeably from the scaling relations followed by the trusted measurements, as it will be shown in \sect{results}.}

\begin{table}
\caption{Average and intrinsic scatter of the difference between structural parameters of the GCSs of the red and blue GC populations.}
\label{tab:diffs}    
\centering                              
\begin{tabular}{p{2cm}ll}
\hline\hline                
Parameter (red-blue) & Mean & $\sigma_\mathrm{int}$ \\
\hline
$a$  &  $-0.0^{+0.2}_{-0.2}$  &  $0.5^{+0.2}_{-0.1}$  \\
$b$  &  $-0.4^{+0.2}_{-0.2}$  &  $0.01^{+0.3}_{-0.01}$  \\
$a-b$  &  $0.0^{+0.3}_{-0.3}$  &  $0.4^{+0.4}_{-0.4}$  \\
\rbr [kpc]  &  $-0.3^{+0.2}_{-0.2}$  &  $0.03^{+0.3}_{-0.02}$  \\

\hline                                  
\end{tabular}
\end{table}

\section{Calculation of $a_0$ radii}
\label{sec:ao}

{The $a_0$ radii were calculated on the basis of the S\'ersic fits of the galaxies, and the estimates of their stellar masses (\citealp{su21}, see \sect{data} for details).} We were interested in two types of $a_0$ radii: those predicted by Newtonian gravity and those predicted by MOND gravity. Assuming that the galaxies are spherically symmetric, we could use the approximate analytic  formulas for the density  S\'ersic spheres of \citet{limaneto99} (as updated by \citealp{marquez00}) to calculate the profiles of Newtonian gravitational acceleration, $g_\mathrm{N}(r)$. The profiles of MOND acceleration, $g_\mathrm{M}$, were obtained by transforming $g_\mathrm{N}$ using the formula \equ{mond}. We adopted the observationally motivated interpolation function $\nu$:
\begin{equation}
    \nu(x) = \left[ 1-\exp{(-\sqrt{x})}\right]^{-1}
\end{equation}
and value of $a_0 = 1.2\times 10 ^{-10}$\,m\,s$^{-2}$ \citep{mcgaugh16,li18}. Once the radial profiles of gravitational accelerations were known, we could find  the $a_0$ radii  numerically. 

Finding the $a_0$ radii for the stacked galaxies required a more elaborate approach. First, we calculated the acceleration profile for each of the stacked galaxies individually. We then assigned to each stack a final acceleration profile that was calculated as the median acceleration profile of all objects contributing to the stack. Then we could solve for the $a_0$ radii.

In the following, we denote by \reqN the Newtonian $a_0$ radius and by \reqM the MOND one. In the case that we do not need to distinguish between them or the statements hold true for both, we denote the $a_0$ radius by \req.

Uncertainties on the $a_0$ radii were derived only from the uncertainties of stellar masses. These were tabulated by \citet{su21} for several stellar mass bins. We used linear interpolation to obtain the uncertainties of the stellar mass of individual galaxies. We neglected the uncertainties in distance. The total line-of-sight distance scatter for the galaxies in the Fornax cluster is about 0.5\,Mpc \citep{blakeslee09}, which at the assumed distance of the center of the cluster of 20\,Mpc, would result in the difference of stellar mass of 0.02\,dex. This is negligible compared to the uncertainty in masses caused by the uncertainty in the mass-to-light ratio (\tab{gals}). 

For NGC\,1399 we considered including the contribution of the hot intergalactic gas to the gravitational field of the galaxy. This could potentially influence the $a_0$ radius of the galaxy. The profile of cumulative mass of the hot gas for this galaxy was presented in \citet{paolillo02} and \citet{samur06}. It turned out that including the gas mass has only very little effect on the position of the $a_0$ radii. At the position of the MOND $a_0$ radius the cumulative mass of hot gas is just $\sim10^9\,M_\sun$, which is negligible compared to the mass of the galaxy (see \tab{gals}). For this reason, we neglected the contribution of the gas mass in the gravitational field when estimating the $a_0$ radii.

The resulting $a_0$ radii are tabulated in \tab{gals}. Some galaxies at the low-mass end of our sample, do not have $a_0$ radii. These are low-surface-brightness galaxies inside which the gravitational acceleration does not exceed $a_0$. Interestingly, the GCSs of such galaxies still can show broken power-law profiles as we see below.

\section{ Results }
\label{sec:results}

\subsection{Structural parameters of GCSs for different GC subpopulations}
\label{sec:colors}
We explored how the fitted values of the structural parameters of GCSs, $a$, $b$, $a-b$ and \rbr, differ for  the total, red and blue GC populations. Inspection of plots of these parameters against the stellar masses of the galaxies revealed that the suspicious measurements identified in \sect{resultingpars} often lied far from the reliable measurements, with the exception of the break radii. Therefore, all suspicious measurements were excluded from the subsequent analysis in this section.

For each structural parameter, we counted the number of galaxies for which the parameter is greater for the red subpopulation of GCs than for the blue subpopulation, the number of galaxies for which the situation is opposite, and the number of galaxies for which the parameter is consistent for the blue and red GC populations. The consistency means that the uncertainty intervals of the measurements overlap. The results are listed in \tab{numbers}.  Given that we do not have all parameters for all galaxies, the last column of this table indicates for each  structural parameter the total number of galaxies that this comparison is based on.

The inner slope of the GCS density profile, $a$, shows the most variate behavior. It can be both smaller or greater  for the red GCs than for the blue GCs, or the two GCs populations can have it consistent. All these cases occur approximately in the same number of galaxies. This result could have been influenced by the {problems with the contamination by the light of the host galaxies.}  The $b$ parameter is usually consistent for the red and blue GC populations, but in some cases the parameter is higher for the blue GCs. The difference of the slopes, $a-b$, is consistent for the red and blue GCs for all but one galaxy, for which $a-b$ is greater for the red GCs (i.e., the break is more pronounced). The values of \rbr are in a roughly equal number of galaxies either consistent for the red and blue GC populations or they are lower for the blue population.  

Next, we estimated the mean and intrinsic scatter of the structural parameters for all GC populations. We assumed that the intrinsic distribution of every parameter $p_j$ is Gaussian:
\begin{equation}
    f(p_j) = \frac{1}{\sqrt{2\pi}\sigma_{\mathrm{int},j}} \exp{\left[-\frac{(p_j-\overline{p_j})^2}{2\sigma_{\mathrm{int},j}^2}\right]}, 
\end{equation}
where $\overline{p_j}$  is the mean and $\sigma_{\mathrm{int},j}$ the intrinsic scatter of the distribution. We took into account the asymmetric uncertainty intervals of the structural parameters. The probability distribution of the parameter $p_j$ of the $i$-th galaxy, $p_{j,i}$, was modeled as  $g_{\mathrm{mes},j,i}(p_{j,i},\overline{p_{j,i}}, \sigma^+_{p_{j,i}},\sigma^-_{p_{j,i}})$, where the function $g$ was defined by \equ{joint}. The best-fit values of $\overline{p_j}$  and $\sigma_{\mathrm{int},j}$ were found by maximizing the likelihood function:
\begin{equation}
    \mathcal{L}(\overline{p_j}, \sigma_{\mathrm{int}, j})  = \prod_{i=1}\left[ f(p_{j,i})*\,g_{\mathrm{mes},j,i}(p_{j,i})\right](p_{j,i}),
    \label{eq:lik}
\end{equation}
where the symbol $*$ denotes convolution. The uncertainty limits of $\overline{p_j}$  and $\sigma_{\mathrm{int},j}$ were found through the method of support. In order to avoid numerical difficulties, we required the intrinsic scatter to be at least 0.01. The results are stated in \tab{stat}.

\begin{figure}
        \resizebox{\hsize}{!}{\includegraphics{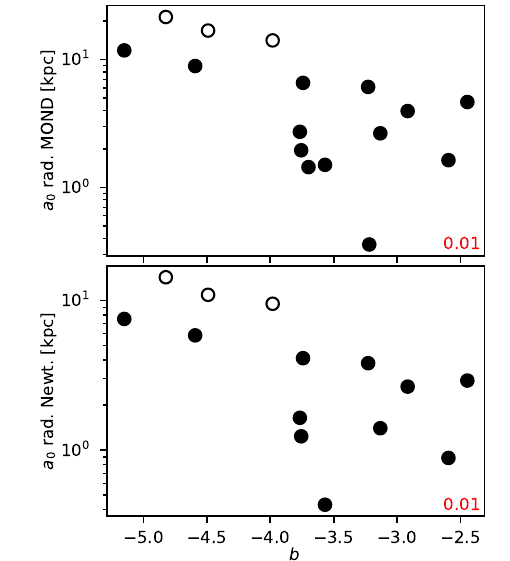}}
        \caption{Plots of the statistically significant correlations of the outer slope of the density profile of the GCSs {$b$}  with the properties of their host galaxies. Top: The MOND $a_0$ radius. Bottom: The Newtonian $a_0$ radius. The open symbol
        indicates a suspicious measurements. The red numbers in the corners indicate the $p$-value of the Spearman's correlations.}
        \label{fig:corr-b}
\end{figure}

\begin{figure}
        \resizebox{\hsize}{!}{\includegraphics{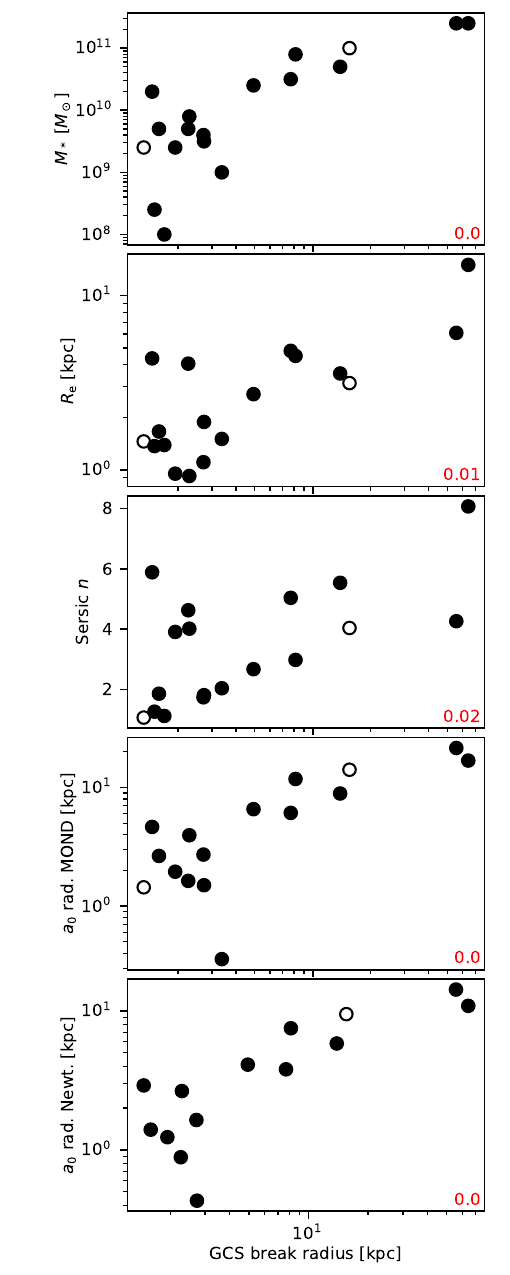}}
        \caption{Plots of the statistically significant correlations of the break radii of GCSs  with the properties of their host galaxies. From top to bottom: Stellar mass, effective radius, S\'ersic index, MOND $a_0$ radius, Newtonian $a_0$ radius. The open symbols indicate suspicious measurements. The red numbers in the corners indicate the $p$-value of the Spearman's correlations.}
        \label{fig:corr-rbr}
\end{figure}

First, the table reveals the typical values of the structural parameters: $a = -1.7$, $b = -3.4$ and $a-b = 1.7$. The parameters $a$, $b$ and their difference $a-b$ do not differ substantially between all GCs, red GCs and blue GCs. The values are consistent within 1$\,\sigma$. The intrinsic scatters of the parameters are consistent with each other for the different GC populations too. It is worth noting that the intrinsic scatter of the prominence of the break, that is $a-b$, is also consistent with being zero for the total and blue GC populations. This means that the intrinsic scatter is smaller than the measurement uncertainties of the individual data points.

We also inspected the statistical distribution of the differences of the structural parameters for red and blue GCs in individual galaxies, that is $p_{j,i,\mathrm{red}}-p_{j,i,\mathrm{blue}}$. We derived the mean and intrinsic scatter as before. The results are summarized in  \tab{diffs}. It confirms that the structural parameters of the red and blue GC populations are in average the same within 1-$2\,\sigma$ uncertainty limits. The outer slope $b$ is at the border of the $2\,\sigma$ uncertainty limit, which suggests that the outer slope might be systematically steeper (i.e., more negative) for the red GCs. The break radii for the red and blue GC populations seem to be remarkably consistent, differing typically just by 0.3\,kpc.

In summary, we found at most marginal evidence that the structural parameters of the GCSs would depend on the color of the GCs. Following the rule of Occam's razor, we consider hereafter  no difference between the structural parameters of the GC subpopulations.

\begin{figure*}[]
        \centering
        \includegraphics[width=17cm]{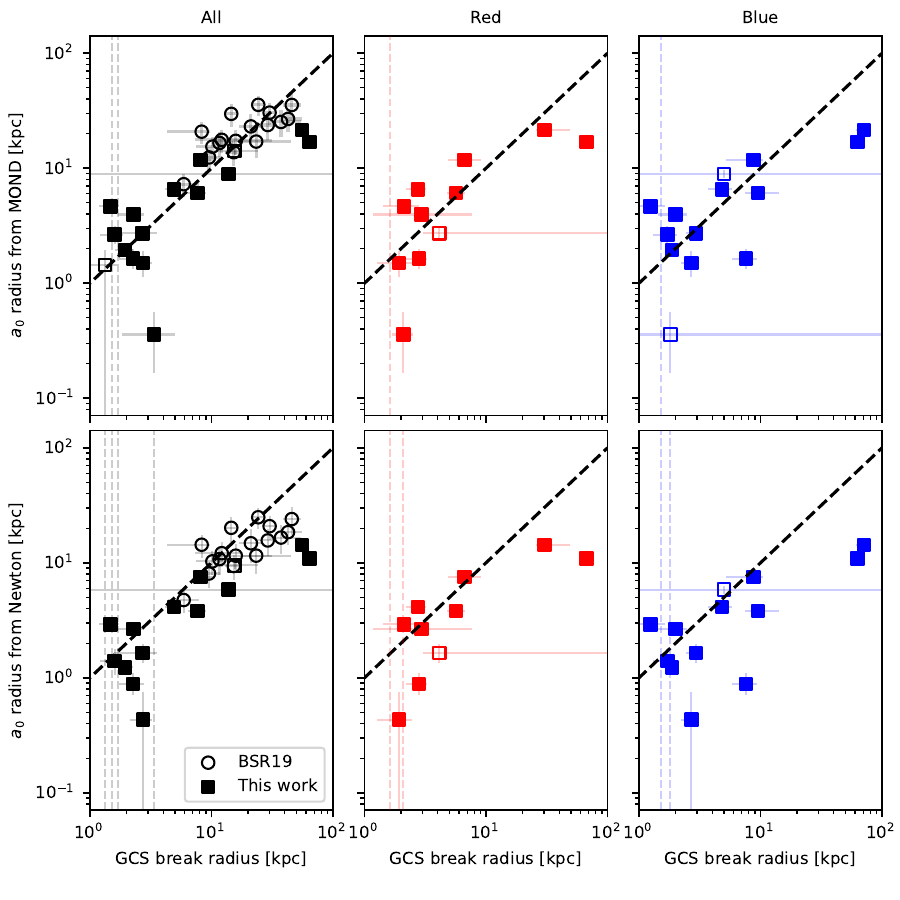}
        \caption{Relation of the $a_0$ radius versus the break radius of the GCS for each galaxy. The upper row corresponds to the MOND $a_0$ radius, the bottom row to the Newtonian $a_0$ radius.   The three columns in the horizontal direction correspond to the whole, red, and blue GC (sub)population, respectively.  The squares in the first column mark the galaxy sample inspected in this paper, the circles mark those from BSR19. The empty squares denote the suspicious measurements. The diagonal dashed black lines represent the one-to-one relation. The vertical gray dashed lines mark the galaxies that do not have $a_0$ radii. }
        \label{fig:radii}
\end{figure*}

\begin{figure*}[h!]
        \centering
        \includegraphics[width=17cm]{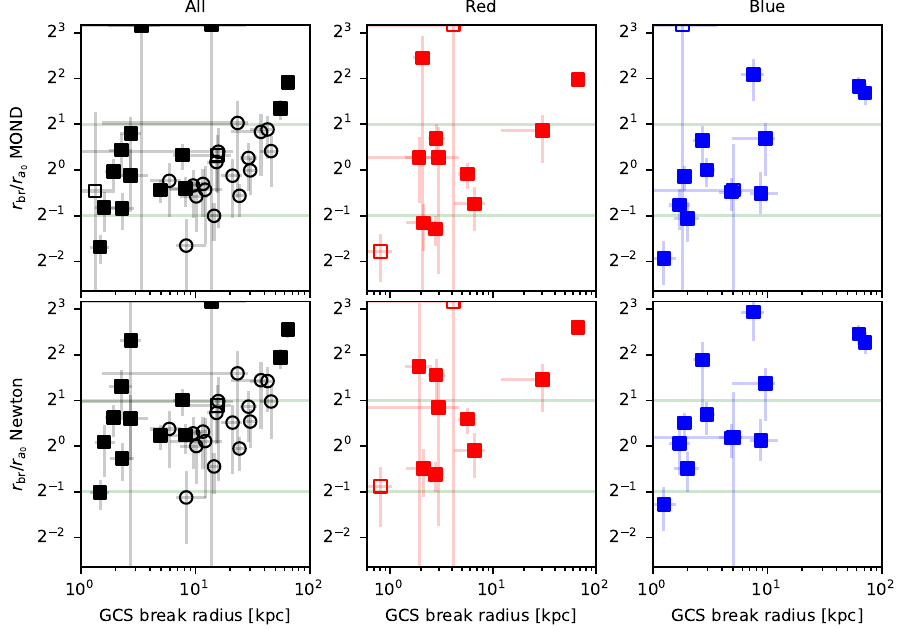}
        \caption{Demonstration of the approximate equality of the break radii of the GCSs and the $a_0$ radii of their host galaxies. {\bf First row}: {The ratio of the break radius and the MOND $a_0$ radius  as a function of the break radius}. The vertical axis is in the base-2 logarithmic scale, while the horizontal axis in the base-10 logarithmic scale. The horizontal line indicates the ratio of one (the break and $a_0$ radii are equal). {\bf Second row}: The same as the first row but for the $a_0$ radii calculated for the Newtonian gravity. {\bf Columns}: From left to right, the rows show the data derived for all GCs and for the red and blue subpopulations of GCs. }
        \label{fig:radii-ratio}
\end{figure*}

\subsection{Correlations of the structural parameters of GCSs}
\label{sec:corr}
We were interested in how the fitted structural parameters of GCSs, $a$, $b$, $a-b$ and \rbr, correlate with the stellar mass, effective radius, S\'ersic index and the Newtonian and MOND $a_0$ radii of the host galaxy. We exploited that through Spearman's correlation coefficient and its $p$-value. The $p$-value expresses the probability that the two quantities under consideration  actually do not correlate and the observed amount of correlation is there because of a coincidence. We comment here only on the pairs of quantities which correlate at least at the 5\% confidence level (i.e., their $p\leq0.05$).

We found that the parameter $b$ correlates significantly with both Newtonian and MOND $a_0$ radii. The correlations are shown in \fig{corr-b}. The red numbers in the corners of the tiles of the figure indicate the $p$-value of the correlations. The open symbols indicate the suspicious measurements.

It is noteworthy that the parameter $b$ correlates better with the $a_0$ radii than with the galaxy stellar mass. The $p$-value of the correlation with the latter is 0.03, after removing of the galaxies for which the $a_0$ radii do not exist. In contrast, the $p$-value of the correlation of $b$ with the $a_0$ radii is  0.008 (0.009) for the Newtonian (MOND) case. This suggests that there is a connection between the distribution of stars in the galaxy and the distribution of GCs beyond the break radius, which typically exceeds the effective radius of the galaxy in our sample twice.

The break radii correlate significantly with all the considered characteristics of the galaxies, as shown in \fig{corr-rbr}. We note, however, that only the correlation with the $a_0$ radii is close to a one-to-one relation, as we describe in more detail in \sect{equality}, in agreement with the finding of BSR19. The mean of the ratio of the break radius and the effective radius $\rbr/R_\mathrm{e} = 2.4$ with a root-mean-square scatter of 0.7. The ratios  $\rbr/\req$ are discussed in \sect{equality}.

We also explored whether the structural parameters $a$, $b$, $a-b$ and \rbr correlate with each other. We found only two significant correlations, that are not surprising in the light of what has been said above. The break radius correlates with the $b$ parameter, as it can be expected because we already found above that $b$ correlates with the $a_0$ radius, which in turn correlates with the break radius. Next, we found that the difference $a-b$ correlates with $b$. This is again not too surprising  given that the absolute value of $b$ is typically larger than the absolute value of $a$ (\tab{stat}).

\subsection{Equality of break radii and $a_0$ radii}
\label{sec:equality}
Here we come to the main part of the paper -- the comparison of the break radii of GCSs with the $a_0$ radii of their host galaxies. The work of BSR19 pointed out their approximate match, nevertheless their sample contained only galaxies with a relatively narrow range of stellar masses and therefore also of the $a_0$ radii. Moreover, they investigated only the total populations of GCs, such that the shapes of profiles of the red and blue GCSs, including the break radii, remained unclear. The sample investigated here rectifies these insufficiencies.

The break and $a_0$ radii are compared in \fig{radii}. The top row corresponds to the $a_0$ radii calculated from the MOND gravity, and the bottom row to those calculated assuming Newtonian gravity. The three columns, from left to right, correspond to the population of all, red and blue GCs, respectively. The squares represent the data from the sample from the present paper, while the circles those from BSR19. The open squares indicate the suspicious measurements. The diagonal black dashed lines indicate the one-to-one relation. The break radii of the galaxies that do not have $a_0$ radii are shown by the vertical dashed lines. In these galaxies, the acceleration generated by their stars, either as predicted by the Newtonian or MOND gravity, is less than $a_0$ everywhere in their extents. The galaxy NGC\,1399 is plotted twice in each panel -- one measurement comes from the data analyzed in this paper, the other comes from the paper BSR19, where it was derived from spectroscopic data. The match of the break and $a_0$ radii for NGC\,1399 became worse in this paper (see \sect{n1399} for more details).

The figure demonstrates a good match of the $a_0$ radii and break radii. The match is better for the $a_0$ radii calculated from MOND. This holds true regardless of the separation of the GCs in red and blue subpopulations. It should be pointed out that an exact match cannot be expected  because of the tidal interactions between the galaxies and galaxy mergers in the dense environment of a galaxy cluster. The tidal interactions can even lead to a loss of GCs from the galaxies or to a transfer of GCs from one galaxy to another \citep{bekki03}. Our data indeed provide an observational indication that galaxy interactions influence the break radius, see \sect{azimuth}.

Figure~\ref{fig:radii-ratio} gives us an alternative view of the same data. It shows the ratios of the break and $a_0$ radii plotted against the break radii. The estimates of the expected values and uncertainty limits of the ratios were based on the fact that the distribution  $h_z(z)$ of the variable $z$, that is a ratio of two independent variables $x$ and $y$,  $z = x/y$, can be calculated from the known distribution functions of $x$, $h_x(x)$ and of $y$, $h_y(y)$ using the so-called ratio distribution formula:
\begin{equation}
h_z(z) = \int_{-\infty}^{\infty}|y|~  h_x\!(zy)~ h_y\!(y)~ \mathrm{d}y.
\end{equation}
The distributions of the break and $a_0$ radii for every individual galaxy were again approximated by \equ{joint}. We estimated the expected value and the 1\,$\sigma$ uncertainty limits of $\rbr/\req$ as the 18, 50 and 84-th percentiles of the ratio distribution.

The top row of \fig{radii-ratio} refers to the $a_0$ radii calculated in the MOND way, the bottom row to those calculated in the Newtonian way.  The points that lie at the top border were shifted downward because they lie out of the displayed radial range, but their error bars are displayed correctly.   We could make several observations from these plots. 1) The break and MOND $a_0$ radii agree with each other within a factor of two for most galaxies -- either in terms of the most likely values or within 1-2 uncertainty limits. This applies to all GC populations. 2) For the total GC population, there is a hint of a correlation of the ratio $\rbr/\req$ with \rbr if the break radius is greater than about 20\,kpc. In this region, however, most data points come from the spectroscopic data analyzed in BSR19, which could be biased by systematic errors. 3) For the blue GC population, there is a correlation of the ratio $\rbr/\req$ with \rbr for the whole galaxy sample.  4) The Newtonian ratios $\rbr/\req$ seem to be systematically more offset from unity than the MOND ratios. {We inspect the points 1) and 4) more in detail below and the points 2) and 3) in \app{linfit}.}

Let us denote $\zeta = \rbr/\req$. We aimed to fit the distribution of $\log_{10}\zeta$ by a normal distribution with the mean of $\log_{10}\mu$ and a scatter of $\sigma_\mathrm{int}$.  {Specifically}, we wanted to fit the distribution of $\zeta$ by the lognormal distribution:
\begin{equation}
    g_\mathrm{LN}(\zeta) = \frac{1}{\sqrt{2\pi}\ln(10)\sigma_\mathrm{int} \zeta}\exp{\left\{ -\frac{\left[\log_{10}(\zeta)-\log_{10}(\mu)\right]^2}{2\sigma_\mathrm{int}^2} \right\}}.
\end{equation}
The best-fit parameters and their uncertainties were found through a likelihood function analogous to \equ{lik}. This was applied to the sets of the total population and the red and blue subpopulations of the Fornax cluster galaxy sample investigated here. In addition, we created a union sample of the galaxy set of BSR19, with NGC\,1399 excluded, and the sample of total GC populations of the Fornax galaxies. This was repeated for the $a_0$ radii calculated in the Newtonian and MOND ways.  The results are listed in \tab{ratiofits}. 

The table indeed shows that the MOND $\rbr/\req$ ratios are indeed close to one for all GC (sub)populations. The intrinsic scatter is around 0.28\,dex (i.e., the factor of 1.9). The Newtonian $\rbr/\req$ ratios are somewhat offset from unity, with a mean of about 1.4. The intrinsic scatter is similar to the MOND case. 

{In \app{linfit}, we fit the relation of the break radius and $a_0$ radius by a linear function with a lognormal intrinsic scatter. {The fitted value of the slope is not consistent with one. Nevertheless, it is possible to argue that the deviation of the slope from one is caused by a few outliers.} The data inspected in this work thus indicate just the fact that the break and $a_0$ radii agree within a factor of two.}

\begin{table}
\caption{Results of the fitting of the distributions of the ratios $\rbr/\req$ by a lognormal distribution  for different GC datasets.}
\label{tab:ratiofits}    
\centering                              
\begin{tabular}{l|cc|cc}
\hline\hline         
 & \multicolumn{2}{c|}{$\rbr/\reqM$} & \multicolumn{2}{c}{$\rbr/\reqN$} \\
Dataset & Mean & $\sigma_\mathrm{int}$ [dex] & Mean & $\sigma_\mathrm{int}$ [dex] \\
\hline
Fornax all  &  $1.0^{+0.2}_{-0.2}$  &  $0.28^{+0.07}_{-0.05}$  &  $1.5^{+0.4}_{-0.3}$  &  $0.29^{+0.08}_{-0.06}$  \\
Union  &  $0.9^{+0.1}_{-0.1}$  &  $0.23^{+0.04}_{-0.03}$  &  $1.4^{+0.2}_{-0.2}$  &  $0.23^{+0.04}_{-0.04}$  \\
Fornax red  &  $1.0^{+0.3}_{-0.2}$  &  $0.35^{+0.1}_{-0.08}$  &  $1.5^{+0.5}_{-0.4}$  &  $0.33^{+0.1}_{-0.07}$  \\
Fornax blue  &  $1.2^{+0.3}_{-0.3}$  &  $0.36^{+0.1}_{-0.07}$  &  $1.7^{+0.5}_{-0.4}$  &  $0.36^{+0.1}_{-0.07}$  \\

\hline                                  
\end{tabular}
\end{table}

\begin{figure}
        \resizebox{\hsize}{!}{\includegraphics{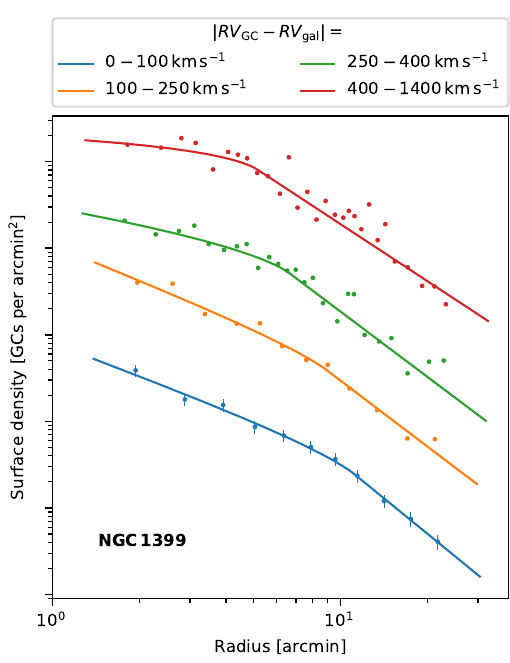}}
        \caption{Radial profiles of surface density of GCs of NGC\,1399 in several bins of radial velocity with respect to the center of the galaxy.}
        \label{fig:velslices}
\end{figure}

\begin{figure*}[]
        \centering
        \includegraphics[width=17cm]{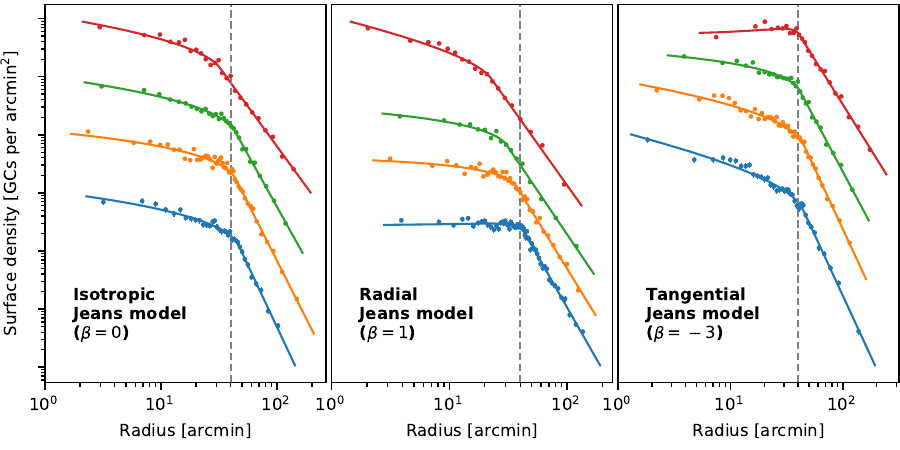}
        \caption{Same as in \fig{velslices}, but for modeled data. The models are Jeans spherical models with the indicated types of anisotropy parameters. The models were based on the gravitational potential of NGC\,1399 derived in \citet{bil19} and the parameters of the GCS used in that work. The vertical dashed line at 30\arcmin is an eye-guide to facilitate the reading of the variations of the break radius.} 
        \label{fig:velslices-models}
\end{figure*}

\section{The break in the GCS of NGC\,1399 under scrutiny}
\label{sec:n1399}

Here we explore in detail the profile of the GCS of NGC\,1399, the central galaxy of the Fornax Cluster.  Apart from that, we chose this galaxy because is has the richest GC system in our sample, and we moreover have multiple datasets for it (FDS, ACS and spectroscopic data). Its GC density profile has been fitted by a broken power law already in BSR19. 

\subsection{Comparison of profiles extracted from different datasets}
We analyzed the positions of GCs for this galaxy from three sources: the photometric catalogs of FDS and ACS and the spectroscopic data of \cite{Chaturvedi22} and \cite{fahrion20}. In addition, even another spectroscopic dataset \citep{schuberth10} was fitted by a broken power law in BSR19 (and without the uncertainty limits already in \citealp{bil19}). The comparison of all the profiles is shown in \fig{g12}.

The inner slope of the broken power law is virtually the same for all the datasets analyzed here. This suggests that {the presence of NGC\,1399 in the center of the GC system did not affect our ability to detect the GCs}. This might be counterintuitive, given that this is the brightest galaxy analyzed here. {This is probably because the relation between magnitude and surface brightness of elliptical galaxies shows a peak at intermediate magnitudes \citep{graham03}.} The match of the inner slope of the profile derived from the spectroscopic GCs of \citet{Chaturvedi22} and \citet{fahrion20} with the photometric samples indicates that it has a good spatial coverage near the center of the galaxy.  

The outer slope and the break radius of the spectroscopic data do not agree with the FDS data that well. While the outer slopes for the total population of GCs agree within the $2\,\sigma$ uncertainty limits, the break radii do not (\tab{allfits}). This might indicate either issues with a geometric incompleteness of the spectroscopic sample or an imprecise estimation of the surface density of contaminant sources when extracting the profile from the photometric data. Indeed, the GCS of NGC\,1399 is extended and the extraction of the data might have be affected by the variations of the sensitivity of OmegaCAM near the edges of its FOV (\sect{fds}). The outer slope of the FDS profile  is rather bumpy, whereas that of the spectroscopic sample looks cleaner. This is probably because of contaminants in the FDS sample, such as background galaxy clusters. These would be excluded from the spectrocopic sample on the basis of radial velocity. 

The profile analyzed in BSR19 deviates most from the other profiles. We attribute this to a geometric incompleteness of the survey \citep[see Fig.\,1 in][]{schuberth10}.
The position of the break radius is however not affected too much: BSR19 found $7.3^{+2\arcmin}_{-0.4}$, while we accepted here $11.0\arcmin\pm0.6$ (\tab{finalfits}). The match of the break radius with the $a_0$ radius, located at about $2.4\arcmin$, got worse in this paper. We note that the surface density profile of the GCs from the catalog of \citet{schuberth10} was fitted by a broken power law already in \citet{samur06}. Their results are in good agreement with those of BSR19.

\subsection{Dependence of the profile parameters on radial velocity cuts}
\label{sec:velcuts}
The large number of GCs around NGC\,1399 allowed us to explore how the radial profile of the density of the GCs depends on the line-of-sight velocity of the GCs with respect to the center of the galaxy. We assumed a radial velocity of NGC\,1399 to be 1424.9\,km\,s$^{-1}$. The profiles for different velocity cuts are displayed in \fig{velslices}. The fitted values of the parameters of the  broken power-law profiles are presented in \tab{velslices}.

The figure shows that the break radius shifts toward the center of the galaxy for the GCs that have larger velocities with respect to the galaxy. The outer slope, $b$, remains constant. The inner slope, $a$, becomes steeper for GCs having lower radial velocities with respect to the galaxy center. {As a consequence, the GC density profile is almost a simple, that is unbroken, power law for the lowest radial velocity bin.}

\subsubsection{Attempt for explanation}
\label{sec:attempt}
As the first step toward the understanding of this observation, we constructed simple dynamical models of the GCS of NGC\,1399 and explored how the profile of GC density depends on the chosen radial velocity bin. To this end, we solved the spherical Jeans equation, which gave us the profile of velocity dispersion of the GCS, see \citet{bil19} for details. We assumed the dark matter halo and the stellar mass-to-light ratio that were derived in that paper from the best-fit isotropic Jeans model of the GCS of this galaxy. For our present model, we assumed the same broken-power-law  profile of the density of the GC system as in \citet{bil19}. We solved the Jeans equation for three different choices of the anisotropy parameter: $\beta = 0$ (isotropic), $\beta = 1$ (purely radial) and $\beta = -3$ (highly tangential). The $\beta$ parameter specifies the typical shapes of orbits of the GCs around the galaxy. Having solved the Jeans equation, we generated a three-dimensional model of the GC system. Each GC was randomly assigned a position and velocity according to the assumptions described above. The velocities were drawn from an ellipsoidal Gaussian distribution, according to the solution of the Jeans equation and the assumed anisotropy parameter. There were $10^4$ GCs in each model. We eventually created a catalog of the projected positions and line-of-sight velocities of the artificial GCs. That was analyzed like the real data.

The results are shown in \fig{velslices-models}. For the isotropic and radial models, the position of the break indeed depends on the radial velocity of the GCs with respect to the galaxy, such that the GCs that are slower have their break radius further away from the galaxy, in agreement with the observed data. The tangential model does not show any obvious dependence of the break radius on the GC velocity range. This hints at a radial or isotropic anisotropy of the real GCS of NGC\,1399. Next, the slopes of the broken power law depend on the inspected range of radial velocities of GCs in all models. The difference is most pronounced for the inner slopes of the radial and tangential models. The inner slopes of the radial model become more shallow toward the low radial velocities of GCs. The trend is the opposite for the tangential model. The tangential model thus resembles the real NGC\,1399 in this regard.

We explored many other values of the anisotropy parameters, allowing them even to be a function of the galactocentric distance. We were never able to fully reproduce the trends observed in NGC\,1399. Most notably, we never got close to equalizing the inner and outer slopes of the GC density profile that is observed for the GCs that have low radial velocities with respect to the center of NGC\,1399. It might be necessary to model the GC system by several GC populations with different spatial distributions and anisotropy parameters. They would correspond to GCs formed in situ and accreted from possibly several galaxies. This more realistic modeling is beyond the scope of this paper. It should also be pointed out that in all of our models, including the non-isotropic ones, we used a gravitational potential that was derived assuming an isotropic and spherical GCS. 

 As a side note, the values of the parameters recovered from the projected positions of the artificial GCs agreed, within the uncertainty limits with the parameters that were used to generate the three-dimensional positions of the GCs. This demonstrates the correctness of our methods.

\begin{figure}
        \resizebox{\hsize}{!}{\includegraphics{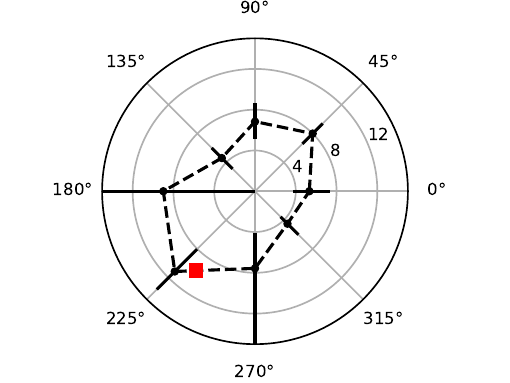}}
        \caption{Break radii of the GCS of NGC\,1399 in different sectors centered on the galaxy. The radial coordinate is in arcminutes. The red point indicates the position of NGC\,1404. }
        \label{fig:sectors}
\end{figure}

\subsection{Dependence of the break radius on azimuth: elongation toward the interacting neighbor NGC\,1404}
\label{sec:azimuth}
We inspected also how the break radius varies in azimuth in NGC\,1399. We defined eight sectors of the same angular widths centered on the galaxy. The first sector was centered on the western photometric major semi-axis of NGC\,1399. This semi-axis points to the west almost exactly. We initially experimented with the GC candidates selected from the FDS data for this exercise, but it turned out that the profiles contained spikes. By inspecting images of the sky  from the Digitized Sky Survey at the position of the spikes in the Aladin software \citep{aladin1,aladin2}, it seemed that the spikes were caused by clusters of galaxies in the background, that is by contaminating sources. This is why we eventually decided to utilize the spectroscopic data. This also is the reason why we could not inspect in detail the second largest GCS in the Fornax cluster that is possessed by NGC\,1316.

The fitted parameters of the individual sectors are presented in \tab{sectors}. The found break radii are plotted as a function of azimuth in \fig{sectors}. The azimuth 0\degr  coincides with the photometric western major semi-axis of NGC\,1399; the azimuth 90\degr coincides with the northern minor semi-axis. We can see that the break radii form approximately an ellipse. It is interesting that the ellipse is inclined substantially with respect to the stellar body of the galaxy. Instead, one of the major semi-axes of the ellipse formed by the break radii points toward the galaxy NGC\,1404. It is well known that both galaxies are tidally interacting \citep{bekki03, sheardown18}. This suggests that galaxy interactions can influence the distribution of GCs, and thus the values of break radii.


\section{Interpretation}
\label{sec:interpretation}

BSR19 proposed several potential explanations for the existence of the breaks in the  profiles of the densities of GCS and  of the coincidence of the break radii with the $a_0$ radii.  In the following subsections, we develop these ideas one step further and discuss them in the light of the new data. New ideas are added. Some of the interpretations are specific for a MOND universe, because in MOND, the constant $a_0$ enters naturally in many phenomena. It is in part because $a_0$ marks the boundary between the two regimes of the law of gravity or inertia. Some of our interpretations are applicable also in the \lcdm cosmology. Some of the explanations are based on the validity of the radial acceleration relation, which is an empirical fact. In MOND, the radial acceleration is a trivial implication of the theory, while the \lcdm cosmology still is finding its way toward its full explanation through the galaxy formation theory \citep[e.g.,][]{dicintio16,santos16,navarro17}.

\subsection{Consequence of two regimes of gravitational potential and of the accretion of GCs in mergers}
\label{sec:tworegimes}

Supposing Newtonian gravity, the gravitational field in the inner region of a high-surface-brightness galaxy is dominated by the contribution of baryons. On the other hand, far from the galaxy center the gravitational field is dominated by the contribution of the dark matter halo. Given that according to the radial acceleration {significant amounts of dark matter are needed}  only beyond the $a_0$ radius, the strengths of the gravitational fields generated by the baryons and the dark matter halo have to be equal roughly at the $a_0$ radius. In the inner region of the galaxy, the gravitational potential is steep, since it  can roughly be approximated by that of a point mass that represents the baryonic component. In the outer region, the potential is more shallow. According to the radial acceleration relation it can be approximated by a logarithmic potential.

A large fraction of the baryonic mass of massive galaxies is expected to be gained by accretion of smaller galaxies. These bring also their GCs into the system. There are multiple pieces of evidence that a large fraction of blue GCs gets into giant galaxies through the accretion of dwarf galaxies \citep{cote98, hilker99}. 

The GCs that are brought in with the satellites are tidally stripped from the satellites preferentially when the satellites are close to their orbital pericenter with respect to the host. Various accreted satellites have various pericentric velocities. If the stripped GCs have a large enough velocity at the moment of stripping, they reach the outer shallow part of the gravitational potential of the host and spread over a large range of apocentric distances. In contrast, if the GCs have a low radial velocity with respect to the host when stripped,  they reach in the apocenters of their orbits only the inner steep part of the gravitational potential. Therefore their apocentric distances span only a narrow range (illustrated in Fig.~2 of BSR19). A break in the profile of the GC system is then expected at the border between the steep an shallow part of the gravitational potential of the host, that is near the $a_0$ radius.

The same reasoning can be applied to MOND. The only difference is that the change of the slope of the gravitational potential in the inner and outer regions of the galaxy is not because of a dark matter halo, but because of the strong and weak field regimes of MOND.

If such an interpretation was true, then the breaks in the density of GCSs would be very useful for investigation of dark matter halos under the assumption of the \lcdm cosmology. The breaks would mark the radius at which the gravitational field changes from the baryon-dominated to the dark-mater dominated regime. This would allow estimating the effective radius of the dark matter halos. It is already known that the masses of dark matter halos of galaxies can be estimated from the number of GCs that the galaxies have \citep{spitler09,harris15}. 

It would be ideal to explore if this mechanism of formation of breaks in GCS density profiles actually works through simulations. This is beyond the scope of this paper. Here we instead resort to looking for observational signatures of this scenario.

For giant galaxies, most blue GCs are deemed to be accreted while many of the red GCs are deemed to be formed in situ \citep{cote98,hilker99}. Under this scenario of the origin of the breaks, we would thus expect that the breaks would be more pronounced for the blue GCs. Our data do not show this -- the broken number density profiles of GCSs are observed for both red and blue subpopulations and the magnitude of the break, $a-b$, does not seem to be different for the two subpopulations. A quantification through simulations however remains desirable.

Next, the galaxy sample presented here contains, unlike the sample of BRS19, also relatively low-mass galaxies, with the masses of the Small Magellanic Cloud. Low-mass galaxies are expected to form mostly by in situ star formation, without a substantial growth through mergers. From this point of view, it is unexpected that we detected breaks also in the low-mass galaxies. On the other hand, the early-type galaxies of our sample, that are supported mainly by velocity dispersion, might have richer merging histories than the typical dwarf galaxies, which are rotating. It is known, however, that dwarf galaxies in cluster environments can transform from disky, rotating dwarfs to spheroidal, dispersion dominated dwarfs via tidal stirring \citep[e.g.,][]{mayer01,mayer01b}. In total, we found tentative evidence against the origin of the breaks of GCSs through this mechanism. 

\begin{figure}
        \resizebox{\hsize}{!}{\includegraphics{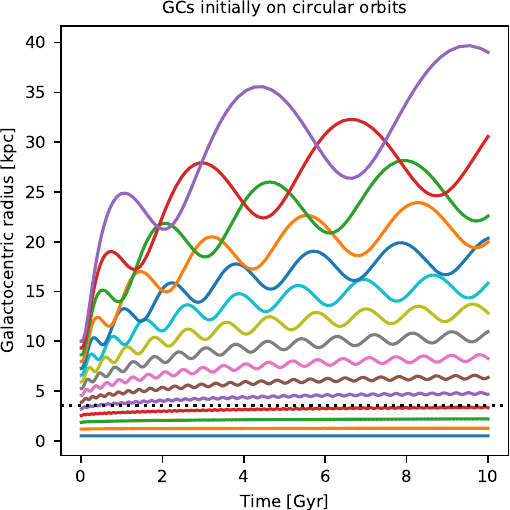}}
        \caption{Impact of the increasing of the  external field on the trajectories of GCs that were originally on circular orbits of different radii, assuming the MOND gravity. The horizontal line indicates the MOND $a_0$ radius. }
        \label{fig:efe}
\end{figure}

\subsection{MOND external field effect and stripping of dark halos}
\label{sec:efe}
In MOND, the dynamics of an object follows the radial acceleration relation only if the object is isolated. If the object is located in an  external gravitational field (for example a galaxy in a galaxy cluster), the apparent enhancement of gravity compared to the Newtonian gravity diminishes because of the nonlinearity of the theory. Once the strength of the external field surpasses the constant $a_0$ substantially, then the dynamics of the object behaves as in Newtonian dynamics without dark matter. This is called the external field effect \citep{milg83a,bm84}. If interpreted in the Newtonian way, an object that is exerted to a gradually stronger external field behaves as it was losing its dark matter halo. Observational evidence for the external field effect has been reported \citep{mcgaugh13b,mcgaugh16b, caldwell17,chae20}, even if the galaxies in clusters might be an exception \citep{freundlich22}. 

Galaxy clusters assemble by accreting individual galaxies from less dense environments. The external field effect then reduces the gravitational fields of the galaxies beyond their $a_0$ radii. At smaller radii, the gravitational field remains Newtonian, as it was when the galaxies were far from the cluster. Therefore, one expects that a break will develop at the $a_0$ radius.  

We made a simple model to explore the impact of this process. The GCs were initiated as orbiting a point source with a mass of $10^{10}\,M_\sun$ on circular orbits and a zero external field was assumed.  Then the model was evolved for 10\,Gyr, increasing the magnitude of the external field linearly with time, such that the external field eventually reached the value of $2a_0$, a value typical for the cores of galaxy clusters \citep{milg09}. The magnitude of the gravitational force in the presence of an external field was calculated using the so-called 1-D approximation of QUMOND \citep{famaey12}. 

The trajectories of the modeled GCs are shown in \fig{efe}. The horizontal dotted black line indicates the MOND $a_0$ radius. While the GCs below the $a_0$ radius stay at their initial orbits, the distant GCs recede as the consequence of the external field effect. This causes a dilution of the GCS beyond the $a_0$ radius, such that the profile bends down.

It is possible to estimate the impact of the external field effect on the distribution of GCs analytically. If the gravitational potential of the host remains spherical when it is changing, then there is no tidal torque acting on a GC moving on a circular orbit. The angular momentum of the GC is then conserved. The angular momentum of a GC orbiting the host galaxy well  beyond the $a_0$ radius when the galaxy is far from the cluster, is $r_0\sqrt[4]{GMa_0}$. Once the GC and its host appear in a strong external field, compared to $a_0$, the gravitational field of the host becomes Newtonian and thus the angular momentum of a GC on a circular orbit becomes $r_1\sqrt{\frac{GM}{r_1}}$. From the conservation of angular momentum we get:
\begin{equation}
r_1 = r_0^2\sqrt{\frac{a_0}{GM}}.
\end{equation}

{If the Newtonian gravity is assumed,} galaxies that enter galaxy clusters can reduce their gravitational fields by stripping of their dark matter halos in the consequence of tidal interactions with the other galaxies \citep{lee18,mitrasinovic22}. This can mimic the external field effect to a certain degree. 

Even this mechanism of creating breaks in GCS density profiles through the reduction of the gravitational fields of galaxies is not perfect. The central galaxies of clusters, that spent their whole lives in the clusters, are not expected to reduce their gravitation fields. Yet NGC\,1399 shows a break in the GCs density profile. Moreover, given that the intensity of the external field or tidal stripping is probably different for every galaxy, it is not clear why the galaxies would end up with a relatively narrow range of the external slope of the GCSs, $b$. Finally, some of the galaxies investigated in BSR19 were isolated (e.g., NGC\,821 and NGC\,3115), and thus never had an opportunity to develop a break through the external field effect or the tidal stripping of their halo.

\subsection{Two regimes of dynamical friction}
\label{sec:friction}
When a GC orbits its host galaxy, it attracts  the stars, gas, or dark matter particles of the host galaxy gravitationally and gives them kinetic energy and angular momentum. As consequence of the laws of conservation of these quantities, the orbital angular momentum and energy of the GCs decrease. This is called dynamical friction. Dynamical friction manifests itself as a force acting on the GC against the direction of the velocity of the GC with respect to its environment. In Newtonian gravity, the magnitude of the dynamical friction force can be estimated by Chandrasehkar's formula \citep{chandrasekhar43}:
\begin{equation}
\begin{split}
   F_\mathrm{DF, NWT} & = \frac{ 2\pi\ln \Lambda G^2\rho m^2}{\sigma^2 X^2}\left[ \erf(X)-\frac{2X}{\sqrt\pi}\exp\left(-X^2\right)\right], \\ X & = \frac{v}{\sqrt{2}\sigma},
   \label{eq:chandra}
\end{split}
\end{equation}
where $m$ stands for the mass of the GC, $v$ its velocity with respect to its local environment, $\rho$ the density of the local environment and  $\sigma$ the velocity dispersion of the environment. The expression $\ln \Lambda$ is called the Coulomb logarithm. Its value depends on the exact configuration of the problem under consideration, but it is on the order of a few.

A MOND analog of Chandrasekhar's formula was proposed by \citet{sanchezsalcedo06} on the basis of heuristic arguments and theoretical results by \citet{ciotti04}:
\begin{equation}
F_\mathrm{DF, MOND}^\mathrm{WF} = \frac{a_0^2}{\sqrt{2}a^2}F_\mathrm{DF, NWT}.
\end{equation}
Here $a$ denotes the gravitational acceleration exerted by the host galaxy on the GC.  S\'anchez-Salcedo's formula has recently been verified by simulations of GCs orbiting ultra-diffuse galaxies by \citet{bil21}. It is however supposed to work only in the weak-field regime of MOND, that is for  $a\ll a_0$. If $a\gg a_0$, the dynamical friction force should reduce back to the one given by Chandrasekhar's formula. 

\begin{figure}[t!]
        \resizebox{\hsize}{!}{\includegraphics{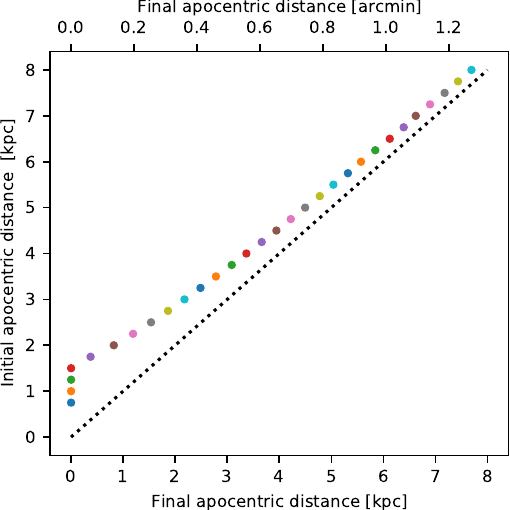}}
        \caption{Effect of dynamical friction on the apocentric radii of GCs that initially move in apocenter with a velocity that is equal to 0.1 of the local circular velocity. }
        \label{fig:dynfrict}
\end{figure}

Therefore here we heuristically propose a universal formula for dynamical friction in MOND:
\begin{equation}
F_\mathrm{DF, MOND} = \xi\left(\frac{a}{a_0}\right)F_\mathrm{DF, NWT},
\end{equation}
where $\xi(x)$ is a function satisfying the limit behavior:
\begin{equation}
    \begin{split}
        \xi(x) & \rightarrow 1~~ \textrm{for}~~ x\gg 1, \\
        \xi(x) & \rightarrow \frac{1}{\sqrt{2}x^2}~~ \textrm{for}~~ x\ll 1.
    \end{split}
\end{equation}
It was proposed in BSR19 that these two regimes of dynamical friction in MOND might give rise to the observed breaks in the GCS density profiles at the $a_0$ radii. For the purpose of the exercise below, we arbitrarily chose
\begin{equation}
    \xi(x) = 1+\frac{1}{\sqrt{2}x^2}. 
\end{equation}

We explored if dynamical friction can influence the distribution of GCs noticeably through simplistic models of the dynamics of the GCs of NGC\,1399 and NGC\,1373. These two are representatives of the most and the least massive galaxies in our sample, respectively. We consider models assuming either Newtonian gravity with dark matter or the MOND gravity.

The distribution of stars of the galaxies was modeled by S\'ersic spheres described by the parameters stated in \tab{gals}. It also was important to include the circumgalactic media in our models given the appearance of $\rho$ in \equ{chandra}. The break radii of the GCSs in our galaxy sample exceed 2-3 times the effective radii of the host galaxies and thus the contribution of the circumgalactic gas to the local baryonic density might be substantial.  The masses of circumgalactic media are generally difficult to determine. They could possibly be comparable to the stellar masses of galaxies \citep{gupta12,werk14,keeney17,das19}. Here we modeled the distribution of gas around NGC\,1399 according to the observed density of hot gas presented in \citet{paolillo02} in their Fig.~17. As central cluster galaxy, NGC\,1399 is not expected to contain much cold gas \citep{yoon17,burchett18}. The extended gaseous component of NGC\,1373 was modeled by a S\'ersic sphere whose mass and S\'ersic index are the same as of the stellar component but the effective radius is four times larger. 

For the Newtonian models, we assumed the NFW dark matter halo derived for NGC\,1399  by \citet{bil19}. It was deduced from an isotropic model of the kinematics of the GCS of the galaxy. For NGC\,1373, we assumed a NFW dark matter halo of the virial mass of $10^{11.35}\,M_\sun$ and a scale radius 13.4\,kpc according to the stellar-to-halo mass relation \citep{behroozi13} and the halo mass-concentration relation \citep{diemer15}.

In the MOND models, it was necessary to consider the external field effect coming from the Fornax cluster. For  the MOND model of NGC\,1373, we included the external field effect such that the galaxy was assigned the Newtonian gravitational field produced solely by the stars of the galaxy. For NGC\,1399, as for a central galaxy of the cluster, a zero external field was assumed. 

The velocity dispersion that appears in \equ{chandra} was obtained by solving the spherical Jeans equation  with a zero anisotropy parameter. All material was approximated as consisting of a single population of collisionless particles. \citet{bil21} found that for GCs on circular orbits, the value of the Coulomb potential is around 10 and it is around 3 for radial orbits. In the following calculations,  we assumed the value of 10. The GCs in our models had a mass of  $10^5\,M_\sun$. Their orbits were integrated for 10\,Gyr.

For NGC\,1399, we found that neither for Newtonian or MOND gravity  dynamical friction is able to affect the orbits of GCs noticeably. We explored various orbital eccentricities of the GCs and different apocentric distances in the range between 10 and 200\,kpc. Dynamical friction started to influence the orbit of the modeled GC noticeably only after the mass of the GC was increased to at least about $10^{8-9}\,M_\sun$, depending on the orbit shape. Dynamical friction is thus rather expected to affect the spatial distribution of the satellite galaxies. With the mass of the GC of $10^5\,M_\sun$, the situation did not change for the MOND model even if we multiplied the density of matter of the galaxy by a factor of 30 to model the hypothesized cluster baryonic dark matter \citep{milg09}.

For NGC\,1373, we explored various orbital eccentricities of the GCs and different apocentric distances in the range between 0.6 to 8 kpc. The friction came out to be the strongest near the center of the galaxy. Radial orbits were affected more than circular. The GCs that moved within about one effective radius of the galaxy eventually settled in the center of the galaxy, both in the Newtonian and MOND models. The central deficit of GCs was immediately balanced by sinking of the GCs that originally moved on bigger orbits. As result, generally no substantial breaks in the GCS density profiles were induced. We show in \fig{dynfrict} an example of orbits that create a very strong break, compared to the other orbits we explored. The break is interestingly located near the $a_0$ radius. The figure shows the initial versus the final apocentric distances of GCs that move in the Newtonian model and were launched with velocities that are equal to 0.1 of the local circular velocity. The sinking of the GCs caused a mild bend of the density profile at around 2\,kpc, which is where the break in the real galaxy is observed. The black dotted line indicates the one-to-one relation. Once the GCs were put on circular orbits, the break stopped being noticeable. In a real galaxy, where the GCs orbits have various eccentricities, the break induced by dynamical friction would thus be even milder than in \fig{dynfrict}. We verified that there is no change in these conclusions even for the intermediate-mass galaxy NGC\,1379.

This demonstrates that dynamical friction on a GC is alone probably not responsible for the observed breaks in GCS density profiles. In low-mass galaxies, it is important for the dynamics of GCs within the stellar body of the galaxy, but the contribution to the formation of the breaks seems to be insignificant.

\subsection{Change of mass or $a_0$}
Galaxies can increase their baryonic mass, either by forming stars in situ or by accreting other galaxies. Early-type galaxies, that are investigated in this paper, are also suspected to loose their baryonic mass through baryonic outflows \citep{fan08,damjanov09,ragone11}. Let us approximate the galaxy by a point source that  has a mass of $M$ initially and $\epsilon M$  after the mass change.  According to the radial acceleration relation, the gravitational acceleration in the strong field regime will change by a factor of $\epsilon$, while in the weak field regime by a factor of $\sqrt{\epsilon}$. Using the argument with the conservation of angular momentum from \sect{efe}, we found that a circular orbit with an initial radius of $r_0$ will change, after the mass change, to $r_1 = r_0/\epsilon$ in the strong field regime and to $r_1 = r_0/\epsilon^{1/4}$ in the weak field regime. This suggests that the change of the baryonic mass could introduce only a different normalization of the GCS density profile inside and beyond the $a_0$ radius, but not the observed change of slope. On the other hand, GCs in GCSs move on trajectories with different eccentricities, which might possibly change the situation. In total, the change of mass does not seem to be responsible for the formation of the observed breaks, but simulations are necessary to exclude this option definitively. 

It was proposed that in a MOND universe, the value of $a_0$ could vary with time \citep{milg83a}. \citet{milga0var} showed that the variation would affect the orbits of objects in a similar fashion as the change of mass described above, but the orbits would be expanding. Again, the above arguments can be used against this mechanism to be responsible for the formation of the observed breaks of density of GCSs.

\subsection{Other influences shaping the density profile of a GCS}
The radial profiles of number density of GCSs can be also shaped by other processes than those listed above. Here we mention {some examples, even if it is not currently clear why they would introduce GCS density breaks near  the $a_0$ radii.} They should be considered in more complex models of the formation of the density profiles of GCSs.

Early-type galaxies are known to grow their effective radius with time \citep{daddi05, trujillo06, vandokkum09}, even if they do not form new stars or do not increase their average stellar masses substantially. A promising explanation of the growth of their radii are repeated minor mergers \citep{naab09}. The premerger potential energies of the accreted satellites are transformed into kinetic energy of the stars and dark matter particles of the merger remnant. It is plausible that GCs would absorb some of the energy of the satellites too. This  would make the distribution of the GCs more extended, as already explored for major mergers in \citet{bekki06}. 

The distribution of the GCs can further be influenced by the tidal destruction of the GCs. Many GCs of the Milky Way are known to have  stellar streams \citep[e.g.,][]{ibata21}. It is possible that the mechanisms listed above will direct some GCs on radial orbits and once the GCs appear close to the center of the galaxy, they would be destroyed by tidal forces \citep{brockamp14}. {This process would cause the a decrease of the density of the GCS near its center.}

{The density profiles of dark halos derived from observations under the assumption of Newtonian dynamics tend to show central cores \citep{kormendy04,oh08,donato09,salucci12}. The proposed explanations include the baryonic feedback \citep[e.g.,][]{governato10,dicintio14}. Given that the feedback influences the motion and distribution of the dark matter particles, it will most probably influence the distribution of GCs in the same way. This would explain the observed flattening of the GC density profile toward the center of the galaxy. It is however unclear whether the core would have the $a_0$ radius. A mere presence of the dark matter core, formed by any mechanism, could affect the distribution of the GCs in the way described in \sect{tworegimes}. }

\section{Summary and conclusions}
\label{sec:sum}
It was found in BSR19 that the number density profiles of GCSs follow broken power laws (\equ{bpl}) whose break radii coincide with the $a_0$ radii of their host galaxies. An $a_0$ radius is defined as the radius at which the acceleration generated by the baryons of the galaxy equals the galactic acceleration constant $a_0$. It was shown in BSR19 that the $a_0$ radii coincide with the break radii better than other characteristic length scales of galaxies, such as the effective radii or the dark matter halo characteristic radii. The galaxy sample of BRS19 nevertheless {spans} only a relatively narrow range of baryonic masses of the galaxies, namely about one decade. They investigated the distribution of GCs only on the basis of spectroscopic catalogs that can suffer from geometrical incompleteness. They investigated only the total population of GCs. It was unclear, for example, whether the density profiles of the blue and red GC subpopulations follow the broken power-law as well and, if so, whether the break radii of the subpopulations are located at the $a_0$ radius. In the current contribution, we aimed to overcome these deficiencies. We analyzed two  catalogs of photometric GC candidates in the Fornax cluster, one based on the ground-based Fornax Deep Survey data and the other on the ACS Fornax cluster data. The profiles of density of GCSs for the lowest-mass galaxies were derived from stacks of the GC candidates over many galaxies of a similar stellar mass. Additionally, we inspected a new spectroscopic catalog of GCs in the vicinity of the central galaxy of the cluster, NGC\,1399. We were investigating only the GCSs of ETGs, since for LTGs, it is not possible to distinguish GCs from the numerous compact regions of star formation in the disks. The galaxy sample studied here spans logarithmic stellar masses from {8.0} to 11.4\,$M_\sun$. The fitted parameters of the GCS density profiles are listed in \tab{finalfits}.

Our observational findings can be summarized as follows:
\begin{enumerate}
    \item We were able to detect breaks in the GCS profiles of virtually all galaxies (\app{fits}). The exceptions are only the galaxies that have too few GCs. The breaks were found in the entire GC population as well as in the blue and red GC subpopulations (see the figures in \app{fits}).
    \item In the cases when the outer part of the broken power-law profile was observed in both the FDS and ACS catalogs, the outer slopes, $b$, generally agreed well, just as the break radii (\app{fits}). The inner slope in the FDS data was usually shallower than in the ACS data. We attribute this difference to the greater difficulty of detecting faint GCs near galactic centers in the ground-based data because of the {contamination by the light of the host galaxy} (\sect{fitting}). 
    \item The break radii of the total GC population and the red and blue subpopulations are rather similar (\sect{colors}). There is a marginal tendency for the blue GCs to have systematically higher break radii than the red GCs at the $1.5\sigma$ confidence level, namely by $0.3\pm0.2\,$kpc, on average.
    \item We calculated the $a_0$ radii in two ways: assuming the Newtonian gravity and the MONDian gravity, which provides a clear theoretical understanding of the existence of the acceleration scale $a_0$. We found that the break and $a_0$ radii agree typically within a factor of two (\sect{equality}). The $a_0$ radii calculated from the MOND gravity are less offset from the break radii than the $a_0$ radii calculated from the Newtonian gravity. 
    \item The break radii also show significant correlations with the stellar masses of the galaxies, their effective radii,  and S\'ersic indices (\sect{corr}). None of these correlations is however close to the one-to-one relation.
    \item Gravitational fields of some galaxies are weaker than the constant $a_0$ in the whole extents of the galaxies (\sect{ao}). Such galaxies thus do not have  $a_0$ radii. They still show broken power-law density profiles for their GCSs.    
    \item The outer slope of the GCSs profiles, $b$,  strongly correlates with the $a_0$ radii (both Newtonian and MOND, \sect{corr}). The correlations of $b$  with the $a_0$ radii are more significant than the correlation of $b$ with the stellar mass of the host galaxy. This suggests that the mechanism that sets the profile of the GCS is causally linked rather with the spatial distribution of mass in the galaxy than with its total stellar mass.
    \item The parameter $b$ for the blue GCs is higher (i.e., the profile is less steep) than for the red GCs at the $2\,\sigma$ confidence level (\sect{colors}). 
    \item We inspected in more detail the galaxy with the highest number of GCs, which is  NGC\,1399, the central galaxy of the Fornax cluster. We used the new catalog of spectroscopic GCs \citep{Chaturvedi22}. We divided the GCs into groups according to a similar absolute value of the radial velocity with respect to the center of the galaxy. The shape of the profile shows systematic trends with the mean velocity of the selected group (\sect{velcuts}, \fig{velslices}). 
    \item We divided the spectroscopic GCs around NGC\,1399 in azimuthal sectors  centered on the galaxy and derived the radial profiles of the density of GCs in each of the sectors. When the break radii are plotted in polar coordinates according to the angle of the middle line of the sector, they form an ellipse in the plane of the sky (\sect{azimuth}, \fig{sectors}). The major axis of the ellipse points toward the neighboring galaxy NGC\,1404. It is well known that these two galaxies are undergoing a tidal interaction. This demonstrates that break radii are influenced by galaxy interactions.
\end{enumerate}

BSR19 proposed several explanations for the approximate match of the break and $a_0$ radii. We explored those and a few others in more detail (\sect{interpretation}). We made use of simple models and observational arguments. None of them explains our findings  completely satisfactory. More elaborate models and simulations are desirable. More data, ideally coming from a larger variety of environments and galaxy morphological types, could give us hints as to why the $a_0$ and break radii are similar.

\begin{acknowledgements}
{We thank the referees for their comments that helped to improve the quality of the manuscript. 

MB acknowledges the support by the ESO SSDF grant 21/10.  

FR acknowledges support from the Knut and Alice Wallenberg Foundation, and from the University of Strasbourg Institute for Advanced Study (USIAS) within the French national program Investment for the Future (Excellence Initiative) IdEx-Unistra.

SS acknowledges the financial support of the Ministry of Education, Science and Technological Development of the Republic of Serbia through the contract No.~451-03-68/2022-14/200002.

 The Digitized Sky Surveys were produced at the Space Telescope Science Institute under U.S. Government grant NAG W-2166. The images of these surveys are based on photographic data obtained using the Oschin Schmidt Telescope on Palomar Mountain and the UK Schmidt Telescope. The plates were processed into the present compressed digital form with the permission of these institutions.

The National Geographic Society - Palomar Observatory Sky Atlas (POSS-I) was made by the California Institute of Technology with grants from the National Geographic Society.

The Second Palomar Observatory Sky Survey (POSS-II) was made by the California Institute of Technology with funds from the National Science Foundation, the National Geographic Society, the Sloan Foundation, the Samuel Oschin Foundation, and the Eastman Kodak Corporation.

The Oschin Schmidt Telescope is operated by the California Institute of Technology and Palomar Observatory.

The UK Schmidt Telescope was operated by the Royal Observatory Edinburgh, with funding from the UK Science and Engineering Research Council (later the UK Particle Physics and Astronomy Research Council), until 1988 June, and thereafter by the Anglo-Australian Observatory. The blue plates of the southern Sky Atlas and its Equatorial Extension (together known as the SERC-J), as well as the Equatorial Red (ER), and the Second Epoch [red] Survey (SES) were all taken with the UK Schmidt.

All data are subject to the copyright given in the copyright summary. Copyright information specific to individual plates is provided in the downloaded FITS headers.

Supplemental funding for sky-survey work at the ST ScI is provided by the European Southern Observatory. }
\end{acknowledgements}

\bibliographystyle{aa}
\bibliography{literature}


\clearpage
\onecolumn

\begin{appendix}
\section{Detailed results of the GC surface density fitting}
\label{app:fits}

\begin{figure*}[h!]
        \centering
        \includegraphics[width=17cm]{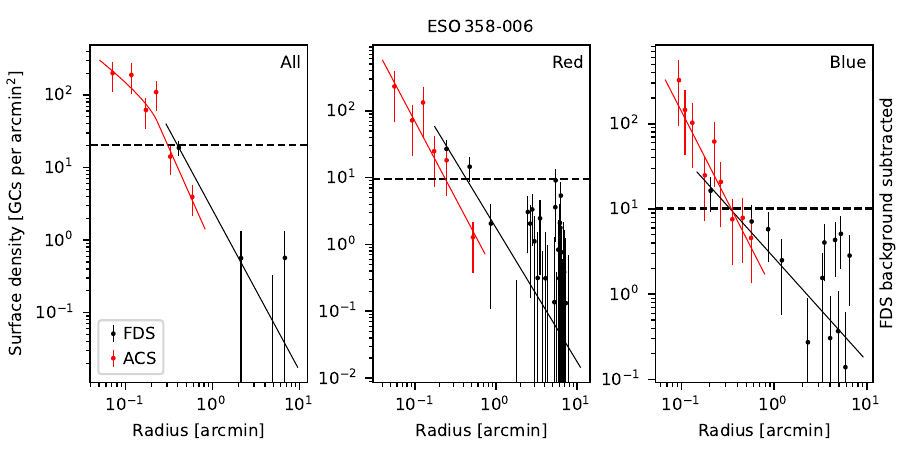}
        \caption{Fits of the surface density profile of the sources centered on ESO\,358-006. } 
        \label{fig:g1}
\end{figure*}

\begin{figure*}[h!]
        \centering
        \includegraphics[width=17cm]{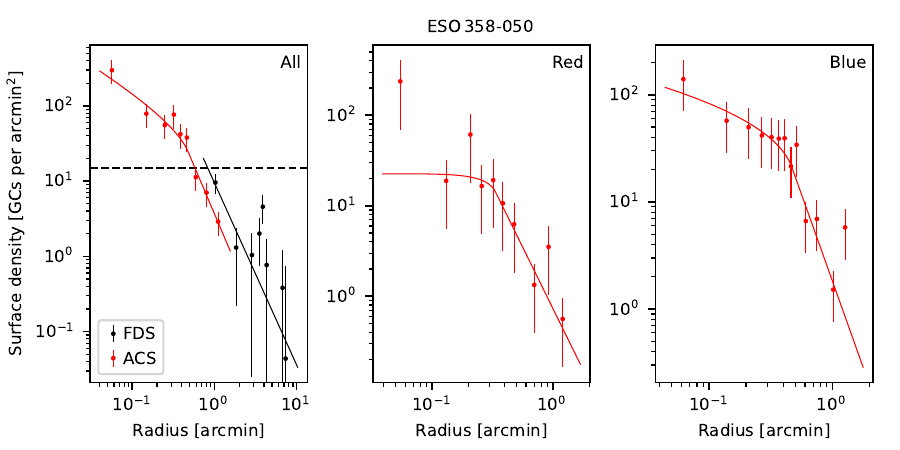}
        \caption{Fits of the surface density profile of the sources centered on ESO\,358-050. } 
        \label{fig:g2}
\end{figure*}

\begin{figure*}[h!]
        \centering
        \includegraphics[width=17cm]{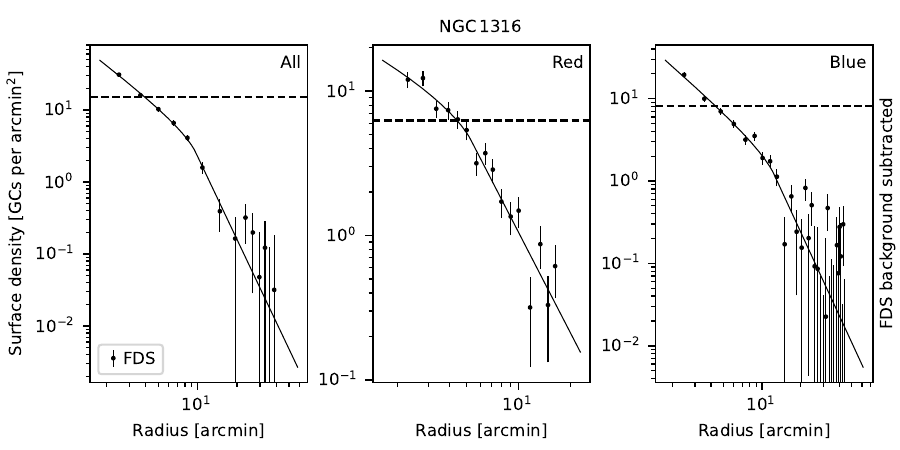}
        \caption{Fits of the surface density profile of the sources centered on NGC\,1316. } 
        \label{fig:g3}
\end{figure*}

\begin{figure*}[h!]
        \centering
        \includegraphics[width=17cm]{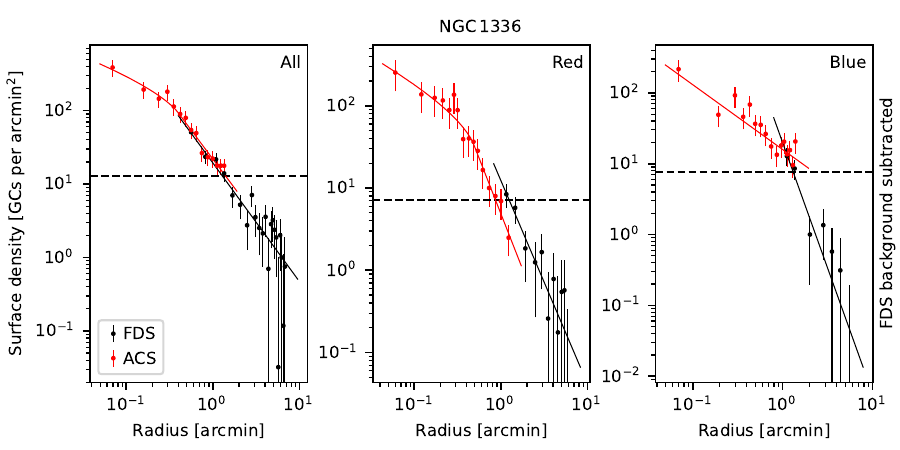}
        \caption{Fits of the surface density profile of the sources centered on NGC\,1336. } 
        \label{fig:g4}
\end{figure*}

\begin{figure*}[h!]
        \centering
        \includegraphics[width=17cm]{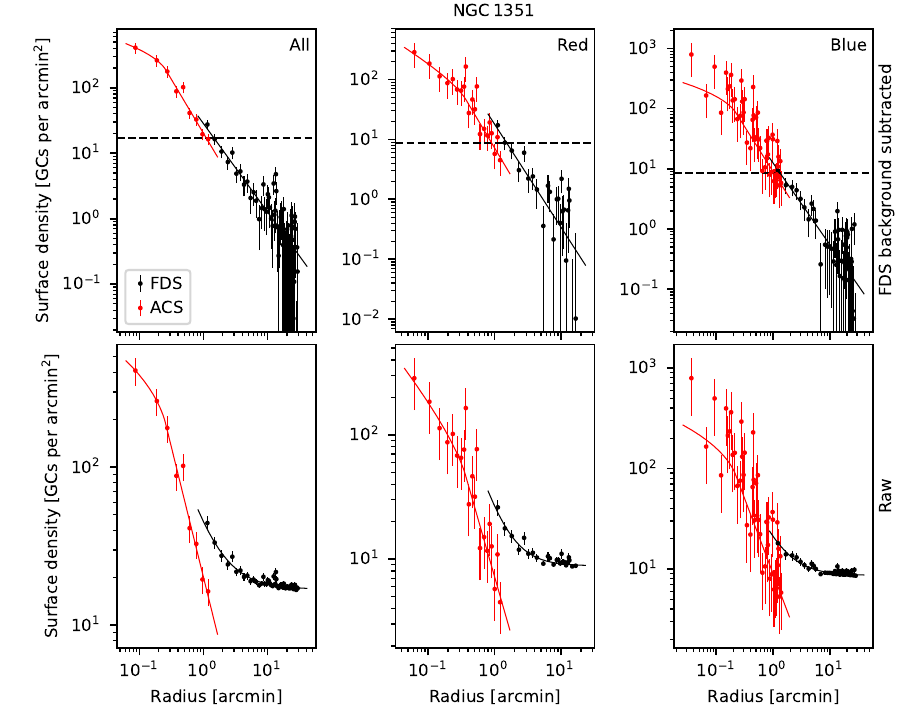}
        \caption{Fits of the surface density profile of the sources centered on NGC\,1351. } 
        \label{fig:g5}
\end{figure*}

\begin{figure*}[h!]
        \centering
        \includegraphics[width=17cm]{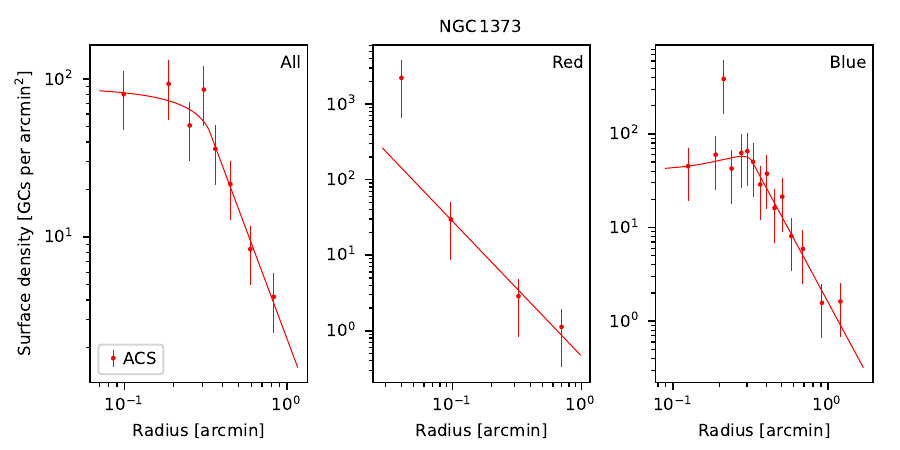}
        \caption{Fits of the surface density profile of the sources centered on NGC\,1373. } 
        \label{fig:g6}
\end{figure*}

\begin{figure*}[h!]
        \centering
        \includegraphics[width=17cm]{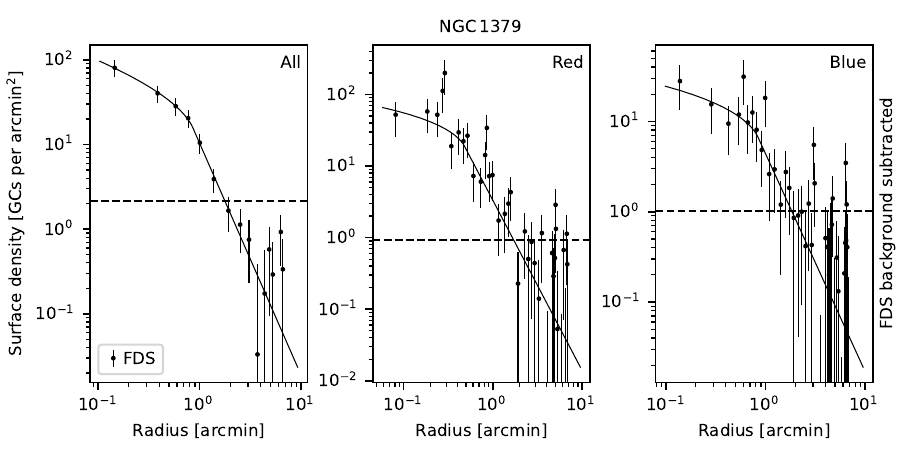}
        \caption{Fits of the surface density profile of the sources centered on NGC\,1379. } 
        \label{fig:g7}
\end{figure*}

\begin{figure*}[h!]
        \centering
        \includegraphics[width=17cm]{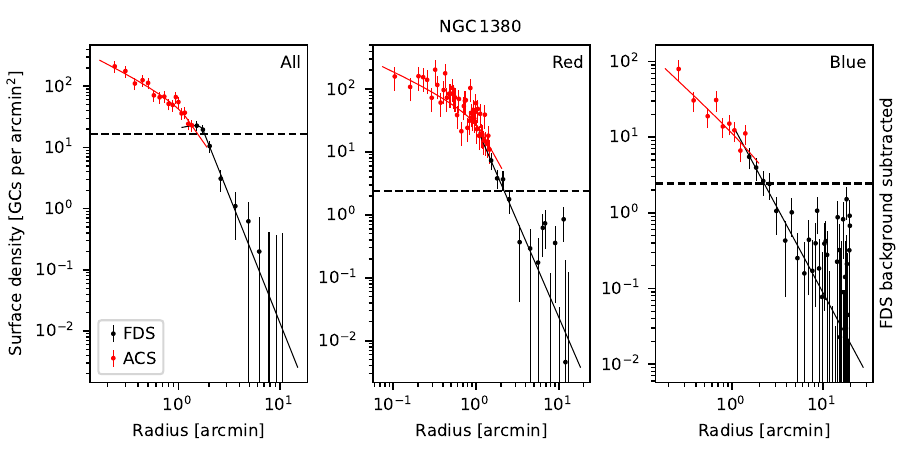}
        \caption{Fits of the surface density profile of the sources centered on NGC\,1380. } 
        \label{fig:g8}
\end{figure*}

\begin{figure*}[h!]
        \centering
        \includegraphics[width=17cm]{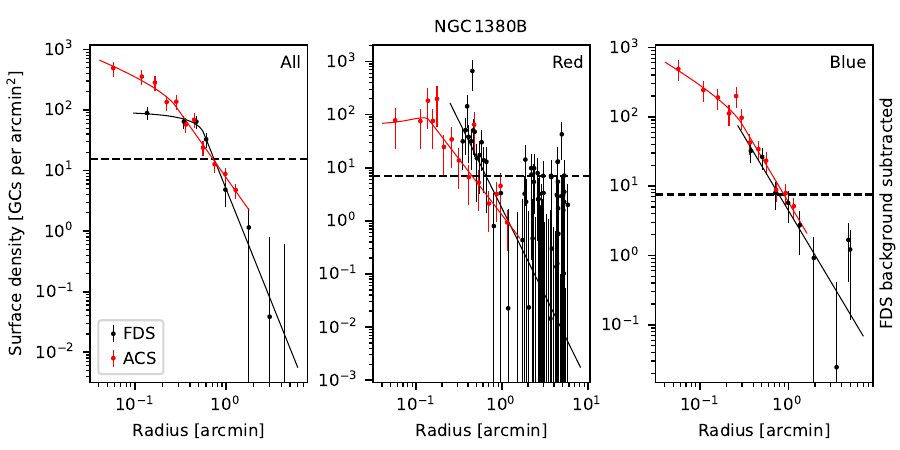}
        \caption{Fits of the surface density profile of the sources centered on NGC\,1380B. } 
        \label{fig:g9}
\end{figure*}

\begin{figure*}[h!]
        \centering
        \includegraphics[width=17cm]{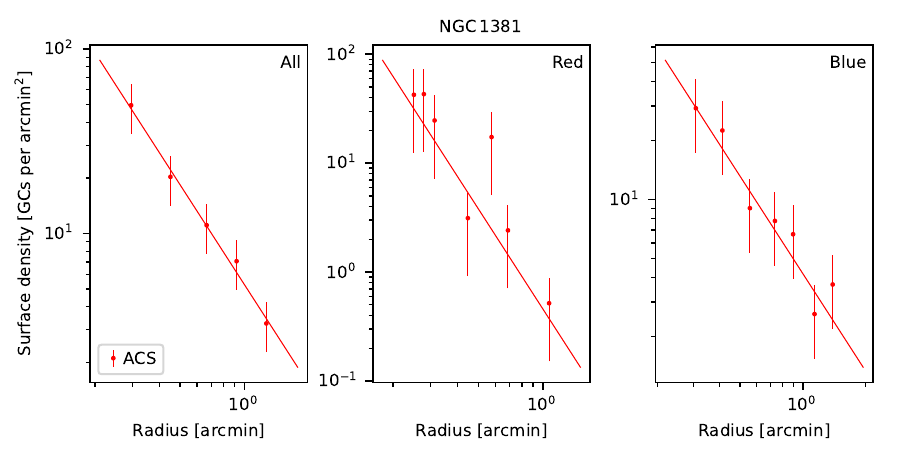}
        \caption{Fits of the surface density profile of the sources centered on NGC\,1381. } 
        \label{fig:g10}
\end{figure*}

\begin{figure*}[h!]
        \centering
        \includegraphics[width=17cm]{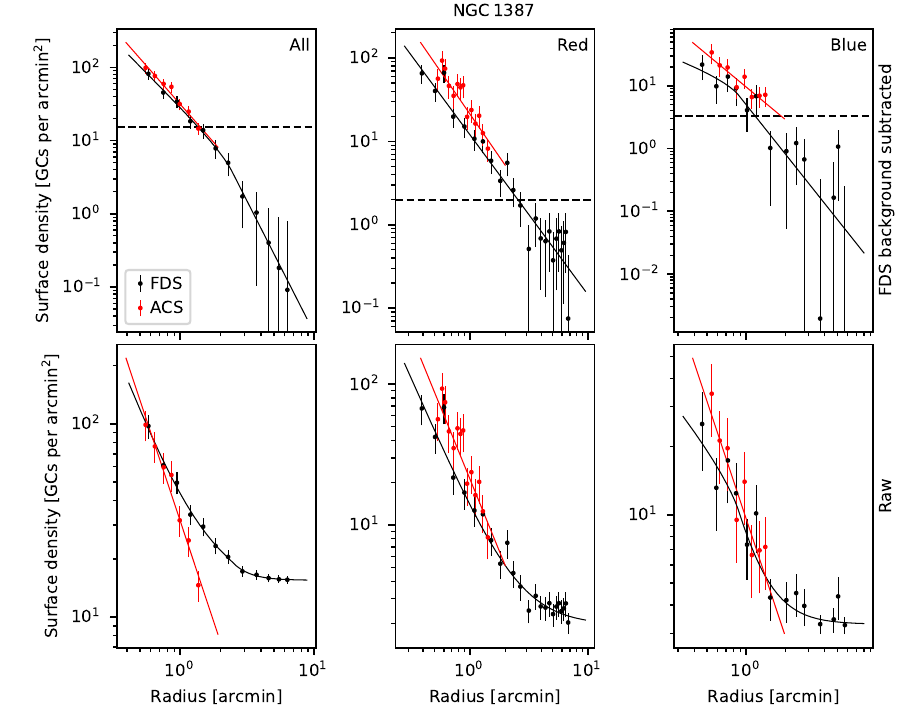}
        \caption{Fits of the surface density profile of the sources centered on NGC\,1387. } 
        \label{fig:g11}
\end{figure*}

\begin{figure*}[h!]
        \centering
        \includegraphics[width=17cm]{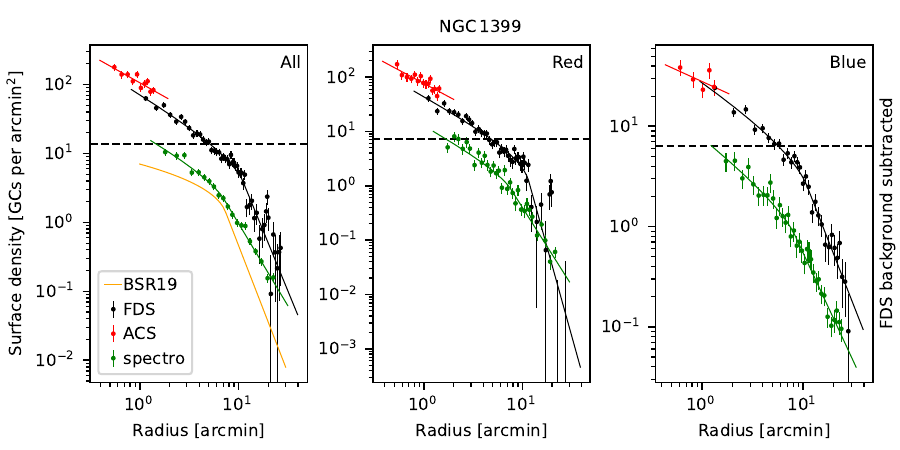}
        \caption{Fits of the surface density profile of the sources centered on NGC\,1399. } 
        \label{fig:g12}
\end{figure*}

\begin{figure*}[h!]
        \centering
        \includegraphics[width=17cm]{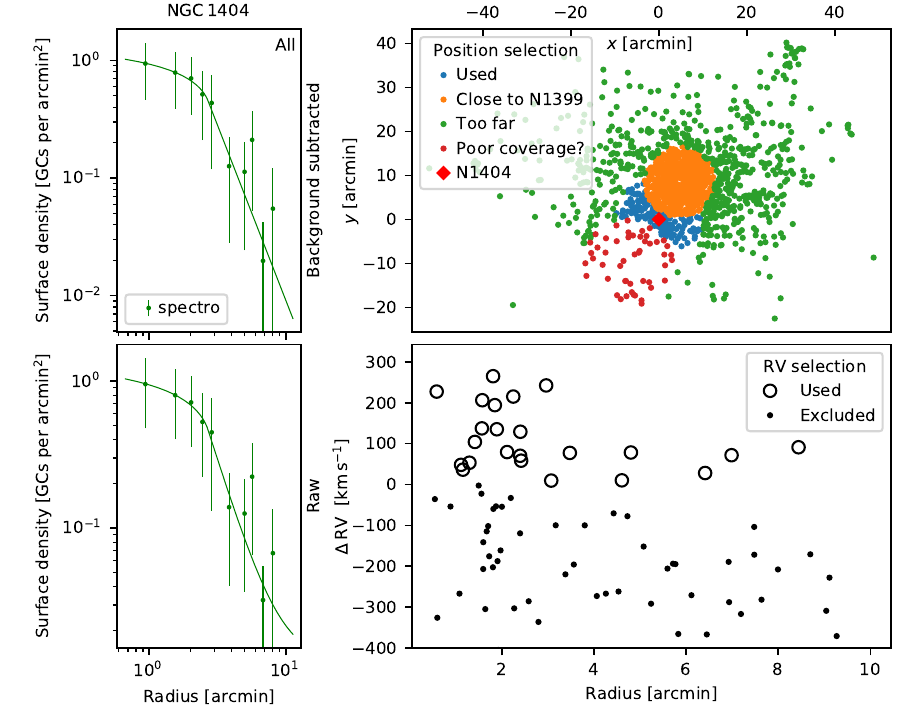}
        \caption{Fits of the surface density profile of the sources centered on NGC\,1404.}
        \label{fig:g13}
\end{figure*}

\begin{figure*}[h!]
        \centering
        \includegraphics[width=17cm]{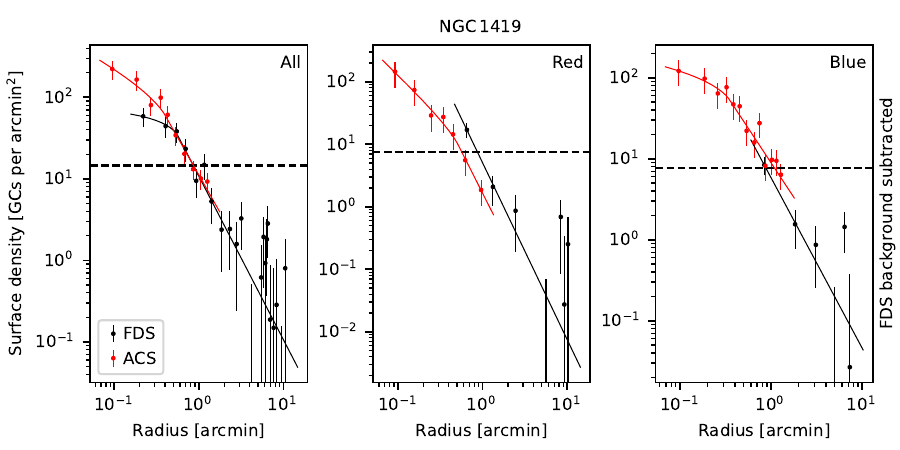}
        \caption{Fits of the surface density profile of the sources centered on NGC\,1419. } 
        \label{fig:g14}
\end{figure*}

\begin{figure*}[h!]
        \centering
        \includegraphics[width=17cm]{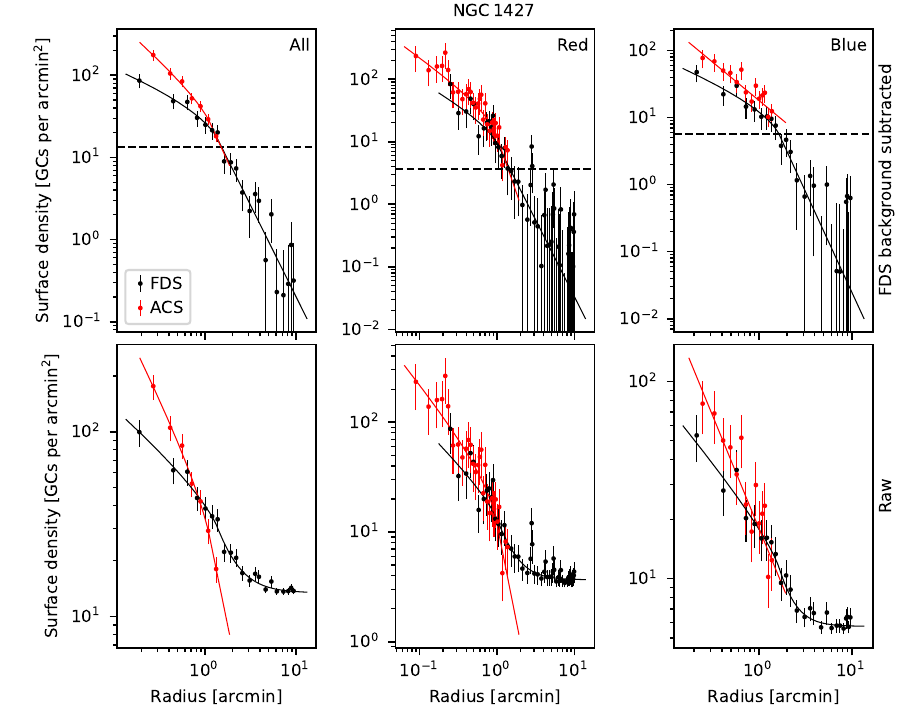}
        \caption{Fits of the surface density profile of the sources centered on NGC\,1427. } 
        \label{fig:g15}
\end{figure*}

\begin{figure*}[h!]
        \centering
        \includegraphics[width=17cm]{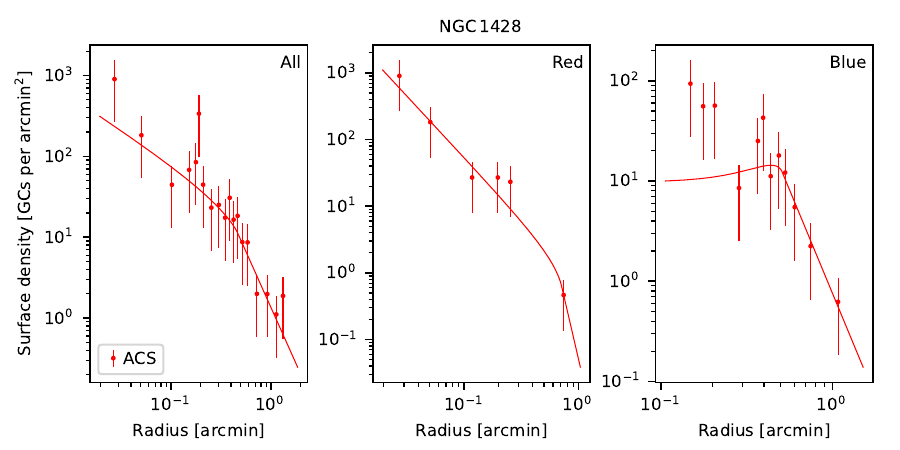}
        \caption{Fits of the surface density profile of the sources centered on NGC\,1428. } 
        \label{fig:g16}
\end{figure*}

\begin{figure*}[h!]
        \centering
        \includegraphics[width=17cm]{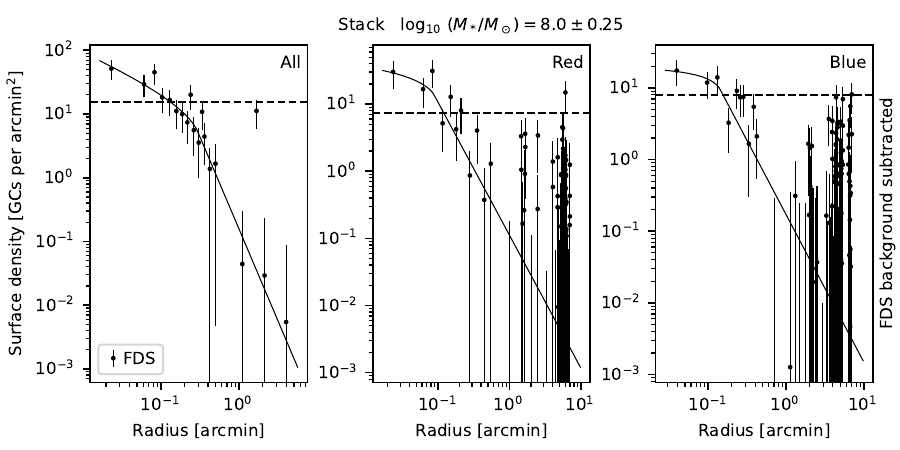}
        \caption{Fits of the surface density stacked profile of sources centered on galaxies with stellar masses $\log_{10} M_*/M_\sun = 8\pm0.25$. } 
        \label{fig:g17}
\end{figure*}

\begin{figure*}[h!]
        \centering
        \includegraphics[width=17cm]{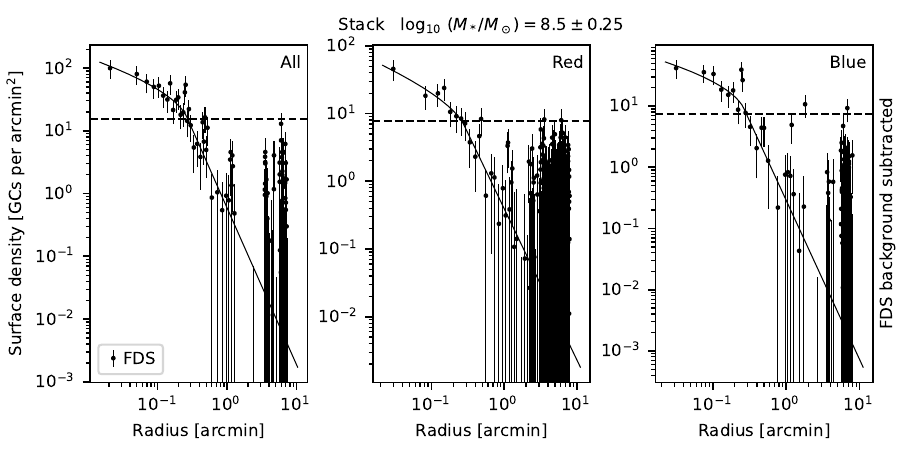}
        \caption{Fits of the surface density stacked profile of sources centered on galaxies with stellar masses $\log_{10} M_*/M_\sun = 8.5\pm0.25$. } 
        \label{fig:g18}
\end{figure*}

\begin{figure*}[h!]
        \centering
        \includegraphics[width=17cm]{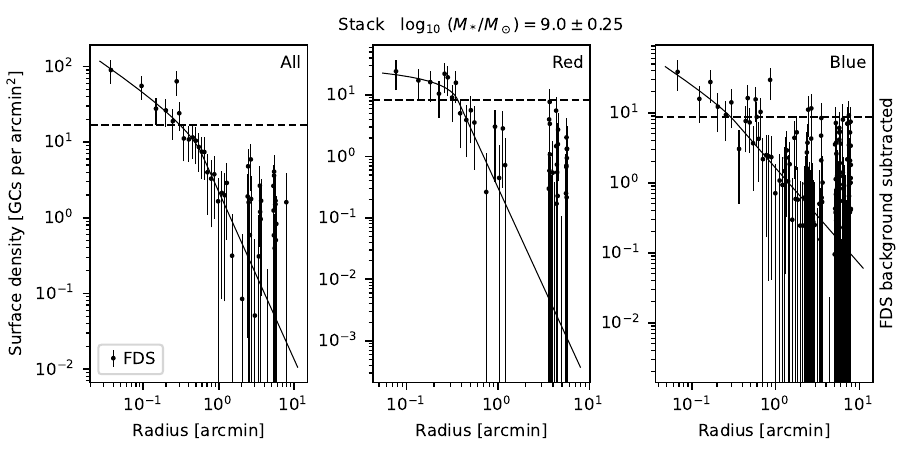}
        \caption{Fits of the surface density stacked profile of sources centered on galaxies with stellar masses $\log_{10} M_*/M_\sun = 9\pm0.25$. } 
        \label{fig:g19}
\end{figure*}

\begin{landscape}
\begin{longtable}{p{2cm}llllllllllp{4cm}}
        \caption{\label{tab:fits} Fits of volume number densities of GCS of all galaxies and all datasets investigated in this paper.}\\ \hline\hline
    Galaxy & GC type & Data & $\rho_0$ & $a$ & $b$ & $r_\mathrm{br}$ & $\gamma$ & $r_\mathrm{min}$ & $r_\mathrm{max}$ & $g_\mathrm{max}$  & Notes \\
      &   &   & [arcmin$^{-3}$] &   &   & [arcmin] &   & [arcmin] & [arcmin] & [mag]  & \\
\hline
\endfirsthead
\caption{ Fits of volume number densities of GCS of all galaxies and all datasets investigated in this paper,  continued}\\ \hline\hline
    Galaxy & GC type & Data & $\rho_0$ & $a$ & $b$ & $r_\mathrm{br}$ & $\gamma$ & $r_\mathrm{min}$ & $r_\mathrm{max}$ & $g_\mathrm{max}$  & Notes \\
      &   &   & [arcmin$^{-3}$] &   &   & [arcmin] &   & [arcmin] & [arcmin] & [mag] &   \\
\hline
\endhead
\hline
\endfoot
ESO\,358-006  &  All  &  FDS  &  $1.39^{+0.8}_{-1}$  &  $-3.22^{+0.8}_{-10}$  &  --  &  --  &  $20.59^{+0.5}_{-0.6}$  &  0.0  &  8  &  --  &  --   \\
           &  Red  &  FDS  &  $0.89^{+0.4}_{-0.5}$  &  $-3.00^{+0.5}_{-0.7}$  &  --  &  --  &  $9.48^{+0.3}_{-0.3}$  &  0.0  &  8  &  --  &  --   \\
           &  Blue  &  FDS  &  $0.97^{+0.4}_{-0.4}$  &  $-2.21^{+0.4}_{-0.5}$  &  --  &  --  &  $10.15^{+0.5}_{-0.9}$  &  0.0  &  6.6  &  --  &  --   \\
           &  All  &  ACS  &  $7.6^{+300}_{-6}$  &  $-1.86^{+0.2}_{-0.2}$  &  $-3.70^{+0.7}_{-1}$  &  $0.228^{+0.06}_{-0.1}$  &  --  &  0  &  1.5  &  --  &  4 GC per bin   \\
           &  Red  &  ACS  &  $0.19^{+0.1}_{-0.1}$  &  $-3.29^{+0.3}_{-0.4}$  &  --  &  --  &  --  &  0  &  1.5  &  --  &  1 GC per bin   \\
           &  Blue  &  ACS  &  $0.54^{+0.5}_{-0.3}$  &  $-3.13^{+0.6}_{-0.6}$  &  --  &  --  &  --  &  0  &  1.5  &  --  &  1 GC per bin   \\
\hline
ESO\,358-050  &  All  &  FDS  &  $5.3^{+4}_{-3}$  &  $-3.4^{+1}_{-2}$  &  --  &  --  &  $14.94^{+0.4}_{-1}$  &  0.8  &  8  &  --  &  --   \\
           &  All  &  ACS  &  $9.1^{+10}_{-4}$  &  $-1.67^{+0.2}_{-0.1}$  &  $-3.57^{+0.5}_{-0.6}$  &  $0.469^{+0.09}_{-0.1}$  &  --  &  0  &  1.5  &  --  &  --   \\
           &  Red  &  ACS  &  $30^{+400}_{-30}$  &  $0.12^{+0.6}_{-0.5}$  &  $-3.71^{+0.8}_{-1}$  &  $0.331^{+0.09}_{-0.1}$  &  --  &  0.03  &  1.5  &  --  &  1 GC per bin   \\
           &  Blue  &  ACS  &  $13.2^{+60}_{-8}$  &  $-1.14^{+0.2}_{-0.2}$  &  $-4.23^{+0.7}_{-1}$  &  $0.463^{+0.08}_{-0.08}$  &  --  &  0  &  1.5  &  --  &  3 GCs per bin   \\
\hline
NGC\,1316  &  All  &  FDS  &  $45.2^{+10}_{-9}$  &  $-2.387^{+0.03}_{-0.03}$  &  $-4.82^{+0.7}_{-2}$  &  $9.6^{+1}_{-1}$  &  $15.248^{+0.05}_{-0.06}$  &  2  &  45  &  --  &  --   \\
           &  Red  &  FDS  &  $6.2^{+10}_{-3}$  &  $-1.52^{+0.2}_{-0.1}$  &  $-3.32^{+0.5}_{-20}$  &  $5.23^{+3}_{-0.8}$  &  $6.26^{+0.3}_{-0.4}$  &  2  &  17  &  --  &  --   \\
           &  Blue  &  FDS  &  $25.2^{+6}_{-5}$  &  $-2.350^{+0.02}_{-0.04}$  &  $-4.37^{+0.8}_{-2}$  &  $12.2^{+2}_{-2}$  &  $8.199^{+0.05}_{-0.07}$  &  2  &  45  &  --  &  --   \\
\hline
NGC\,1336  &  All  &  FDS  &  $8.4^{+1}_{-1}$  &  $-2.62^{+0.2}_{-0.2}$  &  --  &  --  &  $13.00^{+0.7}_{-0.9}$  &  0.45  &  7  &  --  &  --   \\
           &  Red  &  FDS  &  $7.0^{+5}_{-3}$  &  $-3.5^{+1}_{-1}$  &  --  &  --  &  $7.07^{+0.4}_{-0.7}$  &  1.0  &  7  &  --  &  --   \\
           &  Blue  &  FDS  &  $13.9^{+8}_{-6}$  &  $-4.54^{+0.9}_{-0.6}$  &  --  &  --  &  $7.54^{+0.3}_{-0.3}$  &  1.0  &  7  &  --  &  Break near the excluded inner region   \\
           &  All  &  ACS  &  $29^{+50}_{-10}$  &  $-1.39^{+0.1}_{-0.1}$  &  $-2.58^{+0.2}_{-0.2}$  &  $0.389^{+0.09}_{-0.09}$  &  --  &  0  &  1.5  &  --  &  --   \\
           &  Red  &  ACS  &  $13.6^{+200}_{-6}$  &  $-1.56^{+0.2}_{-0.1}$  &  $-3.68^{+0.5}_{-0.6}$  &  $0.483^{+0.08}_{-0.1}$  &  --  &  0  &  1.5  &  --  &  --   \\
           &  Blue  &  ACS  &  $4.74^{+0.4}_{-0.4}$  &  $-1.92^{+0.1}_{-0.1}$  &  --  &  --  &  --  &  0  &  1.5  &  --  &  Break likely at the border of ACS field   \\
\hline
NGC\,1351  &  All  &  FDS  &  $11.0^{+2}_{-2}$  &  $-2.35^{+0.1}_{-0.1}$  &  --  &  --  &  $16.94^{+0.1}_{-0.1}$  &  1  &  30  &  --  &  --   \\
           &  Red  &  FDS  &  $8.2^{+3}_{-2}$  &  $-2.72^{+0.3}_{-0.3}$  &  --  &  --  &  $8.82^{+0.2}_{-0.2}$  &  1  &  30  &  --  &  --   \\
           &  Blue  &  FDS  &  $4.9^{+2}_{-1}$  &  $-2.36^{+0.2}_{-0.2}$  &  --  &  --  &  $8.62^{+0.1}_{-0.1}$  &  1  &  30  &  --  &  --   \\
           &  All  &  ACS  &  $83^{+200}_{-60}$  &  $-1.03^{+0.1}_{-0.1}$  &  $-2.65^{+0.2}_{-0.2}$  &  $0.252^{+0.05}_{-0.05}$  &  --  &  0  &  1.5  &  --  &  --   \\
           &  Red  &  ACS  &  $11.9^{+40}_{-7}$  &  $-1.65^{+0.1}_{-0.1}$  &  $-2.87^{+0.3}_{-0.4}$  &  $0.36^{+0.1}_{-0.1}$  &  --  &  0  &  1.5  &  --  &  4 GCs per bin   \\
           &  Blue  &  ACS  &  $35^{+300}_{-30}$  &  $-1.04^{+0.2}_{-0.2}$  &  $-2.49^{+0.2}_{-0.2}$  &  $0.213^{+0.07}_{-0.06}$  &  --  &  0  &  1.5  &  --  &  2 GCs per bin   \\
\hline
NGC\,1373  &  All  &  ACS  &  $70^{+300}_{-60}$  &  $-0.18^{+0.3}_{-0.2}$  &  $-3.76^{+0.6}_{-0.7}$  &  $0.333^{+0.05}_{-0.06}$  &  --  &  0  &  1.5  &  25.5  &  --   \\
           &  Red  &  ACS  &  $0.21^{+0.2}_{-0.1}$  &  $-2.79^{+0.8}_{-0.7}$  &  --  &  --  &  --  &  0  &  1.5  &  --  &  1 GC per bin   \\
           &  Blue  &  ACS  &  $1970^{+100}_{-200}$  &  $2.59^{+0.5}_{-0.4}$  &  $-4.07^{+0.5}_{-0.6}$  &  $0.321^{+0.04}_{-0.04}$  &  --  &  0  &  1.5  &  25.5  &  2 GCs per bin   \\
\hline
NGC\,1379  &  All  &  FDS  &  $9.3^{+5}_{-3}$  &  $-1.37^{+0.2}_{-0.2}$  &  $-3.74^{+0.5}_{-0.7}$  &  $0.85^{+0.1}_{-0.1}$  &  $2.15^{+0.2}_{-0.2}$  &  0  &  7  &  24  &  --   \\
           &  Red  &  FDS  &  $11.4^{+300}_{-6}$  &  $-0.94^{+0.3}_{-0.2}$  &  $-3.38^{+0.4}_{-0.5}$  &  $0.474^{+0.08}_{-0.1}$  &  $0.92^{+0.1}_{-0.1}$  &  0  &  7  &  24  &  3 GCs per bin   \\
           &  Blue  &  FDS  &  $3.9^{+30}_{-2}$  &  $-0.98^{+1}_{-0.8}$  &  $-3.41^{+0.6}_{-1}$  &  $0.83^{+0.2}_{-0.2}$  &  $1.03^{+0.1}_{-0.2}$  &  0  &  7  &  24  &  --   \\
\hline
NGC\,1380  &  All  &  FDS  &  $2.4^{+50}_{-2}$  &  $2.3^{+20}_{-2}$  &  $-5.15^{+0.8}_{-2}$  &  $1.74^{+0.2}_{-0.2}$  &  $16.67^{+0.2}_{-0.2}$  &  1.4  &  20  &  --  &  --   \\
           &  Red  &  FDS  &  $16.8^{+10}_{-8}$  &  $-4.05^{+0.3}_{-0.2}$  &  --  &  --  &  $2.413^{+0.07}_{-0.08}$  &  1.4  &  20  &  24.5  &  Break near the excluded inner region   \\
           &  Blue  &  FDS  &  $7.9^{+5}_{-3}$  &  $-3.23^{+0.2}_{-0.4}$  &  --  &  --  &  $2.436^{+0.06}_{-0.06}$  &  1.4  &  20  &  24.5  &  Break near the excluded inner region   \\
           &  All  &  ACS  &  $20.1^{+3}_{-3}$  &  $-1.76^{+0.2}_{-0.2}$  &  $-3.36^{+0.8}_{-20}$  &  $1.17^{+0.6}_{-0.3}$  &  --  &  0.2  &  1.5  &  --  &  --   \\
           &  Red  &  ACS  &  $13.0^{+2}_{-2}$  &  $-1.54^{+0.2}_{-0.1}$  &  $-3.37^{+0.8}_{-10}$  &  $1.14^{+0.4}_{-0.3}$  &  --  &  0.0  &  1.5  &  --  &  Break likely at the border of ACS field   \\
           &  Blue  &  ACS  &  $4.29^{+1}_{-0.7}$  &  $-2.10^{+0.3}_{-0.3}$  &  $-8.0^{+7}_{-9}$  &  $4.0^{+900}_{-2}$  &  --  &  0.2  &  1.5  &  --  &  Break likely at the border of ACS field   \\
\hline
NGC\,1380B  &  All  &  FDS  &  $48^{+100}_{-30}$  &  $-0.23^{+0.2}_{-0.2}$  &  $-4.76^{+0.9}_{-1}$  &  $0.563^{+0.05}_{-0.04}$  &  $15.42^{+0.5}_{-0.5}$  &  0.0  &  6  &  --  &  --   \\
           &  Red  &  FDS  &  $1.14^{+0.6}_{-0.5}$  &  $-4.28^{+0.6}_{-0.8}$  &  --  &  --  &  $7.09^{+0.3}_{-0.3}$  &  0.3  &  6  &  --  &  2 GCs per bin   \\
           &  Blue  &  FDS  &  $2.32^{+0.6}_{-0.6}$  &  $-3.14^{+0.4}_{-0.5}$  &  --  &  --  &  $7.58^{+0.4}_{-0.5}$  &  0.3  &  6  &  --  &  --   \\
           &  All  &  ACS  &  $32^{+200}_{-20}$  &  $-1.51^{+0.1}_{-0.1}$  &  $-3.13^{+0.2}_{-0.2}$  &  $0.274^{+0.05}_{-0.06}$  &  --  &  0  &  1.5  &  --  &  --   \\
           &  Red  &  ACS  &  $30240^{+200}_{-100}$  &  $2.34^{+0.6}_{-0.6}$  &  $-3.10^{+0.3}_{-0.4}$  &  $0.139^{+0.03}_{-0.04}$  &  --  &  0  &  1.5  &  --  &  1 GC per bin   \\
           &  Blue  &  ACS  &  $18^{+40}_{-10}$  &  $-1.706^{+0.1}_{-0.09}$  &  $-3.15^{+0.3}_{-0.3}$  &  $0.294^{+0.06}_{-0.07}$  &  --  &  0  &  1.5  &  --  &  --   \\
\hline
NGC\,1381  &  All  &  ACS  &  $2.47^{+0.3}_{-0.3}$  &  $-2.80^{+0.3}_{-0.3}$  &  --  &  --  &  --  &  0  &  1.5  &  --  &  --   \\
           &  Red  &  ACS  &  $0.30^{+0.2}_{-0.2}$  &  $-4.05^{+0.9}_{-1}$  &  --  &  --  &  --  &  0  &  1.5  &  --  &  1 GC per bin   \\
           &  Blue  &  ACS  &  $1.86^{+0.3}_{-0.3}$  &  $-2.65^{+0.3}_{-0.3}$  &  --  &  --  &  --  &  0  &  1.5  &  --  &  --   \\
\hline
NGC\,1387  &  All  &  FDS  &  $14.0^{+1}_{-1}$  &  $-2.82^{+0.4}_{-0.3}$  &  $-4.6^{+2}_{-20}$  &  $2.4^{+700}_{-2}$  &  $15.52^{+0.6}_{-1}$  &  0.5  &  7  &  --  &  --   \\
           &  Red  &  FDS  &  $5.81^{+0.6}_{-0.6}$  &  $-2.92^{+0.2}_{-0.2}$  &  --  &  --  &  $1.96^{+0.2}_{-0.3}$  &  0.3  &  7  &  24.5  &  --   \\
           &  Blue  &  FDS  &  $3.9^{+300}_{-4}$  &  $-1.6^{+2}_{-2}$  &  $-3.62^{+0.9}_{-6}$  &  $0.86^{+400}_{-0.9}$  &  $3.29^{+0.2}_{-0.3}$  &  0.3  &  7  &  24.5  &  --   \\
           &  All  &  ACS  &  $16.1^{+1}_{-1}$  &  $-3.08^{+0.2}_{-0.2}$  &  --  &  --  &  --  &  0.5  &  1.5  &  --  &  --   \\
           &  Red  &  ACS  &  $10.7^{+1}_{-1}$  &  $-3.05^{+0.2}_{-0.2}$  &  --  &  --  &  --  &  0.5  &  1.5  &  --  &  --   \\
           &  Blue  &  ACS  &  $4.36^{+0.7}_{-0.9}$  &  $-2.73^{+0.5}_{-0.5}$  &  --  &  --  &  --  &  0.5  &  1.5  &  --  &  --   \\
\hline
NGC\,1399  &  All  &  FDS  &  $21.6^{+2}_{-2}$  &  $-1.856^{+0.01}_{-0.01}$  &  $-4.49^{+0.6}_{-0.9}$  &  $11.03^{+0.6}_{-0.6}$  &  $13.52^{+0.1}_{-0.2}$  &  1  &  30  &  --  &  --   \\
           &  Red  &  FDS  &  $14.4^{+2}_{-2}$  &  $-2.003^{+0.02}_{-0.03}$  &  $-7.5^{+3}_{-7}$  &  $11.51^{+0.4}_{-0.5}$  &  $7.150^{+0.05}_{-0.06}$  &  1  &  30  &  --  &  --   \\
           &  Blue  &  FDS  &  $7.2^{+1}_{-1}$  &  $-1.647^{+0.05}_{-0.09}$  &  $-3.58^{+0.6}_{-0.8}$  &  $10.79^{+0.9}_{-1}$  &  $6.34^{+0.2}_{-0.2}$  &  1  &  30  &  --  &  --   \\
           &  All  &  ACS  &  $29.0^{+3}_{-4}$  &  $-1.81^{+0.1}_{-0.1}$  &  --  &  --  &  --  &  0.5  &  1.5  &  --  &  --   \\
           &  Red  &  ACS  &  $23.7^{+3}_{-3}$  &  $-1.97^{+0.2}_{-0.2}$  &  --  &  --  &  --  &  0.5  &  1.5  &  --  &  --   \\
           &  Blue  &  ACS  &  $5.0^{+2}_{-3}$  &  $-1.46^{+0.3}_{-0.3}$  &  --  &  --  &  --  &  0.5  &  1.5  &  --  &  --   \\
           &  All  &  spectro  &  $5.7^{+2}_{-2}$  &  $-1.78^{+0.2}_{-0.2}$  &  $-3.391^{+0.09}_{-0.1}$  &  $7.05^{+0.8}_{-0.9}$  &  --  &  1.5  &  25  &  --  &  --   \\
           &  Red  &  spectro  &  $4.3^{+1}_{-2}$  &  $-2.01^{+0.6}_{-0.2}$  &  $-3.86^{+0.2}_{-0.2}$  &  $7.5^{+1}_{-2}$  &  --  &  1.5  &  25  &  --  &  --   \\
           &  Blue  &  spectro  &  $2.22^{+0.8}_{-0.7}$  &  $-1.77^{+0.2}_{-0.2}$  &  $-3.18^{+0.1}_{-0.1}$  &  $8.2^{+3}_{-2}$  &  --  &  1.5  &  25  &  --  &  --   \\
\hline
NGC\,1404  &  All  &  spectro  &  $0.18^{+0.2}_{-0.1}$  &  $-0.5^{+6}_{-1}$  &  $-4.0^{+2}_{-20}$  &  $2.67^{+1}_{-0.9}$  &  $0.013^{+0.03}_{-0.03}$  &  0  &  10  &  --  &  --   \\
\hline
NGC\,1419  &  All  &  FDS  &  $23^{+200}_{-10}$  &  $-0.48^{+0.4}_{-0.3}$  &  $-3.01^{+0.3}_{-0.5}$  &  $0.566^{+0.09}_{-0.1}$  &  $14.57^{+0.3}_{-0.3}$  &  0  &  14  &  --  &  --   \\
           &  Red  &  FDS  &  $3.12^{+0.7}_{-0.7}$  &  $-3.83^{+0.6}_{-0.9}$  &  --  &  --  &  $7.44^{+0.2}_{-0.2}$  &  0.5  &  14  &  --  &  --   \\
           &  Blue  &  FDS  &  $2.99^{+0.9}_{-0.9}$  &  $-3.10^{+0.7}_{-1}$  &  --  &  --  &  $7.65^{+0.3}_{-0.4}$  &  0.5  &  14  &  --  &  --   \\
           &  All  &  ACS  &  $18.5^{+200}_{-9}$  &  $-1.50^{+0.1}_{-0.1}$  &  $-2.84^{+0.3}_{-0.3}$  &  $0.394^{+0.09}_{-0.1}$  &  --  &  0  &  1.5  &  --  &  --   \\
           &  Red  &  ACS  &  $2.5^{+6}_{-1}$  &  $-2.30^{+0.5}_{-0.5}$  &  $-3.68^{+0.8}_{-10}$  &  $0.51^{+0.8}_{-0.3}$  &  --  &  0  &  1.5  &  --  &  4 GCs per bin   \\
           &  Blue  &  ACS  &  $27^{+200}_{-20}$  &  $-0.93^{+0.2}_{-0.2}$  &  $-2.71^{+0.3}_{-0.3}$  &  $0.342^{+0.08}_{-0.09}$  &  --  &  0  &  1.5  &  --  &  --   \\
\hline
NGC\,1427  &  All  &  FDS  &  $10.3^{+3}_{-1}$  &  $-1.37^{+0.3}_{-0.2}$  &  $-3.23^{+0.4}_{-0.5}$  &  $1.35^{+0.2}_{-0.2}$  &  $13.43^{+0.4}_{-0.5}$  &  0  &  10  &  --  &  --   \\
           &  Red  &  FDS  &  $5.4^{+4}_{-2}$  &  $-1.67^{+0.8}_{-0.7}$  &  $-3.45^{+0.5}_{-1}$  &  $0.99^{+0.6}_{-0.2}$  &  $3.67^{+0.2}_{-0.2}$  &  0.2  &  10  &  25  &  --   \\
           &  Blue  &  FDS  &  $4.75^{+0.9}_{-0.9}$  &  $-1.54^{+0.4}_{-0.4}$  &  $-4.01^{+0.8}_{-7}$  &  $1.64^{+0.8}_{-0.4}$  &  $5.73^{+0.2}_{-0.2}$  &  0  &  10  &  --  &  --   \\
           &  All  &  ACS  &  $17.0^{+10}_{-3}$  &  $-1.97^{+0.2}_{-0.2}$  &  $-3.42^{+0.8}_{-20}$  &  $1.11^{+0.7}_{-0.4}$  &  --  &  0.2  &  1.5  &  --  &  --   \\
           &  Red  &  ACS  &  $9.1^{+7}_{-2}$  &  $-1.87^{+0.1}_{-0.2}$  &  $-4.6^{+1}_{-3}$  &  $0.96^{+0.2}_{-0.3}$  &  --  &  0  &  1.5  &  --  &  4 GCs per bin   \\
           &  Blue  &  ACS  &  $6.29^{+0.6}_{-0.6}$  &  $-2.15^{+0.2}_{-0.1}$  &  --  &  --  &  --  &  0.2  &  1.5  &  --  &  --   \\
\hline
NGC\,1428  &  All  &  ACS  &  $3.7^{+80}_{-2}$  &  $-1.81^{+0.4}_{-0.4}$  &  $-3.77^{+0.8}_{-1}$  &  $0.47^{+0.1}_{-0.2}$  &  --  &  0  &  1.5  &  --  &  1 GC per bin   \\
           &  Red  &  ACS  &  $0.37^{+4}_{-0.3}$  &  $-2.85^{+1}_{-0.6}$  &  $-8.4^{+7}_{-8}$  &  $0.71^{+400}_{-0.2}$  &  --  &  0  &  1.5  &  --  &  Break defined by a single point, 1 GC per bin   \\
           &  Blue  &  ACS  &  $130^{+300}_{-100}$  &  $2.82^{+1}_{-0.8}$  &  $-5.1^{+1}_{-2}$  &  $0.507^{+0.06}_{-0.09}$  &  --  &  0  &  1.5  &  --  &  1 GC per bin   \\
\hline
Stack\_8.0  &  All  &  FDS  &  $1.78^{+3}_{-0.9}$  &  $-1.54^{+0.2}_{-0.3}$  &  $-3.90^{+0.9}_{-10}$  &  $0.292^{+0.1}_{-0.05}$  &  $15.439^{+0.08}_{-0.07}$  &  0.0  &  8  &  --  &  Stacked objects: FDS10\_0014, FDS10\_0023, FDS10\_0077, FDS10\_0228, FDS10\_0302, FDS11\_0327, FDS12\_0197, FDS13\_0302, FDS14\_0073, FDS15\_0232, FDS15\_0245, FDS16\_0027, FDS16\_0075, FDS16\_0253, FDS17\_0167, FDS17\_0304, FDS17\_0343, FDS18\_0045, FDS1\_0145, FDS20\_0144, FDS20\_0212, FDS21\_0333, FDS31\_0129, FDS4\_0030, FDS4\_0053, FDS4\_0337, FDS6\_0170, FDS6\_0462, FDS7\_0158, FDS9\_0255, FDS9\_0492   \\
           &  Red  &  FDS  &  $13^{+100}_{-10}$  &  $-0.73^{+0.2}_{-0.1}$  &  $-2.99^{+0.4}_{-0.6}$  &  $0.088^{+0.02}_{-0.02}$  &  $7.410^{+0.04}_{-0.04}$  &  0.0  &  8  &  --  &  --   \\
           &  Blue  &  FDS  &  $18^{+100}_{-20}$  &  $-0.36^{+0.2}_{-0.2}$  &  $-3.06^{+0.4}_{-0.6}$  &  $0.140^{+0.03}_{-0.03}$  &  $7.952^{+0.04}_{-0.04}$  &  0.0  &  8  &  --  &  --   \\
\hline
Stack\_8.5  &  All  &  FDS  &  $6.6^{+5}_{-3}$  &  $-1.286^{+0.07}_{-0.08}$  &  $-3.50^{+0.4}_{-0.6}$  &  $0.260^{+0.04}_{-0.02}$  &  $15.463^{+0.07}_{-0.07}$  &  0.0  &  8  &  --  &  Stacked objects: FDS10\_0189, FDS11\_0069, FDS11\_0079, FDS11\_0155, FDS11\_0458, FDS12\_0194, FDS12\_0367, FDS13\_0042, FDS13\_0299, FDS15\_0384, FDS16\_0024, FDS16\_0417, FDS17\_0188, FDS19\_0380, FDS20\_0138, FDS31\_0042, FDS4\_0002, FDS4\_0061, FDS6\_0414, FDS6\_0455, FDS7\_0326   \\
           &  Red  &  FDS  &  $2.2^{+10}_{-1}$  &  $-1.42^{+0.2}_{-0.1}$  &  $-3.24^{+0.5}_{-0.8}$  &  $0.276^{+0.07}_{-0.07}$  &  $7.769^{+0.05}_{-0.04}$  &  0.0  &  8  &  --  &  --   \\
           &  Blue  &  FDS  &  $4.2^{+9}_{-2}$  &  $-1.214^{+0.1}_{-0.09}$  &  $-3.60^{+0.5}_{-0.7}$  &  $0.261^{+0.03}_{-0.03}$  &  $7.507^{+0.05}_{-0.05}$  &  0.0  &  8  &  --  &  --   \\
\hline
Stack\_9.0  &  All  &  FDS  &  $2.77^{+3}_{-0.9}$  &  $-1.65^{+0.2}_{-0.2}$  &  $-3.22^{+0.6}_{-1}$  &  $0.58^{+0.3}_{-0.3}$  &  $16.88^{+0.2}_{-0.2}$  &  0.0  &  8  &  --  &  Stacked objects: FDS11\_0279, FDS14\_0144, FDS15\_0417, FDS16\_0159, FDS20\_0334, FDS31\_0196, FDS4\_0000   \\
           &  Red  &  FDS  &  $8.8^{+200}_{-7}$  &  $-0.59^{+0.2}_{-0.2}$  &  $-4.24^{+0.9}_{-2}$  &  $0.358^{+0.07}_{-0.07}$  &  $8.24^{+0.1}_{-0.1}$  &  0.04  &  8  &  --  &  --   \\
           &  Blue  &  FDS  &  $1.31^{+100}_{-0.8}$  &  $-1.75^{+0.6}_{-0.5}$  &  $-2.37^{+0.3}_{-10}$  &  $0.31^{+500}_{-0.3}$  &  $8.63^{+0.2}_{-0.2}$  &  0.0  &  8  &  --  &  --   \\

\label{tab:allfits}
\end{longtable}
\end{landscape}
\twocolumn


\begin{table*}[h!]
\caption{Fits of sectors of the GCS of NGC\,1399.}              
\label{tab:sectors} 
\centering            
\begin{tabular}{lllllll}
\hline\hline
Sector &  $\rho_0$ & $a$ & $b$ & $r_\mathrm{br}$ & $r_\mathrm{min}$ & $r_\mathrm{max}$  \\
 & [arcmin$^{-3}$] &   &   & [arcmin] &  [arcmin] & [arcmin]   \\
\hline
$0\pm 22.5\degr$  &  $6.3^{+7}_{-4}$  &  $-1.73^{+0.9}_{-0.2}$  &  $-2.78^{+0.1}_{-0.2}$  &  $5.3^{+2}_{-2}$  &  1.5  &  25   \\
$45\pm 22.5\degr$  &  $6.6^{+6}_{-4}$  &  $-1.92^{+0.6}_{-0.3}$  &  $-4.28^{+0.4}_{-0.6}$  &  $8.0^{+1}_{-1}$  &  1.5  &  25   \\
$90\pm 22.5\degr$  &  $7.0^{+7}_{-4}$  &  $-2.03^{+0.6}_{-0.2}$  &  $-3.62^{+0.3}_{-0.3}$  &  $6.8^{+2}_{-2}$  &  1.5  &  25   \\
$135\pm 22.5\degr$  &  $7.6^{+10}_{-7}$  &  $-1.79^{+1}_{-0.3}$  &  $-3.24^{+0.2}_{-0.2}$  &  $4.6^{+1}_{-1}$  &  1.5  &  25   \\
$180\pm 22.5\degr$  &  $14.0^{+8}_{-6}$  &  $-2.534^{+0.08}_{-0.09}$  &  $-2.78^{+0.2}_{-0.3}$  &  $9.0^{+800}_{-9}$  &  1.5  &  25   \\
$225\pm 22.5\degr$  &  $10.1^{+9}_{-6}$  &  $-2.20^{+0.5}_{-0.2}$  &  $-4.58^{+0.7}_{-1}$  &  $11.1^{+2}_{-3}$  &  1.5  &  25   \\
$270\pm 22.5\degr$  &  $4.1^{+4}_{-2}$  &  $-1.76^{+2}_{-0.4}$  &  $-3.24^{+0.3}_{-0.5}$  &  $7.6^{+10}_{-3}$  &  1.5  &  18   \\
$315\pm 22.5\degr$  &  $0.89^{+3}_{-0.8}$  &  $-0.3^{+3}_{-1}$  &  $-2.99^{+0.2}_{-0.2}$  &  $4.5^{+2}_{-1}$  &  1.5  &  25   \\

\hline      
\end{tabular}
\end{table*}

\begin{table*}[h!]
\caption{Fits of radial velocity slices of the GCS of NGC\,1399.}              
\label{tab:velslices} 
\centering            
\begin{tabular}{lllllll}
\hline\hline
 $\left| RV_\mathrm{GC}- RV_\mathrm{gal}\right|$ &  $\rho_0$ & $a$ & $b$ & $r_\mathrm{br}$ & $r_\mathrm{min}$ & $r_\mathrm{max}$  \\\relax
[km\,s$^{-1}$] & [arcmin$^{-3}$] &   &   & [arcmin] &  [arcmin] & [arcmin]   \\
\hline
$0-100$  &  $3.00^{+1}_{-0.9}$  &  $-2.24^{+0.2}_{-0.2}$  &  $-3.72^{+0.3}_{-0.3}$  &  $11.0^{+2}_{-2}$  &  1.5  &  25   \\
$100-250$  &  $4.3^{+1}_{-1}$  &  $-2.33^{+0.2}_{-0.2}$  &  $-3.52^{+0.3}_{-0.3}$  &  $8.8^{+2}_{-3}$  &  1.5  &  25   \\
$250-400$  &  $0.71^{+0.6}_{-0.4}$  &  $-1.45^{+0.6}_{-0.4}$  &  $-3.50^{+0.2}_{-0.3}$  &  $6.6^{+1}_{-1}$  &  1.5  &  25   \\
$400-1400$  &  $0.25^{+0.5}_{-0.2}$  &  $-0.66^{+1}_{-0.8}$  &  $-3.18^{+0.1}_{-0.1}$  &  $5.09^{+2}_{-0.7}$  &  1.5  &  25   \\

\hline      
\end{tabular}
\end{table*}

\clearpage
\section{The influence of magnitude cut on the measurement of the break radius}
\label{app:magcut}
\begin{figure}
        \resizebox{\hsize}{!}{\includegraphics{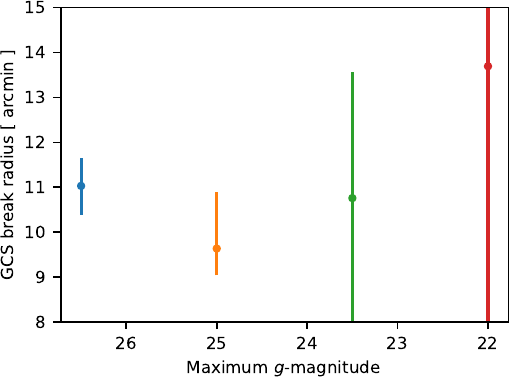}}
        \caption{Demonstration  that the break radius does not depend on the applied magnitude cut. The data for the galaxy NGC\,1399 were used.}
        \label{fig:magcut}
\end{figure}

{Before fitting the GC density profiles, we sometimes had to remove from the analyzed sample of GC candidates the faint objects, such that we mitigate the problems with the contamination by the light of the host galaxy. In this appendix, we investigate whether imposing the magnitude cut has any impact on the break radius. 

We used the galaxy NGC\,1399 for this test. Since it hosts the highest number of GC candidates, it is possible to measure the break radius relatively precisely even if most of them are excluded from the fit. 

The results are shown in \fig{magcut}. The values of the break radius agree with each other within the uncertainty limits. The faintest GC candidates in the sample are around 26 mag.}

\section{The effect of ellipticity of the GCS on the extracted profile of its surface density}
\label{app:ellipticity}

In our work, we used circular annuli to extract the radial profiles of surface density of GC candidates. We neglected the fact that the distribution of GCs around a galaxy can be flattened. It has been found that the ellipticity of GCs, particularly of the red ones, follows the ellipticity of the host galaxy (see \citealp{brodie06} for a review). For example the galaxy NGC\,1380, one of the galaxies of our sample, has a rather high ellipticity of 0.5 \citep{kissler97b}. We thus explore here how the  profile of the surface denstity of GC candidates, extracted in our way, would change if the GC system would actually be elliptical.

The real data for the galaxy NGC\,1380 are not suitable for this task. The break radius is located at the border of the ACS FOV. In the FDS data the break is located  inside the region that was excluded because of the {problems with the contamination by the light of the host galaxy}, see \fig{g8}. We therefore decided to use artificial data for the test.

We constructed a circular and an elliptical GCS. The positions of 500 GCs were initially generated in three dimensions. The positions were drawn from a spherical distribution whose volume density followed a broken power law described by the parameters $a=-1.7$, $b= -3.4$, $\rbr = 1\arcmin$. These are the typical parameters for the real galaxies (\tab{stat}). For constructing the circular GCS, the Cartesian coordinates $x_c$ and $y_c$ of the modeled GCs were then directly considered the on-sky positions of the GCs. For the elliptical GCS, the on-sky coordinates of the GCs were constructed as $x_e= \sqrt{\epsilon}x_c$ and $y_c/\sqrt{\epsilon}$, where $\epsilon$ is the ellipticity of the GCS, chosen to be $\epsilon = 0.5$. Finally, we added contaminants in the list of the modeled GCs candidates. The contaminants were assumed to have a uniform distribution on the sky with a mean density of 1\,arcsec$^{-2}$. 

\begin{figure}
        \resizebox{\hsize}{!}{\includegraphics{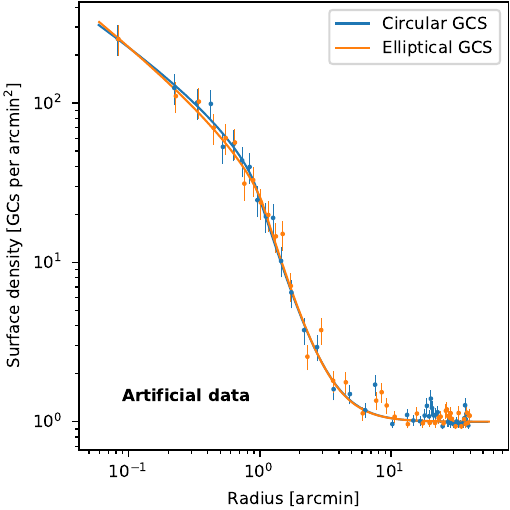}}
        \caption{Demonstration  that the ellipticity of the distribution of the GCs in the GCS does not influence the  extracted profile of surface density of GC candidates substantially. Artificial data. The flattened GCS has the on-sky ellipticity of 0.5. }
        \label{fig:ellipticity}
\end{figure}

Both the circular and elliptical GCSs were then analyzed in the same way as the observational data. Figure~\ref{fig:ellipticity} shows the comparison of the extracted profiles of surface density of the modeled GC candidates and fits by the broken power laws (\equ{bplb}). Both the extracted surface density profiles and the best-fit curves are very similar. This is confirmed by consistency of the values of the fitted parameters within their uncertainty limits, as shown in \tab{ellipticity}. The last column of the table, listing the true values of parameters of the GCS density profile and the density of contaminants,  confirms that our method is able to recover these parameters correctly. 

\begin{table}
\caption{Fitted parameters of the density profile of the GCS in our test of the influence of the ellipticity of the GCSs.}
\label{tab:ellipticity}    
\centering                              
\begin{tabular}{lp{1.5cm}p{1.5cm}l}
\csname @@input\endcsname
"table_ellipticitytest.txt"
\end{tabular}
\end{table}

\section{Inhomogeneity of the sensitivity of the FDS survey}
\label{app:sensitivity}
As we mentioned in \sect{fds}, the catalog of FDS GC candidates shows spatial inhomogeities that form a tile-like pattern, see \fig{tiles}. The pattern seems to be arising because of a varying sensitivity of the survey both between the individual tiles of the mosaic and inside of the individual tiles. We were not able to remove the tile pattern even if excluding all but the brightest sources. Here we investigate whether the sensitivity variations could affect our measurements of the GCS density profiles. We focused on the cases of the galaxy with the richest and most extended GCS, that is NGC\,1399, because it would be affected most by the large-scale sensitivity variations.  This galaxy is located close to the center of one of the tiles.

\begin{figure}
        \resizebox{\hsize}{!}{\includegraphics{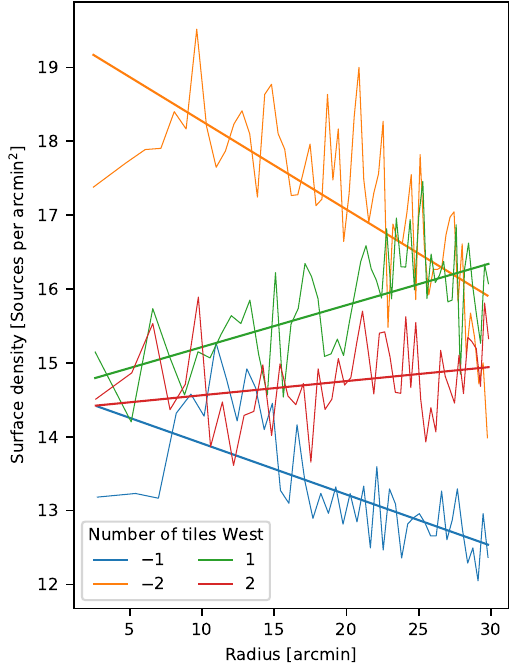}}
        \caption{Variation of the density of sources with distance from the centers of four tiles of the FDS mosaic. These are the two tiles to the west of the tile containing NGC\,1399 and the two tiles to the east. }
        \label{fig:sensitivityprofiles}
\end{figure}

To this end, we constructed radial profiles of the density of sources in four tiles close to NGC\,1399, centered approximately on the centers of the respective tiles. In particular, we used the two tiles adjacent from the west to the tile containing the galaxy, and the two tiles adjacent from the east, along the line of a constant declination, see \fig{tiles}. These tiles do not contain any galaxies with substantial GCSs. The profiles were extracted up to the distance of 30\arcmin, which is the distance that was used for constructing the GCS profile of NGC\,1399. The regions occupied by the GCSs of intermediate galaxies were masked in the way described in \sect{extracting} before extracting the profiles. The radial bins were chosen such that each contains 800 sources. The measured profiles are shown in \fig{sensitivityprofiles} as the thin jiggles lines. The figure also shows by the thicker lines linear fits to the extracted profiles. The figure shows that the sensitivity at the outermost radius is different by around 10\% from the central sensitivity. The sensitivity can both increase and decrease with the distance from the center of a given tile. The central surface densities of sources can differ by several tens of percent between the individual tiles. 

We further investigated quantitatively what is the effect of the sensitivity variations on our estimates of the parameters of the broken power-law profiles of the GCSs. We assumed that the sensitivity variations visible in \fig{tiles} affect GCs and contaminating sources in the same way. As a first approximation, we assumed that the sensitivity is a linear function of the distance from the center of the tile. We thus took the measured  profile of the source density of NGC\,1399 $\Sigma(R)$ and transformed it to:
\begin{equation}
    \Sigma_2(R) = \Sigma (R)\,(1+vR/R_\mathrm{max}),     
\end{equation}
where $R_\mathrm{max}=30\arcmin$ is the radius of the last bin measured for NGC\,1399. The parameter $v$ quantifies the magnitude of the variation of sensitivity. It is the ratio of the sensitivities in the center of the tile (which is the same as the center of NGC\,1399) and at $R_\mathrm{max}$. The profile $\Sigma_2$ was then fitted by the projected broken power law, in the same way we did with the real data.

The results are presented in \tab{sensitivityvar}. The table reveals that the position of the break radius is not very sensitive to this type of spatial variation of sensitivity, even for very strong variations of sensitivity. On the other hand, a realistic variation of the sensitivity of plus or minus 10\%  already has a substantial effect on the measured outer slope $b$. 

This sheds doubt on the measured value of the slopes $b$ for the two galaxies with the most extended GCSs, that is NGC\,1399 and NGC\,1316. This is why we marked in \tab{finalfits} the values of these parameters for these two galaxies as suspicious. The GCSs of the other galaxies in our sample are much smaller (\fig{tiles} and the values of break radii in \tab{finalfits}). The profiles of these galaxies would not be affected much.

\begin{table}
\caption{Simulation of the effect of the large-scale variations of the sensitivity of the FDS survey on the derived parameters of the GCS of NGC\,1399.}
\label{tab:sensitivityvar}    
\centering                              
\begin{tabular}{l|lllll}\hline
$v$ &  $\rho_0$ & $a$ & $b$ & $r_\mathrm{br}$ & $\gamma$ \\\hline\hline
\csname @@input\endcsname
"table_backgroundtest.txt"
\hline
\end{tabular}
\end{table}


\section{Linear fits of the relation between the break radii and $a_0$ radii}
\label{app:linfit}
{We made linear fits of the relation between the $a_0$ and break radii, and recover the intrinsic scatter of the correlation for several datasets.} More specifically, we assumed that at a given \rbr, $\log_{10}\req$ follows a Gaussian distribution with a mean of $C_1\log_{10}\rbr+C_0$ and a standard deviation of $\sigma_\mathrm{int}$. This implies that at a given \rbr, $\req$ follows the lognormal distribution
\begin{equation}
\begin{split}
    G(\rbr, \req) = & \frac{1}{\sqrt{2\pi}\ln(10)\sigma_\mathrm{int}\rbr}\\ & \exp\left[ - \frac{\left\{\log_{10}[\req]-\left(C_1\log_{10}[\rbr]+C_0\right)\right\}^2 }{2\sigma_\mathrm{int}^2} \right].
\end{split}
\end{equation}
We had to take into account the uncertainty in the estimates of the break and $a_0$ radii. For the $i$-th galaxy, we assumed for the pair of variables $(\rbr,\req)$ the distribution function:
\begin{equation}
    g_{\mathrm{mes},i}(\rbr, \req) = g(\rbr, \overline{r_{\mathrm{br},i}}, \sigma^+_{r_{\mathrm{br},i}}, \sigma^-_{r_{\mathrm{br},i}}) g(\req, \overline{r_{\mathrm{eq},i}}, \sigma^+_{r_{\mathrm{eq},i}}, \sigma^-_{r_{\mathrm{eq},i}}),
\end{equation}
where $g$ was defined in \equ{joint}. The estimated slope, intercept and intrinsic scatter of the correlation between \rbr and \req were found by maximizing the likelihood function:
\begin{equation}
    \mathcal{L}(C_1,C_0,\sigma_\mathrm{int}) = \prod_i [G(\rbr,\req)*g_{\mathrm{mes},i}(\rbr, \req)](\overline{r_{\mathrm{br},i}}, \overline{r_{\mathrm{eq},i}}),
\end{equation}
where the asterisk denotes convolution over both variables. The intrinsic scatter was restricted to be at least 0.01\,dex, which is much less than the measurement uncertainties of \rbr and \req, in order to avoid numerical problems.

This procedure was applied to several datasets. First, to the Fornax Cluster data presented in this paper, for the total, red and blue GC populations, then it was applied to the union of the data from this paper and from BSR19 for the total GC populations. For NGC\,1399, that appears in both of the samples of this paper and BSR19, we used only the datapoint from this paper. Finally, we divided the union dataset in two parts according to whether the GCS has a break radius greater or lower than 20\,kpc, since, as we {noted in \sect{equality},} the trend might be different in these two regimes. All of this was done for the Newtonian and MOND $a_0$ radii.  The results are stated in \tab{rbrreqfits}. The values are in many cases not consistent with the one-to-one relation between the break and $a_0$ radii.

\begin{figure}
        \resizebox{\hsize}{!}{\includegraphics{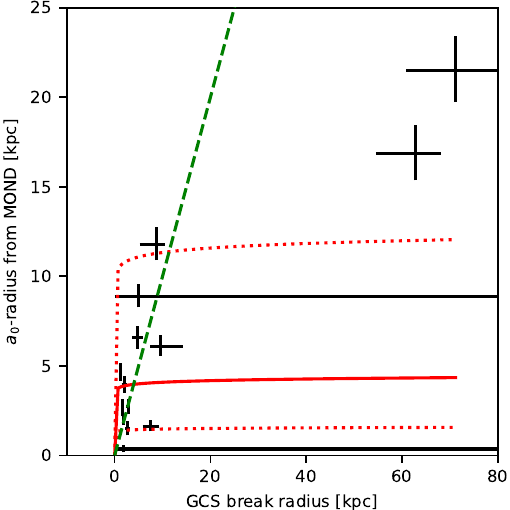}}
        \caption{Fit of the relation of the observed break radius and the MOND $a_0$ radius by a law in the form of $\log_{10}\rbr = C_1\log_{10}\req +C_0$. The black crosses represent the uncertainty intervals of the data. The full red line show the fit and the dotted red lines the fitted intrinsic scatter of the relation. The green dashed line indicates the one-to-one relation. The deviation of the best-fit curve from the one-to-one relation is driven by a few data points.}
        \label{fig:fitradii}
\end{figure}

We demonstrate the origin of this disagreement  on the example of the blue GC population and the MOND $a_0$ radii, for which the fit indicated a zero slope. The fitted data and the best-fit curve are shown in \fig{fitradii}. The full red line indicates the best-fit linear fit of the relation between $\log_{10}\req$ and $\log_{10}\rbr$ and the dotted lines the fitted intrinsic scatter. The green dashed line indicates the one-to-one relation. A likelihood ratio test indeed indicates that the best-fit model is significantly better than the green line. It seems that the deviation of the fitted curve from the one-to-one relation is caused by a few datapoints. At this moment, it is not clear whether these points deviate because they are outliers (either because of observational errors or because the galaxies experienced an unusual event) or the true correlation between the break and $a_0$ radii is not a one-to-one correlation exactly. This will become clearer when more data points become available. For the moment, we were able to prove the match between the break and $a_0$ radii only in the sense that they agree within a factor of a about two, as demonstrated {in \sect{equality}}.

\begin{table}
\caption{Fits of the relation $\log_{10}\rbr = C_1\log_{10}r_\mathrm{eq}+C_0$ and of its intrinsic scatter for different datasets.} 
\label{tab:rbrreqfits}      
\centering                              
\begin{tabular}{llll}
\hline\hline                
 Dataset & $C_1$ & $C_0$ & $\sigma_\mathrm{int}/\mathrm{kpc}$[dex] \\ 
\hline
\csname @@input\endcsname "table_rbr_ra0_loglog_logsca_final.txt"
\hline                                  
\end{tabular}
\end{table}

\end{appendix}

\end{document}